\documentclass[floatfix,letterpaper,longbibliography,nofootinbib,aps,pre,preprint,superscriptaddress,showkeys,showpacs,12pt]{revtex4-1}
\usepackage{amsfonts}
\usepackage{amsmath}
\usepackage{amssymb}
\usepackage{mathtools}
\usepackage{graphicx}
\usepackage{longtable}
\usepackage{hyperref}
\usepackage{bm}
\usepackage{breqn}
\usepackage{url}
\usepackage[caption=false]{subfig}
\usepackage{float}
\usepackage{placeins}
\newcommand\numberthis{\addtocounter{equation}{1}\tag{\theequation}}
\numberwithin{equation}{section}
\begin{document}

\title{Robust Computation of Dipole Electromagnetic Fields in Arbitrarily-Anisotropic, Planar-Stratified Environments}
\date{\today}
\author{Kamalesh Sainath}
\email{sainath.1@osu.edu}
\author{Fernando L. Teixeira}
\email{teixeira@ece.osu.edu}
\affiliation{The Ohio State University: ElectroScience Laboratory}
\altaffiliation[Address: ]{1330 Kinnear Road, Columbus, Ohio, USA 43212}
\author{Burkay Donderici}
\email{burkay.donderici@halliburton.com}
\affiliation{Halliburton: Sensor Physics and Technology}
\altaffiliation[Address: ]{3000 N. Sam Houston Pkwy E, Houston, TX, USA 77032}

\begin{abstract}
\noindent

We develop a general-purpose formulation, based on two-dimensional spectral integrals, for computing electromagnetic fields produced by arbitrarily-oriented dipoles in planar-stratified environments, where each layer may exhibit arbitrary and independent anisotropy in both the (complex) permittivity and permeability. Among the salient features of our formulation are ($i$) computation of eigenmodes (characteristic plane waves) supported in arbitrarily anisotropic media in a numerically robust fashion, ($ii$) implementation of an $hp$-adaptive refinement for the numerical integration to evaluate the radiation and weakly-evanescent spectra contributions, and ($iii$) development of an adaptive extension of an integral convergence acceleration technique to compute the strongly-evanescent spectrum contribution. While other semianalytic techniques exist to solve this problem, none have full applicability to media exhibiting arbitrary double anisotropies in each layer, where one must account for the whole range of possible phenomena such as mode coupling at interfaces and non-reciprocal mode propagation. Brute-force numerical methods can tackle this problem but only at a much higher computational cost. The present formulation provides an efficient and robust technique for field computation in arbitrary planar-stratified environments. We demonstrate the formulation for a number of problems related to geophysical exploration.

\end{abstract}


\maketitle
\section{Introduction}
The study of electromagnetic fields produced by dipole sources in planar-stratified environments with anisotropic layers is pertinent to many applications such as geophysical prospection~\cite{wei,anderson1,howard,zhdanov,wang,moran1,tang}, microwave remote sensing~\cite{jehle}, ground-penetrating radar~\cite{lambot1,lambot2}, optical field focusing~\cite{jain}, antenna design~\cite{pozar,pozar2}, microwave circuits~\cite{mosig2}, and plasma physics~\cite{paulus}. For this problem class, one can exploit the planar symmetry and employ pseudo-analytical approaches based upon embedding spectral Green's Function kernels within Fourier-type integrals to compute the space-domain fields~\cite{sommerfeld,mich3,mich4}. A crucial aspect then becomes how to efficiently compute such integrals~\cite{mich1,mich2,mosig1,mosig4,mich5}. Based on the specific characteristics of the planar-stratified environment(s) considered, efficient, case-specific methods arise. For example, when one assumes isotropic layers so that no TE$_z$/TM$_z$ mode-coupling occurs at the planar interfaces, the original vector problem can be reduced to a set of scalar problems whose mixed domain Green's functions (i.e. those functions having ($k_x,k_y,z$) dependence) are either the primary kernels in integral representations of the Green's dyads (e.g. ``transmission-line"-type Green's functions~\cite{mich1,mich3,mich4}) or the field components themselves (e.g. free-space Green's function~\cite{chewch2}). Alternatively, when each layer exhibits azimuthal symmetry in its material properties, one can transform two-dimensional, infinite-range Fourier integrals into one-dimensional, semi-infinite range Sommerfeld integrals~\cite{mich2,mich3,mosig1,mich4,sommerfeld,mosig4}.
For layers with arbitrary anisotropy, however, neither of the above simplifications apply, and a more general formulation is required.

Irrespective of the integral representation used, the following challenges exist concerning their numerical evaluation~\cite{chewch2,mich1}: (1) The presence of branch-points/branch-cuts associated with semi-infinite and infinite-thickness layers, (2) the presence of poles associated with slab- and interface-guided modes, and (3) an oscillatory integrand that demands adequate sampling and whose exponential decay rate reduces with decreasing source-observer depth separation~\cite{mosig1}.
Among the approaches to address these issues one can cite (1) direct numerical evaluation, possibly combined with integral acceleration techniques \cite{chew2,mich1,mich2,mosig1,mosig3,mosig4}, (2) asymptotic approximation of the space-domain field \cite{chewch2}, and (3) approximation of the mixed-domain integrand via a sum of analytically invertible ``images" \cite{mich1,mich5,caboussat}.
While image-approximation and asymptotic methods exhibit faster solution time, they are fundamentally approximate methods that either (resp.) (1) require user intervention in performing a-priori ``fine-tuning", have medium-dependent applicability, and lack tight error-control \cite{mich1,mosig4}, or (2) have a limited range of applicability in terms of admitted medium classes and source/observer locations \cite{chewch2}.

Since our focus is on the general applicability and robustness of the algorithm (and not on the optimality for a specific class of layer arrangements, medium parameters, and source-observer geometries), we adopt a direct numerical integration methodology based on 2-D, infinite-range Fourier-type integrals. Some key ingredients of the present formulation are:
\begin{enumerate}
\item A numerically-balanced recasting of the state matrix \cite{chewch2} to enable the accurate computation of the eigenmodes supported in media exhibiting \emph{arbitrary} anisotropy (e.g. isotropic, uniaxial, biaxial, gyrotropic).
\item Closed-form eigenmode formulations for isotropic and reciprocal, electrically uniaxial media that significantly reduce eigenmode solution time (versus the state matrix method), obviate numerical overflow, and yield higher-precision results versus prior (canonical) formulations in \cite{chewch2,felsen}.
\item A numerically stable method to decompose degenerate modes produced by sources in isotropic layers.
\item A multi-level, error-controlled, adaptive $hp$ refinement procedure to evaluate the radiation/weakly evanescent spectral field contributions, employing nested Kronrod-Gauss quadrature rules to reduce computation time.
\item Adaptive extension of the original Method of Weighted Averages (MWA) \cite{mich2,mosig3} and its application to accelerating the numerical evaluation of \emph{infinite-range, 2-D Fourier-type integrals} concerning environments containing media with \emph{arbitrary anisotropy and loss}.
\end{enumerate}
 Section \ref{sec2} overviews the formulation. Section \ref{sec3} contains an analytical derivation of the mixed-domain, vector-valued integrand $ \bold{W}_L(k_x,k_y;z)$ of the 2-D Fourier integral. Section \ref{sec4} exhibits an efficient numerical algorithm to compute the (inner) $k_x$ integral in Eq. (\ref{V1}) (note that this discussion applies, in dual fashion, to the $k_y$ integral).

The appendix summarizes the conventions and notation used in this paper.

\section{\label{sec2}Formulation Overview}
   \begin{figure}[H]
       \centering
       \vspace{-20pt}
    \includegraphics[width=4in]{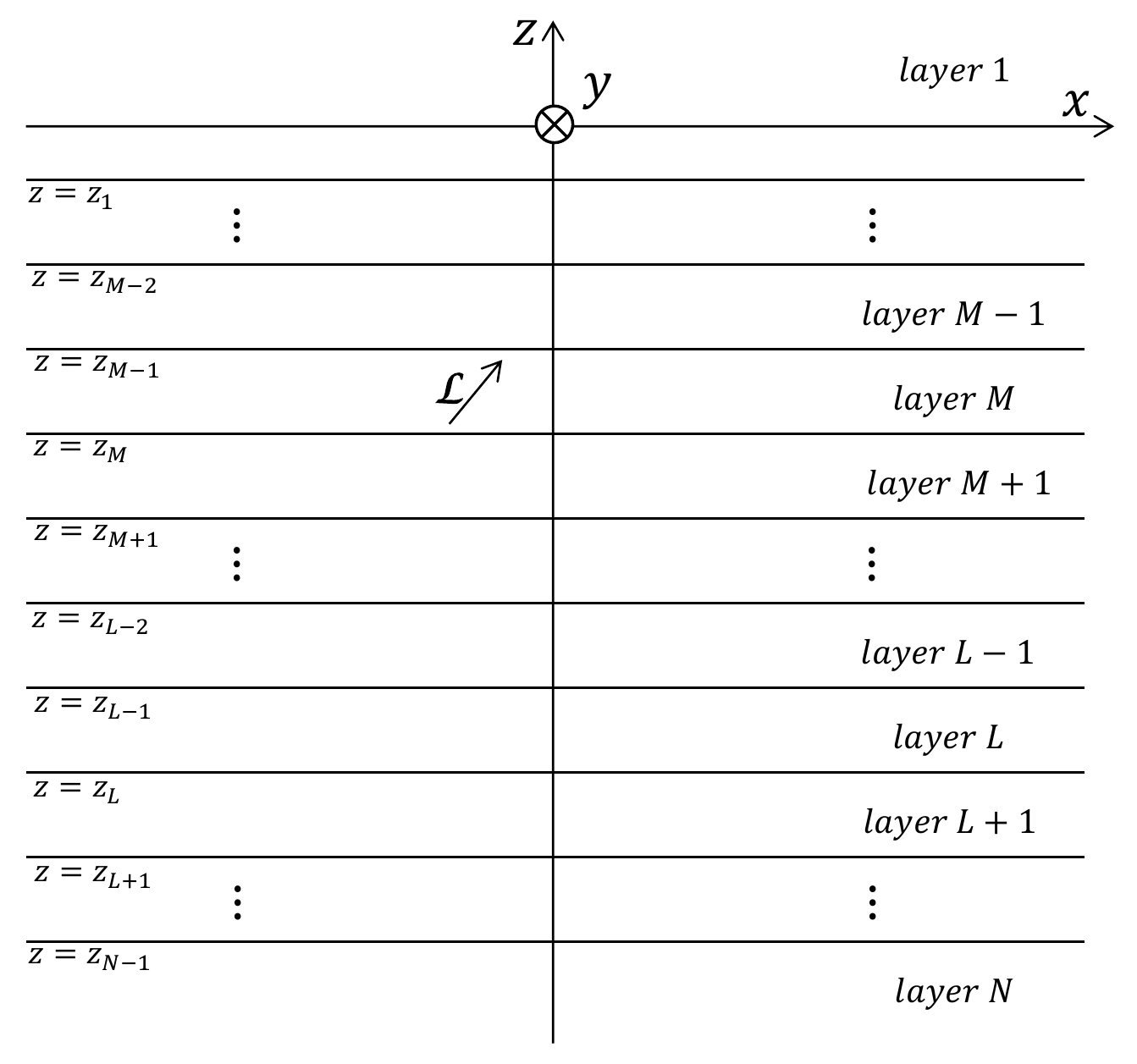}
    \vspace{-20pt}
    \caption{\label{Fig1}\small Layer $M$ contains the source point $\bold{r}'$ and layer $L$ contains the observation point $\bold{r}$. The dipole source $\bm{\mathcal{L}}$ can be either electric or magnetic.}
    \end{figure}
Our problem concerns computing the electromagnetic field at $\bold{r}$ produced by an elementary/Herztian dipole source which radiates at frequency $\omega$ within a planar-stratified, anisotropic environment. We assume $N$ layers stratified along the $z$ axis as depicted in Figure \ref{Fig1}, each with (complex-valued) material tensors $\boldsymbol{\bar{\epsilon}}_c$ and $\boldsymbol{\bar{\mu}}_c$ exhibiting independent and arbitrary anisotropy\footnote{We assume the material tensors to be diagonalizable, as this facilitates using plane wave fields as a basis to synthesize the field solution. Since all naturally occurring media possess diagonalizable material tensors, this constraint is not a practical concern and thus warrants no further discussion.}, that is
\begin{equation} \boldsymbol{\bar{\epsilon}}_c  = \begin{bmatrix} \epsilon_{xx} & \epsilon_{xy} &  \epsilon_{xz}\\  \epsilon_{yx} & \epsilon_{yy} & \epsilon_{yz}  \\
\epsilon_{zx} & \epsilon_{zy} & \epsilon_{zz}
\end{bmatrix}, \ \ \boldsymbol{\bar{\mu}}_c  = \begin{bmatrix} \mu_{xx} & \mu_{xy} &  \mu_{xz}\\  \mu_{yx} & \mu_{yy} & \mu_{yz}  \\
\mu_{zx} & \mu_{zy} & \mu_{zz} \end{bmatrix} \end{equation}
being simultaneous full, complex-valued tensors that can be different for each layer. With this in mind, Maxwell's equations in a homogeneous region with impressed electric and (equivalent) magnetic current densities $\bm{\mathcal{J}}$ and $\bm{\mathcal{M}}$ (resp.), as well as impressed volumetric electric and (equivalent) magnetic charge densities $\rho_v$ and $\rho_m$ (resp.), write as
       \begin{align} \nabla \times \bm{\mathcal{E}}  &=  i\omega \boldsymbol{\bar{\mu}}_c \cdot \bm{\mathcal{H}}-\bm{\mathcal{M}} \numberthis   \label{ME1}\\
       \nabla \times \bm{\mathcal{H}}  &=  \bm{\mathcal{J}}-i\omega \boldsymbol{\bar{\epsilon}}_c \cdot \bm{\mathcal{E}} \numberthis  \label{ME2}\\
        \nabla \cdot ( \boldsymbol{\bar{\epsilon}}_c \cdot \bm{\mathcal{E}} )  &=  \rho_v \numberthis  \label{ME3} \\
        \nabla \cdot ( \boldsymbol{\bar{\mu}}_c \cdot \bm{\mathcal{H}} )  &= \rho_m \numberthis \label{ME4} \end{align}
     After multiplying Eq. (\ref{ME1}) by \( \nabla \times \boldsymbol{\bar{\mu}}_c^{-1} \cdot \) and using Eq. (\ref{ME2}), one has \cite{chewch1}:
     \begin{equation}\label{ME5A} \left[\nabla \times \left(\boldsymbol{\bar{\mu}}_c^{-1} \cdot \nabla \times \right) - \omega^2 \boldsymbol{\bar{\epsilon}}_c \cdot\right]\bm{\mathcal{E}}=i\omega\bm{\mathcal{J}}-\nabla \times \boldsymbol{\bar{\mu}}_c^{-1} \cdot \bm{\mathcal{M}} \end{equation}
     Alternatively, defining the tensor-valued vector wave operator as
     \begin{equation}\label{WaveOp} \bm{\mathcal{\bar{A}}}=\nabla \times \boldsymbol{\bar{\mu}}_r^{-1} \cdot \nabla \times - k_o^2 \boldsymbol{\bar{\epsilon}}_r \cdot \end{equation}
     one can re-express Eq. (\ref{ME5A}) as
     \begin{equation}\label{ME5B}  \bm{\mathcal{\bar{A}}} \cdot \bm{\mathcal{E}} = ik_o\eta_o\bm{\mathcal{J}}-\nabla \times \boldsymbol{\bar{\mu}}_r^{-1} \cdot \bm{\mathcal{M}} \end{equation}
     Now, define a three-dimensional Fourier Transform (FT) pair as:
     \begin{align}\label{FT3D} \ \bold{\tilde{E}}(\bold{k}) \ &= \ \iiint\limits_{-\infty}^{+\infty} \bm{\mathcal{E}}(\bold{r}) \, \mathrm{e}^{-i\bold{k} \cdot \bold{r}} \, \mathrm{d}x \, \mathrm{d}y \, \mathrm{d}z \\
     \label{IFT3D} \ \bm{\mathcal{E}}(\bold{r}) \ &= \ \left(\frac{1}{2\pi}\right)^3\iiint\limits_{-\infty}^{+\infty} \bold{\tilde{E}}(\bold{k}) \, \mathrm{e}^{i\bold{k} \cdot \bold{r}} \, \mathrm{d}k_x \,  \mathrm{d}k_y \, \mathrm{d}k_z \end{align}
with \(\bold{r}=(x,y,z) \ \mathrm{and} \ \bold{k}=(k_x,k_y,k_z) \),
and similarly for all other field and source quantities. Now, assuming an electric or magnetic dipole source (resp.), one has $\bm{\mathcal{J}}=\boldsymbol{\hat{a}}J_o\delta\left(\bold{r}-\bold{r}'\right)$ or $\bm{\mathcal{M}}=\boldsymbol{\hat{a}}M_o\delta\left(\bold{r}-\bold{r}'\right)$ in the space domain and $\bold{\tilde{J}}=\bold{\hat{a}}J_o$ or $\bold{\tilde{M}}=\bold{\hat{a}} M_o$ in the Fourier domain.
To determine the spectral-domain fields, we first write the inverse of $\bold{\tilde{\bar{A}}}$ as
     \(\mathrm{inv}(\bold{\tilde{\bar{A}}})= \mathrm{adj}(\bold{\tilde{\bar{A}}})/\det(\bold{\tilde{\bar{A}}}) \), where
 \(\mathrm{adj}(\bold{\tilde{\bar{A}}})\) is the adjugate matrix (\emph{not} the conjugate-transpose matrix) \cite{adjugate}. The determinant $\det(\bold{\tilde{\bar{A}}})= g_o(k_z-\tilde{k}_{1z})(k_z-\tilde{k}_{2z})(k_z-\tilde{k}_{3z})(k_z-\tilde{k}_{4z})$, where $g_o= \epsilon_{zz}k_o^2(\mu_{xy}\mu_{yx}-\mu_{xx}\mu_{yy})$, is a fourth-order polynomial in $k_z$. Next, define the spectral Green's dyad operators
$\bold{\tilde{\bar{G}}}_{ee}(\bold{k};\bold{r}')=\mathrm{e}^{-i\bold{k} \cdot \bold{r}'}\mathrm{inv}\left(\bold{\tilde{\bar{A}}}\right)$ and $
\bold{\tilde{\bar{G}}}_{em}(\bold{k};\bold{r}')=\mathrm{e}^{-i\bold{k} \cdot \bold{r}'}\mathrm{inv}\left(\bold{\tilde{\bar{A}}}\right)\cdot \tilde{\nabla} \times$
 that (resp.) map electric and magnetic sources to the spectral electric field as follows:
$\bold{\tilde{E}}(\bold{k})=ik_o\eta_o\bold{\tilde{\bar{G}}}_{ee} \cdot \bold{\tilde{J}}$ and
$\bold{\tilde{E}}(\bold{k})=-\bold{\tilde{\bar{G}}}_{em} \cdot \boldsymbol{\bar{\mu}}_r^{-1} \cdot \bold{\tilde{M}}$.

In a homogeneous medium, the integral along $k_z$ in Eq. (\ref{IFT3D}) can be performed analytically using the Residue Theorem. The vector-valued residues are the four supported eigenmode electric fields having propagation constants corresponding to the four roots of $\det(\bold{\tilde{\bar{A}}})$, in terms of which we have the following generic expression for the space-domain (direct) electric field $\bm{\mathcal{E}}_d(\bold{r})$:
\begin{multline} \bm{\mathcal{E}}_d(\bold{r})=\frac{i}{(2\pi)^{2}} \iint \limits_{-\infty}^{+\infty}
\left[u(z-z')\sum_{n=1}^2{\tilde{a}_{n}\bold{\tilde{e}}_{n}\mathrm{e}^{i\tilde{k}_{nz}(z-z')}}+ u(z'-z)\sum_{n=3}^4{\tilde{a}_{n}\bold{\tilde{e}}_{n}\mathrm{e}^{i\tilde{k}_{nz}(z-z')}}
\right] \times \\
 \mathrm{e}^{ik_x(x-x')+ik_y(y-y')} \, \mathrm{d}k_x \, \mathrm{d}k_y \numberthis \label{EHM2a} \end{multline}
where the $\{ \bold{\tilde{e}}_{n}(k_x,k_y) \}$ are unit-norm eigenmode electric field vectors and the $\{ \tilde{a}_{n}(k_x,k_y) \}$ are (source dependent) modal amplitudes associated with the four eigenvalues (i.e. poles of $\mathrm{inv}(\bold{\tilde{\bar{A}}})$) $\{ \tilde{k}_{nz} \}$. In the multi-layer case, with $\bold{r}'$ in layer $M$ and $\bold{r}$ in layer $L$, a scattered field contribution $\bm{\mathcal{E}}^s_L(\bold{r})$ is added to $\bm{\mathcal{E}}_d(\bold{r})$ so that the total electric field in layer $L$ writes as $\bm{\mathcal{E}}_L(\bold{r})=\delta_{LM}\bm{\mathcal{E}}_d(\bold{r})+\bm{\mathcal{E}}^s_L(\bold{r})$, where
\begin{multline}  \bm{\mathcal{E}}^s_L(\bold{r})  =
\frac{i}{(2\pi)^{2}} \iint\limits_{-\infty}^{+\infty}\left[ \left(1-\delta_{LN}\right)\sum_{n=1}^2{\tilde{a}^s_{L,n}\bold{\tilde{e}}_{L,n}\mathrm{e}^{i\tilde{k}_{L,nz}z}}+
      \left(1-\delta_{L1}\right)\sum_{n=3}^4{\tilde{a}^s_{L,n}\bold{\tilde{e}}_{L,n}\mathrm{e}^{i\tilde{k}_{L,nz}z}}
\right]  \times \\
 \mathrm{e}^{ik_x(x-x')+ik_y(y-y')} \, \mathrm{d}k_x \, \mathrm{d}k_y \numberthis \label{Espace1} \end{multline}
an additional subscript is introduced to denote the layer number (e.g. $L$ in this case), $\delta_{pq}$ denotes the Kronecker delta, and the $\{ \tilde{a}^s_{L,n}(k_x,k_y) \}$ represent the (source-dependent) scattered-field modal amplitudes. The four modal terms inside both the direct and scattered field integrals above can be classified into two upward and two downward propagation modes, distinguished according to the signs of $\{ \mathrm{Im}(\tilde{k}_{L,nz}) \}$\footnote{The eigenvalues $\left\{\tilde{k}_{L,1z},\tilde{k}_{L,2z},\tilde{k}_{L,3z},\tilde{k}_{L,4z}\right\}$ correspond to the propagation constants of the (resp.) Type I up-going, Type II up-going, Type I down-going, and Type II down-going plane wave modes of layer $L$, and so on for the other $N-1$ layers \cite{chewch2}.}.

To expedite propagating the source fields to $\bold{r}$, which requires enforcing continuity of the tangential EM field components throughout the environment, instead of working with Eq. (\ref{Espace1}) directly it is more convenient to work with a 4$\times$1 vector composed of the four tangential EM field components (see \cite{chewch2}): $\bm{\mathcal{V}} = \left[\mathcal{E}_{x} \ \mathcal{E}_{y} \ \mathcal{H}_{x} \ \mathcal{H}_{y} \right]$. The two longitudinal field components can be subsequently obtained from the transverse components \cite{chewch2}. An equation analogous to Eq. (\ref{Espace1}) thus arises, with $\bm{\mathcal{E}}_L$ replaced by $\bm{\mathcal{V}}_L$, which writes as
 \begin{equation}\label{V1} \bm{\mathcal{V}}_L(\bold{r}) \ = \ \frac{i}{(2\pi)^{2}} \iint\limits_{-\infty}^{+\infty} \bold{W}_L(k_x,k_y;z) \, \mathrm{e}^{ik_x(x-x')+ik_y(y-y')}\,\mathrm{d}  k_x \,\mathrm{d} k_y \end{equation}
\section{\label{sec3}Integrand Manipulations}
     For some $(k_x,k_y)$ that defines the transverse phase variation exp[$ik_x(x-x')+ik_y(y-y')$] common to all the plane wave modes within the environment, one desires the total modal contribution $\bold{W}_L(k_x,k_y;z)$exp[$ik_x(x-x')+ik_y(y-y')$] at $\bold{r}$. Assuming this transverse phase variation exp$[ik_x(x-x') +ik_y(y-y')]$, Maxwell's equations for a homogeneous medium can be manipulated \cite{chewch2} to yield the state matrix shown in Eq. (\ref{SV3b}). After substituting in a given layer's constitutive properties, its solution yields the four eigenmodes supported in that layer along with the corresponding modal (axial) propagation constants; this process, repeated for all $N$ layers, is the starting point of procuring $\bold{W}_L(k_x,k_y;z)$\footnote{The form of Eq. (\ref{SV3b}) differs slightly from formula (2.10.10) in \cite{chewch2}. The $-i$ factor on both sides of Eq. (\ref{SV3b}), which is embedded into $\bold{\tilde{\bar{H}}}$ on the left side and explicitly shown on the right side, facilitates an eigenvalue/eigenvector problem in which the propagation constants $\{ k_{m,nz}\}$ are the sought-after values rather than the $\{ik_{m,nz}\}$ values procured in~\cite{chewch2}.}. Subsequently, knowledge of the transverse modal fields in each layer combined with enforcement of tangential field continuity across layer interfaces allows one to propagate the direct source fields to $\bold{r}$ in layer $L$. Note that given the transverse EM fields of the $n$th mode, the complete six-component, $z$-independent modal field vector $\{ \bold{\tilde{e}}_n \ \bold{\tilde{h}}_n \}$ is completely determined \cite{chewch2}.
\subsection{Modal Eigenvectors and Eigenvalues}
     The characteristic plane wave modes for an arbitrarily anisotropic layer $m$ are summarily described by the four eigenvalues ($\tilde{k}_{m,1z}, \, \tilde{k}_{m,2z}$, \,
     $\tilde{k}_{m,3z}, \, \tilde{k}_{m,4z}$) and the four corresponding 4 $\times$ 1 eigenvectors  \( \begin{bmatrix} \bold{\tilde{s}}_{m,1} & \bold{\tilde{s}}_{m,2} & \bold{\tilde{s}}_{m,3} & \bold{\tilde{s}}_{m,4} \end{bmatrix} \) of the 4 $\times$ 4 state matrix $\bold{\tilde{\bar{H}}}=\bold{\tilde{\bar{H}}}(k_x,k_y)$. Defining the $n$th eigenvector as
     \begin{equation}\label{SV} \bold{\tilde{s}}_{m,n}=\bold{\tilde{s}}_{m,n}(k_x,k_y)=\begin{bmatrix} \tilde{e}_{m,nx} \\ \tilde{e}_{m,ny} \\ \tilde{h}_{m,nx} \\ \tilde{h}_{m,ny}\end{bmatrix}\end{equation}
     and noting that the corresponding $n$th characteristic solution $\bold{v}_{m,n}$ to \begin{equation}\label{SV3b} \bold{\tilde{\bar{H}}} \cdot \bold{v}_{m,n}\ = \ -i\frac{\partial}{ \partial z} \bold{v}_{m,n} \end{equation}
     has the form $\bold{v}_{m,n}=\bold{\tilde{s}}_{m,n}\mathrm{e}^{i\tilde{k}_{m,nz}(z-z^*)}$,
     one can show that the eigenvector/eigenvalue problem $\bold{\tilde{\bar{H}}} \cdot \bold{\tilde{s}}_{m,n} \ = \  \tilde{k}_{m,nz}\bold{\tilde{s}}_{m,n}$ results.

      To facilitate accurate and rapid numerical eigenmode computation, the following relations comprise analytical changes made to the canonical eigenmode formulations for isotropic media \cite{chewch2}, reciprocal, electrically uniaxial media \cite{felsen}, and generally anisotropic media (i.e. via the state matrix $\bold{\tilde{\bar{H}}}$) \cite{chewch2}:
      \begin{equation} k_x \to k_o (k_x/k_o)=k_ok_{xr}, \ k_y \to k_o (k_y/k_o)=k_o k_{yr}, \ \omega \mu_o \to k_o \eta_o, \ \mathrm{and} \ \omega \epsilon_o \to k_o/\eta_o \end{equation}
       Accurate computation of the eigenvectors and eigenvalues is of paramount importance to achieving high-precision results. This is because, as will be seen throughout this section, \emph{every mixed-domain field quantity is dependent upon the eigenvectors and/or eigenvalues}.
     \subsection{Intrinsic Reflection and Transmission Matrices}
\begin{figure}[H]
    \centering
    \includegraphics[width=4in]{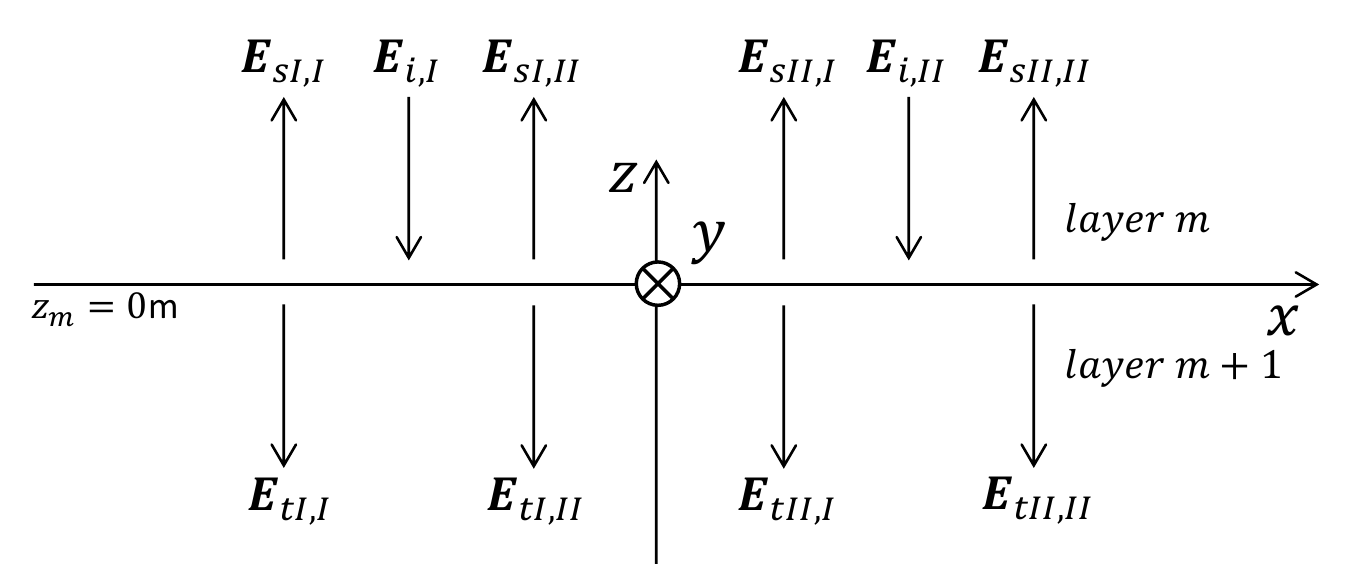}
    \caption{\label{FigC18}\small The incident modes ({$i$,I} and {$i$,II} subscripts), Type I/II reflected modes due to the incident Type I ({$s$I,I} and {$s$I,II} subscripts) and Type II modes ({$s$II,I} and {$s$II,II} subscripts), and Type I/II transmitted modes due to the incident Type I ({$t$I,I} and {$t$I,II} subscripts) and Type II modes ({$t$II,I} and {$t$II,II} subscripts) are shown.}
    \end{figure}

     We next calculate the 2 $\times$ 2 {\it intrinsic} reflection and transmission matrices\footnote{``Intrinsic" refers to reflection/transmission matrix quantities associated with only two media present (see Figure \ref{FigC18}).}. If down-going incident fields in layer $m$ are phase-referenced to $z=z_m$, then $\bold{\bar{R}}_{m,m+1}$ and $\bold{\bar{T}}_{m,m+1}$ are easily procured \cite{chewch2}; similar holds for $\bold{\bar{R}}_{m+1,m}$ and $\bold{\bar{T}}_{m+1,m}$.
     \subsection{\label{secGRM}Generalized Reflection/Three-Layer Transmission Matrices}
\begin{figure}[H]
    \centering
    \includegraphics[width=4in]{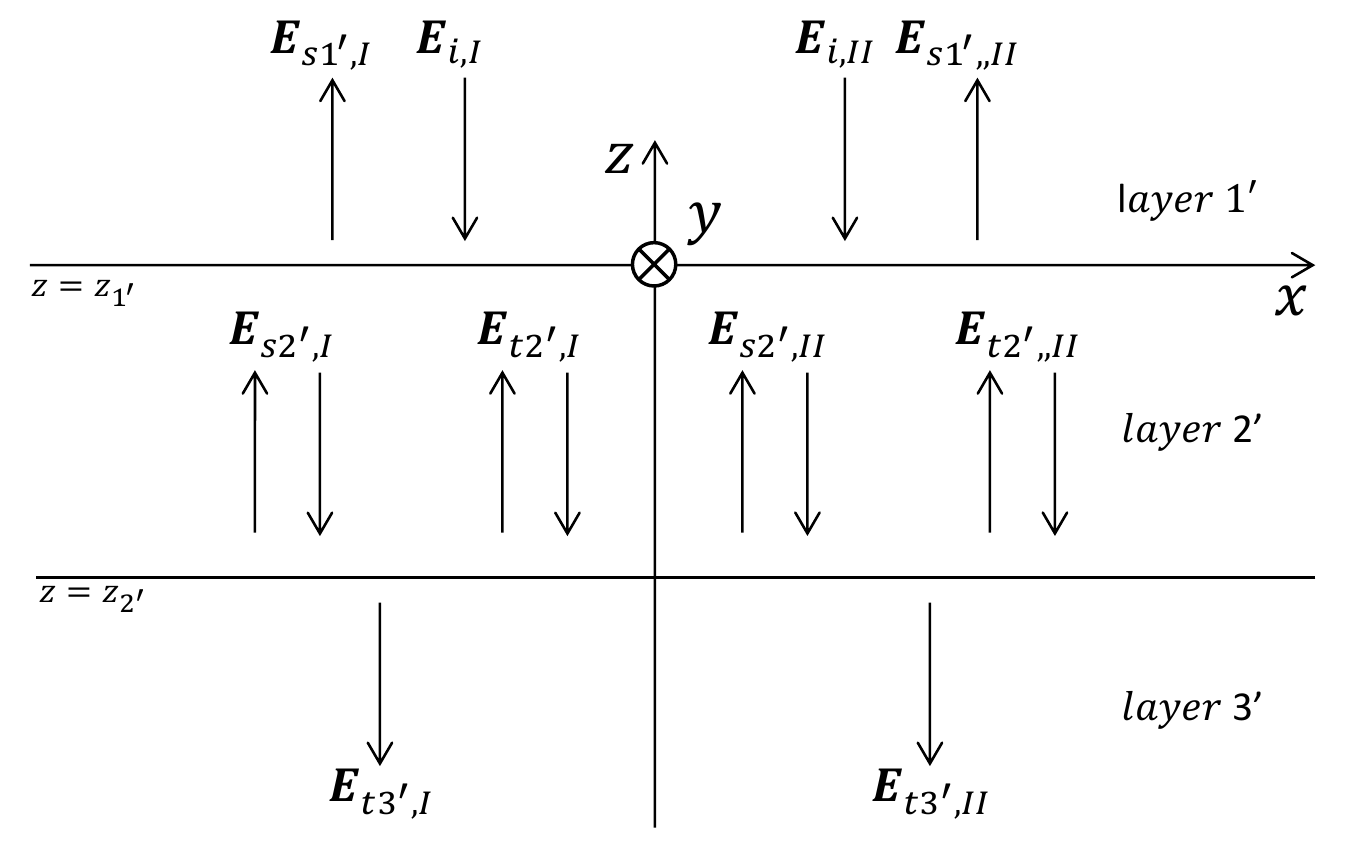}
    \vspace{-5pt}
    \caption{\label{FigC19}\small Schematic depicting the canonical three-layer medium for which the corresponding GRM and 3TM, associated with down-going incident fields in region 1', are calculated.}
    \end{figure}

     With the intrinsic reflection/transmission matrices now available, we derive the generalized reflection matrices (GRM) and three-layer transmission matrices (3TM). The 3TM yields the total down (up) going fields in the slab layer of the canonical three-layer medium problem for incident downward (upward) fields, while the GRM yields the reflected fields in the top (bottom) layer (see Figure \ref{FigC19}).

     The GRM assuming down-going incident fields can be determined by looking down into the three bottom-most layers of an $N$ layer medium (resp. labeled as $1'$ (top), $2'$ (middle), and $3'$ (bottom) in Figure \ref{FigC19}) and assuming that the scattered fields in region $2'$ and down-going incident fields in region $1'$ are phase-referenced to $z_{2'}$ and $z_{1'}$ (resp.). Following \cite{chewch2}, one imposes two ``constraint conditions" that result in two matrix-valued equations
\begin{equation}\label{CE1} \bold{\bar{\Lambda}}_{2'}^-(z_{1'}-z_{2'}) \cdot \bold{\tilde{a}}_{2'}^-  =  \bold{\bar{T}}_{1'2'}  \cdot  \bold{\tilde{a}}_{1'}^- + \bold{\bar{R}}_{2'1'} \cdot \bold{\bar{\Lambda}}_{2'}^+(z_{1'}-z_{2'}) \cdot \bold{\bar{R}}_{2'3'} \cdot \bold{\tilde{a}}_{2'}^- \end{equation}
\begin{equation}\label{CE2} \bold{\tilde{\bar{R}}}_{1'2'} \cdot \bold{\tilde{a}}_{1'}^-  = \bold{\bar{R}}_{1'2'} \cdot \bold{\tilde{a}}_{1'}^-+\bold{\bar{T}}_{2'1'} \cdot \bold{\bar{\Lambda}}_{2'}^+(z_{1'}-z_{2'}) \cdot \bold{\bar{R}}_{2'3'} \cdot \bold{\tilde{a}}_{2'}^- \end{equation}
By rearranging Eqs. (\ref{CE1})-(\ref{CE2}), one has
\begin{equation}\label{Mmat} \bold{\tilde{\bar{M}}}  = \left[\bold{\bar{I}}_2-\bold{\bar{\Lambda}}_{2'}^-(z_{2'}-z_{1'}) \cdot \bold{\bar{R}}_{2'1'} \cdot \bold{\bar{\Lambda}}_{2'}^+(z_{1'}-z_{2'}) \cdot \bold{\bar{R}}_{2'3'}\right]\end{equation}
and the 3TM
\begin{equation} \label{3TM} \bold{\tilde{\bar{T}}}_{1',2'}=\bold{\tilde{\bar{M}}}^{-1} \cdot \bold{\bar{\Lambda}}_{2'}^-(z_{2'}-z_{1'}) \cdot \bold{\bar{T}}_{1'2'}\end{equation}
with which one has
 \begin{equation}\label{Bdown} \bold{\tilde{a}}_{2'}^-  = \bold{\tilde{\bar{T}}}_{1',2'} \cdot \bold{\tilde{a}}_{1'}^- \end{equation}
Substituting the right hand side of Eq. (\ref{Bdown}) for $\bold{\tilde{a}}_{2'}^-$ in Eq. (\ref{CE2}), one obtains the GRM
\begin{equation}\label{Rgen} \bold{\tilde{\bar{R}}}_{1'2'}  =  \bold{\bar{R}}_{1'2'}+ \bold{\bar{T}}_{2'1'} \cdot \bold{\bar{\Lambda}}_{2'}^+(z_{1'}-z_{2'}) \cdot \bold{\bar{R}}_{2'3'} \cdot \bold{\tilde{\bar{T}}}_{1',2'} \end{equation}
This procedure can be repeated for layers $N-3$, $N-2$, and $N-1$ by labeling them as layers $1'$, $2'$, and $3'$ (resp.) and replacing $\bold{\bar{R}}_{2'3'}$ in Eq. (\ref{Rgen}) with \(\bold{\tilde{\bar{R}}}_{2'3'}\) \cite{chewch2}. The process is recursively performed up to the top three layers. A similar procedure can be used to find the GRM and 3TM looking up into each interface, whose expressions are found by using Eq. (\ref{Bdown}) and Eq. (\ref{Rgen}), labeling the bottom, middle, and top layers as $1'$, $2'$, and $3'$ (resp.), and making the following two variable interchanges in the modified GRM/3TM relations:
\begin{align}
     \bold{\bar{\Lambda}}_{2'}^+(z_{1'}-z_{2'})  &\leftrightarrow  \bold{\bar{\Lambda}}_{2'}^-(z_{2'}-z_{1'})  \numberthis \label{T1} \\
\bold{\tilde{a}}^+_{m'} &\leftrightarrow \bold{\tilde{a}}^-_{m'} (m=1,2,3)
\end{align}
While the procedure above is analytically exact, to avoid the risk of numerical overflow one should shift the reference depth of the slab's transmitted fields to the observation point depth $z$ when the slab contains $\bold{r}$. This avoids propagating downward the up-going modes (or vice versa) at the final stage of assembling the total mixed-domain field $\bold{W}_L(k_x,k_y;z)$. Otherwise, exponentially increasing propagators would be present, which may cause numerical overflow. To find the numerically stable 3TM and GRM expressions, we perform similar manipulations as before to obtain:
\begin{align}
 \bold{\tilde{\bar{M}}}&=\left[\bold{\bar{I}}_2-\bold{\bar{\Lambda}}_{2'}^-(z-z_{1'}) \cdot \bold{\bar{R}}_{2'1'} \cdot \bold{\bar{\Lambda}}_{2'}^+(z_{1'}-z_{2'}) \cdot \bold{\bar{R}}_{2'3'} \cdot \bold{\bar{\Lambda}}_{2'}^-(z_{2'}-z)\right] \numberthis \label{Mmat2} \\
 \bold{\tilde{a}}_{2'}^-&=\bold{\tilde{\bar{M}}}^{-1} \cdot \bold{\bar{\Lambda}}_{2'}^-(z-z_{1'}) \cdot \bold{\bar{T}}_{1'2'} \cdot \bold{\tilde{a}}_{1'}^-=\bold{\tilde{\bar{T}}}_{1',2'} \cdot \bold{\tilde{a}}_{1'}^- \numberthis \label{Bdown2} \\
 \bold{\tilde{\bar{R}}}_{1'2'}&=\bold{\bar{R}}_{1'2'}+ \bold{\bar{T}}_{2'1'} \cdot \bold{\bar{\Lambda}}_{2'}^+(z_{1'}-z_{2'}) \cdot \bold{\bar{R}}_{2'3'} \cdot \bold{\bar{\Lambda}}_{2'}^-(z_{2'}-z) \cdot \bold{\tilde{\bar{T}}}_{1',2'} \numberthis \label{Rgen2}
\end{align}
\subsection{Direct Field Modal Amplitudes}
We next procure the direct field modal amplitudes. For simplicity, the layer-number notation is omitted in this sub-section with the understanding that all field quantities are associated with layer $M$.

If the eigenvalues are unique, we first obtain $\bold{\tilde{H}}$ from $\bold{\tilde{E}}$ to form the four-component vector $\bold{\tilde{V}}= \left[ \tilde{E}_x \ \tilde{E}_y \ \tilde{H}_x \ \tilde{H}_y \right]$. With this, we perform the analytic $k_z$ integration of $\bold{\tilde{V}} \mathrm{e}^{i\bold{k} \cdot \bold{r}}$ to obtain
\begin{equation}\label{EFDb2} \mathrm{e}^{ik_x(x-x') +ik_y(y-y')}2 \pi i \sum_{l=l_1}^{l_2}{\left[\left(k_z-\tilde{k}_{lz}\right)\bold{\tilde{V}} \mathrm{e}^{ik_z (z-z')}\right]\bigg|_{k_z=\tilde{k}_{lz}}} \end{equation}
Equivalently, by setting $\bold{\tilde{V}}'=\bold{\tilde{V}} \mathrm{e}^{ik_z(z^*-z')}$, one obtains
\begin{equation}\label{EFDb2B} \mathrm{e}^{ik_x(x-x') +ik_y(y-y')}2 \pi i \sum_{l=l_1}^{l_2}{\left[\left(k_z-\tilde{k}_{lz}\right)\bold{\tilde{V}}' \mathrm{e}^{ik_z (z-z^*)}\right]\bigg|_{k_z=\tilde{k}_{lz}}} \end{equation}
where the sum runs over the two up-going modes (denoted by the substitutions $(l_1,l_2)\to(1,2)$ and $z^* \to z^*_{M-1}$) or two down-going modes (denoted by the substitutions $(l_1,l_2)\to(3,4)$ and $z^* \to z^*_M$), $z^*_{M-1}=\delta_{1M}z'+(1-\delta_{1M})z_{M-1}$, and $z^*_M=\delta_{NM}z'+(1-\delta_{NM})z_M$. Note that Eq. (\ref{EFDb2}) was redefined as Eq. (\ref{EFDb2B}) to facilitate subsequently calculating reflected and transmitted fields.

Next, define for up-going mode $l$ ($l=1,2$) the tangential fields, obtained after $k_z$ integration followed by suppression of the propagators, as
\begin{equation}\label{Vprime2b} \bold{\tilde{u}}_l^*=\bold{\tilde{u}}_l^*(k_x,k_y)=\left[\left( k_z-\tilde{k}_{lz}\right)\bold{\tilde{V}}\right]\Big|_{k_z=\tilde{k}_{lz}}\end{equation}
\begin{equation}\label{Vprime2} \bold{\tilde{u}}_l=\bold{\tilde{u}}_l(k_x,k_y)=\left[\left( k_z-\tilde{k}_{lz}\right)\bold{\tilde{V}}'\right]\Big|_{k_z=\tilde{k}_{lz}} \end{equation}
 one can define the amplitudes $\tilde{a}_{l,D}^*$ and $\tilde{a}_{l,D}$ (the $D$ subscript stands for ``direct"), corresponding to this mode, which satisfy
 \begin{equation}\label{ampstar} \bold{\tilde{u}}_l^*= \tilde{a}_{l,D}^* \bold{\tilde{\hat{s}}}_l \end{equation}
 \begin{equation}\label{ampunstar} \bold{\tilde{u}}_l= \tilde{a}_{l,D} \bold{\tilde{\hat{s}}}_l \end{equation}
If the eigenvalues are degenerate (i.e. when layer $M$ is isotropic), one instead uses the \emph{analytically} simplified spectral Green's Dyads devoid of double-poles \cite{chewch1,chewch7,mathematica} when employing Eqs. (\ref{EFDb2})-(\ref{ampunstar}). Since the resulting degenerate field is a linear combination of the TE$_z$ and TM$_z$ modes, one follows its evaluation with a $\mathrm{TE}_z$/$\mathrm{TM}_z$ modal decomposition. One decomposition example is
 \begin{equation}\label{Split} \begin{bmatrix} \tilde{e}_x^{I+} & \tilde{e}_x^{II+} \\ \tilde{e}_y^{I+} & \tilde{e}_y^{II+} \end{bmatrix} \begin{bmatrix} \tilde{a}^{I+}_{D} \\ \tilde{a}^{II+}_{D} \end{bmatrix}= \begin{bmatrix} \tilde{e}_x^{+} \\ \tilde{e}_y^{+} \end{bmatrix} \end{equation}
where $\tilde{a}^{I+}_{D}$ and $\tilde{a}^{II+}_{D}$ are the up-going TE$_z$ and TM$_z$ modal amplitudes (resp.). If using the transverse components leads to an ill-conditioned system, one can use relations in \cite{chewch2} to find $\tilde{e}_z^{I+}$,$\tilde{e}_z^{II+}$ and then solve Eq. (\ref{Split}) using $\tilde{e}_x$ and $\tilde{e}_z$ (or $\tilde{e}_y$ and $\tilde{e}_z$). Note that since Eq. (\ref{Split}) is a second-rank linear system, its inversion is trivial; therefore, only the system's \emph{conditioning} limits the accuracy of the computed amplitudes \cite{trefethen}.
\subsection{Scattered Mode Calculation and Field Transmission}
Now, the total field impinging upon the interfaces $z=z_{M-1}$ and $z=z_M$ must be calculated; this is done via exhibiting and solving the \emph{vectorial generalization} of relations in \cite{chewch2} accounting for arbitrary anisotropy (i.e. including inter-mode coupling at planar interfaces). All field quantities exhibited below through Eq. (\ref{RFS1}) are associated with layer $M$.

Define $\bold{\tilde{a}}^+_{D}=(\tilde{a}^{I+}_{D},\tilde{a}^{II+}_{D})$, $\bold{\tilde{a}}^+_{S1}$, and $\bold{\tilde{a}}^-_{S1}$ as 2 $\times$ 1 vectors containing (resp.) the amplitudes of the direct up-going, scattered up-going, and scattered down-going modes phase-referenced to $z=z_{M-1}$.
 Similarly, define $\bold{\tilde{a}}^-_{D}=(\tilde{a}^{I-}_{D},\tilde{a}^{II-}_{D})$, $\bold{\tilde{a}}^+_{S2}$, and $\bold{\tilde{a}}^-_{S2}$ for the \emph{same} modes but phase-referenced to $z=z_M$. With this, one defines the following quantities:
\begin{align}\bold{f}^+_{D}(k_x,k_y;z)&=\bold{\tilde{\bar{S}}}_M^+ \cdot \bold{\bar{\Lambda}}_M^+(z-z_{M-1}) \cdot \bold{\tilde{a}}^+_{D}, \ \bold{f}^-_{D}(k_x,k_y;z)=  \bold{\tilde{\bar{S}}}_M^- \cdot \bold{\bar{\Lambda}}_M^-(z-z_M) \cdot \bold{\tilde{a}}^-_{D} \numberthis \label{FSU} \\
\bold{f}^+_{S1}(k_x,k_y;z)&=\bold{\tilde{\bar{S}}}_M^+ \cdot \bold{\bar{\Lambda}}_M^+(z-z_{M-1}) \cdot \bold{\tilde{a}}^+_{S1}, \ \bold{f}^+_{S2}(k_x,k_y;z)=  \bold{\tilde{\bar{S}}}_M^+ \cdot \bold{\bar{\Lambda}}_M^+(z-z_M) \cdot \bold{\tilde{a}}^+_{S2} \\
\bold{f}^-_{S1}(k_x,k_y;z)&= \bold{\tilde{\bar{S}}}_M^- \cdot \bold{\bar{\Lambda}}_M^-(z-z_{M-1}) \cdot \bold{\tilde{a}}^-_{S1}, \ \bold{f}^-_{S2}(k_x,k_y;z)=  \bold{\tilde{\bar{S}}}_M^- \cdot \bold{\bar{\Lambda}}_M^-(z-z_M) \cdot \bold{\tilde{a}}^-_{S2}
\end{align}
Subsequently, in layer $M$ we can represent the tangential fields $\bold{W}_{M}(k_x,k_y;z)$ as
\begin{equation}\label{Wm1} \bold{W}_{M}(k_x,k_y;z)= \begin{cases} \bold{f}^+_{D}+ \bold{f}^+_{S1}+\bold{f}^-_{S1}, & z > z' \\ \bold{f}^-_{D}+\bold{f}^+_{S2}+\bold{f}^-_{S2}, & z < z' \end{cases} \end{equation}
Armed with relations Eqs. (\ref{FSU})-(\ref{Wm1}), one now imposes two ``constraint conditions" \cite{chewch2} that yield the relations (1) $\bold{\tilde{a}}^-_{S1}= \bold{\tilde{\bar{R}}}_{M,M-1} \cdot ( \bold{\tilde{a}}^+_{D}+\bold{\tilde{a}}^+_{S1})$ and (2) $\bold{\tilde{a}}^+_{S2}= \bold{\tilde{\bar{R}}}_{M,M+1} \cdot ( \bold{\tilde{a}}^-_{D}+\bold{\tilde{a}}^-_{S2})$. Using these two constraints along with (1) $\bold{\tilde{a}}^+_{S1}=\bold{\bar{\Lambda}}_M^+(z_{M-1}-z_{M}) \cdot \bold{\tilde{a}}^+_{S2}$ and (2) $\bold{\tilde{a}}^-_{S2}=\bold{\bar{\Lambda}}_M^-(z_{M}-z_{M-1})  \cdot \bold{\tilde{a}}^-_{S1}$, which arise from enforcing continuity of the scattered fields at $z=z'$, upon performing algebraic manipulation one has $\bold{\tilde{a}}^+_{S1}$ and $\bold{\tilde{a}}^-_{S2}$ as functions of $\bold{\tilde{a}}^{+}_{D}$ and $\bold{\tilde{a}}^{-}_{D}$:
\begin{equation}
 \bold{\tilde{\bar{M}}}_1  =  \bold{\bar{\Lambda}}_M^-(z_{M}-z_{M-1}) \cdot \bold{\tilde{\bar{R}}}_{M,M-1}, \ \bold{\tilde{\bar{M}}}_2  =   \bold{\bar{\Lambda}}_M^+(z_{M-1}-z_{M}) \cdot \bold{\tilde{\bar{R}}}_{M,M+1} \end{equation}
 \begin{equation}\label{RFS1}\bold{\tilde{a}}_{S1}^+  =  \left[\bold{\bar{I}}_2-\bold{\tilde{\bar{M}}}_2 \cdot \bold{\tilde{\bar{M}}}_1 \right]^{-1}  \cdot \bold{\tilde{\bar{M}}}_2 \cdot \left[\bold{\tilde{a}}_D^{-}+\bold{\tilde{\bar{M}}}_1 \cdot \bold{\tilde{a}}_D^{+} \right ], \
\bold{\tilde{a}}_{S2}^-  =  \left[\bold{\bar{I}}_2-\bold{\tilde{\bar{M}}}_1 \cdot \bold{\tilde{\bar{M}}}_2 \right]^{-1} \cdot \bold{\tilde{\bar{M}}}_1 \cdot \left[\bold{\tilde{a}}_D^{+}+\bold{\tilde{\bar{M}}}_2 \cdot \bold{\tilde{a}}_D^{-} \right ]
 \end{equation}
For $L \neq M$, one then uses the sum $\bold{\tilde{a}}_D^+ +\bold{\tilde{a}}_{S1}^+$ ($\bold{\tilde{a}}_D^- +\bold{\tilde{a}}_{S2}^-$) and the 3TM matrices to find $\bold{\tilde{a}}^+_L (\bold{\tilde{a}}^-_L)$ for $L<M$ ($L>M$), which write as (resp.)
\begin{equation}\label{aplus} \bold{\tilde{a}}^+_L=\bold{\tilde{\bar{T}}}_{L+1,L}\cdot \cdot \cdot \left[\bold{\bar{\Lambda}}^+_{M-2}(z_{M-3}-z^{\mathrm{ref}}_{M-2}) \cdot \bold{\tilde{\bar{T}}}_{M-1,M-2}\right] \cdot \left[\bold{\bar{\Lambda}}^+_{M-1}(z_{M-2}-z^{\mathrm{ref}}_{M-1}) \cdot \bold{\tilde{\bar{T}}}_{M,M-1}\right] \cdot (\bold{\tilde{a}}_D^+ +\bold{\tilde{a}}_{S1}^+) \end{equation}
\begin{equation}\label{aminus} \bold{\tilde{a}}^-_L=\bold{\tilde{\bar{T}}}_{L-1,L}\cdot \cdot \cdot \left[\bold{\bar{\Lambda}}^-_{M+2}(z_{M+2}-z^{\mathrm{ref}}_{M+2}) \cdot \bold{\tilde{\bar{T}}}_{M+1,M+2}\right] \cdot \left[\bold{\bar{\Lambda}}^-_{M+1}(z_{M+1}-z^{\mathrm{ref}}_{M+1}) \cdot \bold{\tilde{\bar{T}}}_{M,M+1}\right] \cdot (\bold{\tilde{a}}_D^- +\bold{\tilde{a}}_{S2}^-) \end{equation}
where for some intermediate layer $m \neq L$, $z^{\mathrm{ref}}_m$ is the user-defined phase-reference depth\footnote{If layer $L$ corresponds to a slab, we compute the 3TM $\bold{\tilde{\bar{T}}}_{L+1,L}$ in \eqref{aplus} according to the numerically stable 3TM/GRM formulation presented in Section \ref{secGRM}. If instead layer $L$ corresponds to the top layer, $\bold{\tilde{\bar{T}}}_{L+1,L}$ reduces to the intrinsic transmission matrix. Similar holds for $\bold{\tilde{\bar{T}}}_{L-1,L}$ in \eqref{aminus}.}. Given $\bold{\tilde{a}}^+_L$ ($\bold{\tilde{a}}^-_L$) for $L<M$ ($L>M$), one then finds $\bold{\tilde{a}}^-_L$ ($\bold{\tilde{a}}^+_L$) as (resp.)
 \begin{align} \bold{\tilde{a}}^-_L &= \bold{\bar{\Lambda}}_L^-(z-z_{L-1}) \cdot \bold{\tilde{\bar{R}}}_{L,L-1} \cdot \bold{\bar{\Lambda}}_L^+(z_{L-1}-z) \cdot \bold{\tilde{a}}^+_L \\
\bold{\tilde{a}}^+_L &= \bold{\bar{\Lambda}}_L^+(z-z_L) \cdot \bold{\tilde{\bar{R}}}_{L,L+1} \cdot \bold{\bar{\Lambda}}_L^-(z_L-z) \cdot \bold{\tilde{a}}^-_L \end{align}
With the above in mind, we have the following expressions when $L<M$ ($L>M$) (resp.):
   \begin{equation}\label{Case1a} \bold{W}_L(k_x,k_y;z)= \Big( \bold{\bar{\Lambda}}^+_L([z-z_1]\delta_{L1}) \cdot \bold{\tilde{\bar{S}}}^+_L+ (1-\delta_{L1})\bold{\tilde{\bar{S}}}^-_L \cdot \bold{\bar{\Lambda}}_L^-(z-z_{L-1}) \cdot \bold{\tilde{\bar{R}}}_{L,L-1} \cdot \bold{\bar{\Lambda}}_L^+(z_{L-1}-z) \Big) \cdot \bold{\tilde{a}}^+_L\end{equation}
   \begin{equation}\label{Case1b} \bold{W}_L(k_x,k_y;z)= \Big( \bold{\bar{\Lambda}}^-_L([z-z_{N-1}]\delta_{LN}) \cdot \bold{\tilde{\bar{S}}}^-_L+ (1-\delta_{LN})\bold{\tilde{\bar{S}}}^+_L \cdot \bold{\bar{\Lambda}}_L^+(z-z_L) \cdot \bold{\tilde{\bar{R}}}_{L,L+1} \cdot \bold{\bar{\Lambda}}_L^-(z_L-z) \Big) \cdot \bold{\tilde{a}}^-_L\end{equation}
If $L=M$, then for $N < M < 1$, one instead obtains $\bold{\tilde{a}}^-_{S1}$ and $\bold{\tilde{a}}^+_{S2}$ and propagates these to $z$. Note that this method obviates propagating downward (upward) $\bold{\tilde{a}}^+_{S1}$ ($\bold{\tilde{a}}^-_{S2}$), thereby preventing \emph{another} potential source of numerical overflow. The up-going (down-going) direct fields, as phase-referenced to $z'$, can be propagated to $z$ for $z>z'$ ($z<z'$). Now recall Eq. (\ref{ampstar}) and define $\bold{\tilde{a}}_D^{+*}=(\tilde{a}_{1,D}^*,\tilde{a}_{2,D}^*)$ and $\bold{\tilde{a}}_D^{-*}=(\tilde{a}_{3,D}^*,\tilde{a}^*_{4,D})$. Then for $z>z'$ ($z<z'$), $\bold{W}_L(k_x,k_y;z)$ writes as (resp.)
\begin{equation}\label{Case3a}\bold{W}_L(k_x,k_y;z)=\bold{\tilde{\bar{S}}}^+_L \cdot \bold{\bar{\Lambda}}_L^+(z-z') \cdot \bold{\tilde{a}}_D^{+*} + \bold{\tilde{\bar{S}}}^+_L \cdot \bold{\bar{\Lambda}}_L^+(z-z_L) \cdot \bold{\tilde{a}}^+_{S2} + \bold{\tilde{\bar{S}}}^-_L \cdot \bold{\bar{\Lambda}}_L^-(z-z_{L-1}) \cdot \bold{\tilde{a}}^-_{S1} \end{equation}
\begin{equation}\label{Case3b}\bold{W}_L(k_x,k_y;z)=\bold{\tilde{\bar{S}}}^-_L \cdot \bold{\bar{\Lambda}}_L^-(z-z') \cdot \bold{\tilde{a}}_D^{-*} + \bold{\tilde{\bar{S}}}^+_L \cdot \bold{\bar{\Lambda}}_L^+(z-z_L) \cdot \bold{\tilde{a}}^+_{S2} + \bold{\tilde{\bar{S}}}^-_L \cdot \bold{\bar{\Lambda}}_L^-(z-z_{L-1}) \cdot \bold{\tilde{a}}^-_{S1} \end{equation}
If $L=M=1$ or $L=M=N$, one uses $\bold{\tilde{a}}^{-}_{D}$ or $\bold{\tilde{a}}^{+}_{D}$ (resp.) to find $\bold{\tilde{a}}^+_{S2}=\bold{\tilde{\bar{R}}}_{1,2} \cdot \bold{\tilde{a}}^-_{D}$ or $\bold{\tilde{a}}^-_{S1}=\bold{\tilde{\bar{R}}}_{N,N-1} \cdot \bold{\tilde{a}}^+_{D}$ (resp.). Subsequently, the up-going (down-going) reflected fields are propagated to $z$. Furthermore, $\bold{\tilde{a}}_D^{+*}$ ($\bold{\tilde{a}}_D^{-*}$) is propagated to $z$ when $z>z'$ ($z<z'$). With this, for $M=1$ ($M=N$) we have (resp.)
   \begin{equation}\label{Case5a} \bold{W}_L(k_x,k_y;z)= u(z-z')\bold{\tilde{\bar{S}}}^+_L \cdot \bold{\bar{\Lambda}}_L^+(z-z') \cdot \bold{\tilde{a}}_D^{+*}+
   u(z'-z)\bold{\tilde{\bar{S}}}^-_L \cdot \bold{\bar{\Lambda}}_L^-(z-z') \cdot \bold{\tilde{a}}_D^{-*}+
   \bold{\tilde{\bar{S}}}^+_L \cdot \bold{\bar{\Lambda}}_L^+(z-z_1) \cdot \bold{\tilde{a}}^+_{S2}
   \end{equation}
   \begin{equation}\label{Case5b} \bold{W}_L(k_x,k_y;z)= u(z-z')\bold{\tilde{\bar{S}}}^+_L \cdot \bold{\bar{\Lambda}}_L^+(z-z') \cdot \bold{\tilde{a}}_D^{+*}+
   u(z'-z)\bold{\tilde{\bar{S}}}^-_L \cdot \bold{\bar{\Lambda}}_L^-(z-z') \cdot \bold{\tilde{a}}_D^{-*}+
   \bold{\tilde{\bar{S}}}^-_L \cdot \bold{\bar{\Lambda}}_L^-(z-z_{N-1}) \cdot \bold{\tilde{a}}^-_{S1}
   \end{equation}
Note that in all expressions obtained throughout this section, \emph{no exponentially rising terms are present} since down-going (up-going) modes are always propagated downward (upward), leading to a stable numerical implementation.
\section{\label{sec4}Integration Methodology}
In the numerical evaluation of
Eq. (\ref{Espace1}), one repeats the steps in Section \ref{sec3} for every sampled $(k_x,k_y)$ point, approximating Eq. (\ref{Espace1}) as the double sum
 \begin{equation}\label{Espace2} \bm{\mathcal{V}}(\bold{r}) \ \simeq \ \frac{i}{(2\pi)^{2}} \sum_{p=-P_1}^{P_2} \sum_{q=-Q_1}^{Q_2} \bold{W}_L(k_{xq},k_{yp};z) \, \mathrm{e}^{ik_{xq}(x-x')+ik_{yp}(y-y')} \, w(k_{xq}) \, w(k_{yp}) \end{equation}

\begin{figure}[H]
\centering
\includegraphics[width=3.3in]{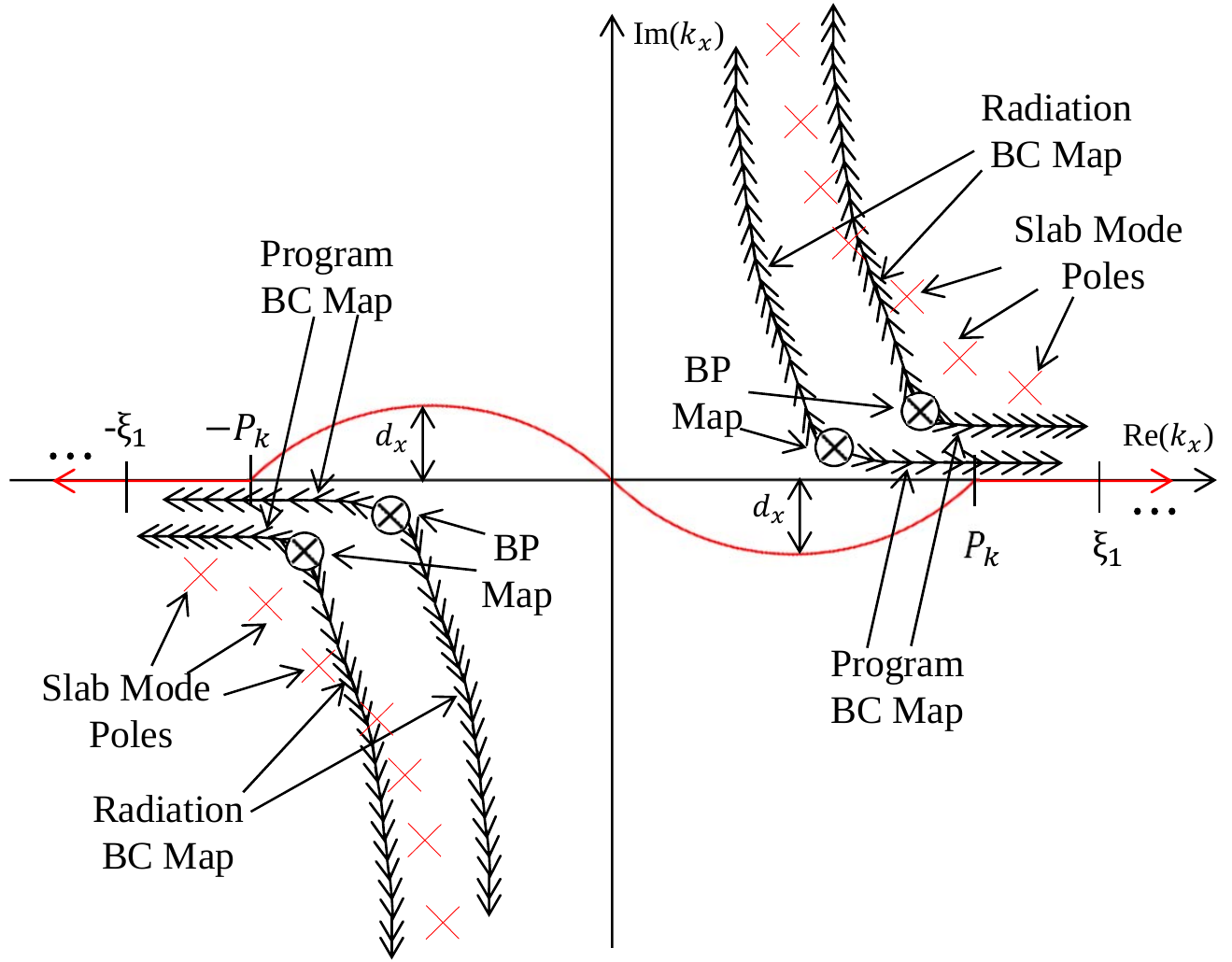}
\caption{\label{Fig3}\small (Color online) Typical $k_x$ plane features present when evaluating Eq. (\ref{Espace1}). ``Radiation BC Map" and ``Program BC Map" refer to the branch cuts associated with the radiation/boundedness condition at infinity and the computer program's square root convention (resp.). The encircled ``X" symbols represent the branch points and the red ``X" symbols represent slab-/interface-guided mode poles. For $K$ extrapolation intervals used, the red contour represents the integration path extending to $k_x=\pm \xi_{K+1}$.}
\end{figure}

In Section \ref{sec4a}, we describe an efficient methodology to compute the contribution from the ``pre-extrapolation" region $-\xi_1 < \mathrm{Re}(k_x) < \xi_1$, (see Figure \ref{Fig3}).
In section \ref{sec4b}, we detail an adaptive implementation of the MWA~\cite{mosig2,mosig3,mich2} tailored for this problem to compute the contribution from the ``extrapolation" region $|k_x| > |\xi_1|$ \cite{mich2}.
\subsection{\label{sec4a}Pre-Extrapolation Region}
The presence of critical points (i.e. branch points/cuts and slab/interface mode poles) near the Re($k_x$) axis in the pre-extrapolation region requires a detoured contour to yield a robust numerical integration~\cite{chewch2}. Furthermore, the oscillatory nature of $\bold{W}_L(k_x,k_y;z) \mathrm{exp}[ik_x(x-x')+ik_y(y-y')]$ and the potentially close proximity of critical points to the detoured contour warrants adaptively integrating to ensure accurate results \cite{mich1,mich2} (see Figure \ref{Fig3}).

First we discuss the integration path's initial sub-division and parameterization. Similar to \cite{mich1}, we define: a maximum detour height $d_{x}$, the two points bounding the detour as $k_x=\pm {P_{k}}$, and the two points within which one adaptively integrates as $k_x=\pm \xi_1$. All these points are indicated in Figure \ref{Fig3}. The detour path can be parameterized, using the real-valued variable $r$, as $k_x=r-i\sin{(\pi r/P_k)}$ and $ \mathrm{d} k_x = (\partial k_x/\partial r) \, \mathrm{d} r$,  where $\partial k_x/\partial r= 1-i(\pi/P_k)\cos{(\pi r/P_k)}$ and $-\xi_1 \leq r \leq \xi_1$ \cite{mich1}.

To compute $\pm P_k$, we adapt the procedure described in \cite{mich1} to arbitrarily anisotropic media. For a layer $p$ ($p$=1,2,...,$N$), we calculate the three eigenvalues of its relative material tensors $\boldsymbol{\bar{\epsilon}}_{r,p}$
 and $\boldsymbol{\bar{\mu}}_{r,p}$ ($\{ \epsilon_{pi} \}$, $ \{\mu_{pi} \}$, $i$=1,2,3),
find $\sqrt{\epsilon_{pi} \mu_{pi}}$ for $i$,$j$=1,2,3,
and take the $p$-th layer ``effective" refractive index $n_p$ ($p=1,2,...,N$) to be the $\sqrt{\epsilon_{pi} \mu_{pj}}$ value having the real part with the largest magnitude but with imaginary part below a user-defined threshold $T$.
Subsequently we compute $n^+=$max($|$Re(\{$n_p$\})$|$), which yields a ``worst-case" scenario for the maximum magnitude of the real part of any poles or branch points near the Re($k_x$) axis. Finally, we set $P_k=l_o k_o(n^++1)$, where $l_o \geq 1$ is a user-defined pre-extrapolation region magnification constant.

Furthermore, defining $\Delta x=|x-x'|$, $\Delta y=|y-y'|$, and $\Delta z=|z-z'|$, we compute the following integration path parameters~\cite{mich1}:
\begin{align}  d_x &=\begin{cases} \frac{1}{\Delta x} &, \Delta x>1 \\ 1 &, \mathrm{otherwise} \end{cases} \numberthis \label{dx} \\
  \Delta \xi_x &=  \begin{cases} \frac{\pi}{\Delta x} &, \Delta x>1 \\ \pi &, \mathrm{otherwise} \end{cases} \numberthis \label{xi} \\
\xi_1 &= \left( \mathrm{Int} \left( \frac{P_k}{\Delta \xi_x} \right)+1 \right) \Delta \xi_x \end{align}
where Int($\cdot$) truncates its argument to an integer number.
Next, we splice the regions $(0,P_k)$ and $(-P_k,0)$ each into $P$ regions. Letting $T_1$ and $T_2$ be two user-defined constants, one has \begin{align} \Delta k &= \begin{cases} \frac{ \pi }{T_1 \mathrm{max}( \Delta x, \Delta z )} &, \Delta x + \Delta z >0 \\ \frac{ \pi }{ T_1\Delta y } &, \mathrm{otherwise} \end{cases} \\
     N_{node} &=\mathrm{Int}\left(\frac{P_k}{\Delta k}\right)+1 \end{align}
resulting in $P$=Int(1+$N_{node}$/$T_2$). This empircal methodology for parameterizing and splicing the pre-extrapolation region relies upon the conservative assumption of equi-distant sampling.

We utilize a nested Patterson-Gauss/Kronrod-Gauss quadrature scheme \cite{mich1} throughout the pre-extrapolation region. Such schemes sacrifice algebraic degrees of precision, yielding only $3n+1$ ($3n+2$) degrees of precision for $n$ odd (even) when adding on $n+1$ nested quadrature nodes \cite{kronrod,pat2}, in contrast to $4n+1$ degrees of precision for a $(2n+1)$-point Gauss quadrature formula \cite{isaacson}. However, considering the extensive calculations involved at each sampled $(k_x,k_y)$ node (see Section \ref{sec3}), a Patterson-Gauss scheme significantly reduces the overall computation time \cite{pat1}.

Finally, one folds the integral results from $(0,\xi_1)$ and $(-\xi_1,0)$ to yield $I_{x0}'=I_{x0}'(k_y)$.
\subsection{\label{sec4b}Extrapolation Region}
Subsequently, one must approximate the integral over the path's tails $(\xi_1,\infty)$ and $(-\infty,-\xi_1)$ along the Re($k_x$) axis. For a robust computation, so that both approximation error and convergence rate are good for different geometries and ranges of layer constitutive properties, an integral acceleration/extrapolation technique is required. Here we adopt the MWA~\cite{mosig2,mosig3,mich2}\footnote{More specifically, we employ the the ``Mosig-Michalski algorithm" variant of MWA \cite{homeier} (MMA for short).} and briefly summarize below the extensions and adaptations made to our problem\footnote{It is assumed that (1) one has detoured sufficiently far past any branch points/poles near to the Re($k_x$) axis~\cite{mich1,mich2,mosig3} and (2) as $|k_x| \to \infty$, $i\tilde{k}_z(z-z') \to -f(k_x) \Delta z$, where $f(k_x)=f(-k_x)$ and Re$(f(k_x))>0$.}:
\begin{enumerate}
\item Splice the path $(\xi_1,\xi_1+N \Delta \xi_x)$ into $N$ sub-intervals\footnote{This $N$ is unrelated to the number of layers.} with bounding break-points $\xi_{xn}=\xi_1+(n-1) \Delta \xi_x $ \cite{mich1,mich2,mosig3}.
\item Integrate each sub-interval using (for example) a 15- or 20-point Legendre-Gauss quadrature rule \cite{mich1,mosig3}.
\item Store the these results as $I_{xp}^{+'}$ ($p=1,2,...N$).
\item Repeat steps 1-3 for the path $(-\xi_1-N \Delta \xi_x,-\xi_1)$ to procure $I_{xp}^{-'}$.
\item Fold $I_{xp}^{+'}$ and $I_{xp}^{-'}$ together to form $I_{xp}'=I_{xp}^{+'} + I_{xp}^{-'}$  ($p$=1,2,...,$N$).
\item Obtain \emph{cumulative} integrals $I_{xp,c}'$ via update: $I_{xp,c}'=I_{xp}'+I_{x(p-1),c}'$ ($p$=2,3,...,$N$) (Note: $I_{x1,c}'=I_{x1}'$).
\item Use the $\{I_{xp,c}'\}$ to estimate the non-truncated tail integral $I^{t'}_x$ as $I^{t'(N)}_x$.
\item  Compute the complete $k_x$ integral $I_x'=I_x'(k_y)=I^{t'(N)}_x+I_{x0}'$.
\end{enumerate}

The MWA accelerates convergence of integrals like Eq. (\ref{Espace1}) via estimating the tail integral's truncation error followed by combining two or more estimates, exemplified by
\begin{equation}\label{Wsum1} I^{t'(N)}_x=\frac{\sum_{n=1}^{n=N}w_nI_{xn,c}'}{\sum_{n=1}^{n=N}w_n}  \end{equation}
to accelerate the truncation error's decay. First, denote the true truncation error of $I_{xn,c}'$ as $R_{xn}$ such that $I^{t'}_x=I_{xn,c}' + R_{xn}$. Then, defining $\gamma_{1,2}=w_2/w_1$ and setting $N=2$, one can re-write Eq. (\ref{Wsum1}) as \cite{mich2,mosig3}:
\begin{equation}\label{Wsum2} I^{t'(2)}_x=\frac{w_1\left[I^{t'}_x-R_{x1}\right]+w_2\left[I^{t'}_x-R_{x2}\right]}{w_1+w_2}= I^{t'}_x-R_{x2}\frac{\frac{R_{x1}}{R_{x2}}+\gamma_{1,2}}{1+\gamma_{1,2}} \end{equation}
Next, setting $\gamma_{1,2}=-R_{x1}/R_{x2}$ yields $I^{t'(2)}_x=I^{t'}_x$ despite using only two finite-length tail integrals. However, in reality one must estimate the $\{ R_{xn} \}$ (thus yielding \emph{estimated} error ratios $\{- \gamma^{est(1)}_{n,n+1}\}$) via approximation of the truncation error integral's asymptotic behavior \cite{mich2}. By folding the asymptotic form of the $k_x$ integral's tail section one has
\begin{multline}\label{IntFold} \int^{\infty}_{\xi_1} k_x^q \mathrm{e}^{-f(k_x)\Delta z} \mathrm{e}^{ik_x(x-x')} \mathrm{d}k_x +\int^{-\xi_1}_{-\infty}
k_x^q \mathrm{e}^{-f(k_x)\Delta z} \mathrm{e}^{ik_x(x-x')} \mathrm{d}k_x = \\ \int^{\infty}_{\xi_1} 2k_x^q
\begin{Bmatrix} \cos{k_x(x-x')} \\ i\sin{k_x(x-x')} \end{Bmatrix} \mathrm{e}^{-f(k_x)\Delta z} \mathrm{d}k_x \end{multline}
with the sine (cosine) factor for $q$ odd (even). Furthermore, the factor $\mathrm{e}^{-f(k_x)\Delta z}k_x^q$ above can be rewritten as $(\mathrm{e}^{-f(k_x)\Delta z}k_x^{q+1})/k_x$ to conservatively ensure that in the multi-layer case, one can
satisfy the assumption \cite{mich2} that the integrand has the form
$h(k_x;z,z')=g(k_x;z,z')p(k_x)$, where $p(k_x)$ is an oscillatory function with period $2T=2\pi/\Delta x$ and (asymptotically) $g(k_x)$ has the form
\begin{equation} g(k_x;z,z') \sim \frac{\mathrm{e}^{-f(k_x)\Delta z}}{k_x^\alpha} \left[C+\mathcal{O}\left(k_x^{-1}\right)\right] \sim \frac{\mathrm{e}^{-f(k_x)\Delta z}}{k_x^\alpha} \sum_{l=0}^{\infty}\frac{c_l}{k_x^l} \end{equation}

Adapted to our problem, the analytic remainder estimate takes the form (for $\Delta x > 0$) $R^{est(1)}_{xn}= (-1)^{n}\mathrm{e}^{-f(k_x)\Delta z}\xi_{n+1}^{q+1}$, where $R_{xn}$ has the asymptotic form $R_{xn,a} \sim R^{est(1)}_{xn} \sum_{l=0}^{\infty} a_l \xi_{n+1}^{-l}$ \cite{mich2}. Subsequently, assuming that $R_{xn}/R_{x(n+1)}$ has the asymptotic form $R_{xn}/R_{x(n+1)}=R_{xn,a}' \sim (R^{est(1)}_{xn}/R^{est(1)}_{x(n+1)})\left[1+\mathcal{O}\left(\xi_{n+1}^{-2}\right)\right]$
one can insert $R_{x1,a}'$ and $\gamma_{1,2}^{est(1)}=-R^{est(1)}_{x1}/R^{est(1)}_{x2}$ (in place of $\gamma_{1,2}$) into Eq. (\ref{Wsum2}) to obtain \cite{mich2}
\begin{equation}\label{Intest1} I^{t'(2)}_x=I^{t'}_x+R_{x2}\frac{\left[1+\mathcal{O}\left(\xi_2^{-2}\right)\right]-1}{1+1/\gamma_{1,2}^{est(1)}}= I^{t'}_x+R_{x2}\frac{\mathcal{O}\left(\xi_2^{-2}\right)}{1+1/\gamma_{1,2}^{est(1)}}=I^{t'}_x-R^{(2)}_{x1} \end{equation}
with remainder $R^{(2)}_{x1}=-R_{x2}\mathcal{O}\left(\xi_2^{-2}\right)/(1+1/\gamma_{1,2}^{est(1)})$.
 It is seen that $R^{(2)}_{x1}$ is asymptotically equal to $R_{x2}$ except for being scaled by the factor $\xi_2^{-2}$; similarly, its corresponding remainder estimate $R^{est(2)}_{x1}$ is also scaled by $\xi_2^{-2}$ \cite{mich2,mosig3}. The above procedure can be applied recursively to estimate $I^{t'}_x$ using $N$ cumulative integrals \cite{mich2,mosig3}.
By defining
\begin{align} \gamma^{est(r-1)}_{n,n+1} &= \gamma^{est(1)}_{n,n+1}(\xi_{n+2}/\xi_{n+1})^{2(r-2)} \, \, \, (r=3,4,...,N+1) \\
 I^{t'(1)}_{xn,c} &= I_{xn,c}' \, \, \, (n=1,2,...,N) \\
I^{t'(N)}_{x1} &= I^{t'(N)}_x \end{align}
the following expression is obtained  in place of Eq. (\ref{Intest1}) \cite{mich2,mosig3}:
\begin{equation} \label{Intest2} I^{t'(r)}_{xn,c}=\frac{I^{t'(r-1)}_{xn,c}+I^{t'(r-1)}_{x(n+1),c}\gamma_{n,n+1}^{est(r-1)}}{1+\gamma_{n,n+1}^{est(r-1)}}, \, \, \, 2 \le r \le N, 1 \le n \le N-r+1 \end{equation}

Note from Eq. (\ref{xi}) that as $|x-x'|$ increases, $\Delta \xi_x$ is reduced. This is
done to keep the interval break-points at the extrema (nulls) of the cosine (sine) function in Eq. (\ref{IntFold}) \cite{mich2}, and sample the integrand at an adequate rate. However, simultaneously shrinking the region $(\xi_1,\xi_{N+1})$ may cause an undesirable degradation in accuracy. This can be solved via adaptive integration of the tail integral, using additional extrapolation intervals combined with successively higher-order weighted average schemes until convergence ensues.

For implementing an adaptive version of the MMA, one could in principle utilize an $N$-tier recursive function call chain to evaluate Eq. (\ref{Intest2}). However, this is not efficient since the number of active, simultaneous calls to the function carrying out extrapolation would peak at $N(N+1)/2$. Instead, pre-computing the weights for each desired $N$-tier scheme prior to integration such that one can simply compute $I^{t'(N)}_{x}=I^{t'(N)}_{x1}$ as
\begin{equation}\label{Wsum3} I^{t'(N)}_{x1}=\sum_{n=1}^{n=N}w_{n,N}I_{xn,c}' \end{equation}
where $w_{n,N}$ is the $n$th weight\footnote{For a given $N$, these weights are related to the weights shown in Eq. (\ref{Wsum1}) via the relation $w_{n,N}=w_n/\sum_{n=1}^{n=N}w_n$, where the $\{ w_n \}$ here tacitly exhibit dependence on $N$.} ($n=1,2,...,N$) of the tier-$N$ MMA scheme, is preferred. The three advantages of this strategy are that it (1) obviates extensive recursive function calls, (2) eliminates the redundancy of re-computing tier $N$ weights for each new $k_y$ node (this is markedly important for 2-D integration), and (3) requires only one weighted average (i.e. Eq. (\ref{Wsum3})), thereby drastically reducing the arithmetic operations associated with each of the $\{ I_{xn,c}' \}$ to one multiplication and one final summation versus $\mathcal{O}\left(2^{N-1}\right)$ total multiplications and additions required to compute $I^{t'(N)}_{x1}$ via the recursive function call chain approach. Assuming $N_{max}>1$ tiers are sought, the pre-computation of the weights proceeds as follows ($N=2,3,...,N_{max}$):
\begin{enumerate}
\item In computing $w_{n,N}$ ($1<n\le N$), admit $n$ ``intermediate" values $\{ w^{(1)}_{n,N},w^{(2)}_{n,N},...,w^{(n)}_{n,N} \}$, where $w^{(1)}_{N,N}=1$.
\item Recall Eq. (\ref{Intest2}) and set $r=2$. Comparing this with Eq. (\ref{Wsum3}), we find $ w_{1,2}=1/(1+\gamma^{est(1)}_{1,2})$ and $w_{2,2}=1/(1+1/\gamma^{est(1)}_{1,2})$.
We also set $w^{(1)}_{2,2}=1$ and $w^{(2)}_{2,2}=w_{2,2}$.
\item Recursively compute the $\{ w_{1,N} \}$ as $w_{1,m}=\frac{w_{1,m-1}}{1+\gamma^{est(m-1)}_{1,2}} \ \ (m=3,4,...,N_{max})$.
\item To compute $w_{n,N}$ ($2 \le n \le N$, $N >2$), first note the $\{ w_{n,N}^{(m)} \}$ initially update as
 \begin{align} w^{(m)}_{n,N}&=\frac{w^{(m)}_{n,N-1}}{1+\gamma_{n-m+1,n-m+2}^{est(N+m-n-1)}} \ \ (m=1,2,...,n; n \ne N)  \numberthis \label{updatecase1} \\
 w^{(1)}_{n,N}&=1, w^{(2)}_{n,N}=w^{(3)}_{n,N}=...=w^{(N)}_{n,N}=0 \ \ (n=N) \numberthis \label{updatecase2} \end{align}
\item Update the $\{ w_{n,N}^{(m)} \}$ again as
\begin{equation} \label{update2} w^{(m)}_{n,N}=w^{(m)}_{n,N}+\frac{w^{(m-1)}_{n,N}}{1+1/\gamma_{n-m+1,n-m+2}^{est(N+m-n-1)}} \ \ (m=2,3,...,n)  \end{equation}
    set $w_{n,N}=w^{(n)}_{n,N}$ to obtain the desired weight, and store the intermediate values for recursive re-application of steps 4-5.
\end{enumerate}
To clarify steps 4-5, let us take a simple example and outline the process of obtaining the third cumulative integral's weights corresponding to the three-tier, four-tier, and five-tier MMA (i.e. $w_{3,3}$, $w_{3,4}$, and $w_{3,5}$). Starting with $N=3$ and noting that $n=N=3$, we apply Eq. (\ref{updatecase2}) to obtain $w^{(1)}_{3,3}=1$ and $w^{(2)}_{3,3}=w^{(3)}_{3,3}=0$. Second, we apply Eq. (\ref{update2}) to obtain $w^{(2)}_{3,3}=0+w^{(1)}_{3,3}/(1+1/\gamma_{2,3}^{est(1)})$ and use this updated $w^{(2)}_{3,3}$ value to compute $w_{3,3}=w^{(3)}_{3,3}=0+w^{(2)}_{3,3}/(1+1/\gamma_{1,2}^{est(2)})$, yielding one of our desired weights. Third, we use these three updated intermediate values as the input to another application of step four with $N=4$, using Eq. (\ref{updatecase1}) to obtain $w^{(1)}_{3,4}=w^{(1)}_{3,3}/(1+\gamma_{3,4}^{est(1)})$, $w^{(2)}_{3,4}=w^{(2)}_{3,3}/(1+\gamma_{2,3}^{est(2)})$, and $w^{(3)}_{3,4}=w^{(3)}_{3,3}/(1+\gamma_{1,2}^{est(3)})$. Finally, use Eq. (\ref{update2}) to obtain $w^{(2)}_{3,4}=w^{(2)}_{3,4}+w^{(1)}_{3,4}/(1+1/\gamma_{2,3}^{est(2)})$ and $w_{3,4}=w^{(3)}_{3,4}=w^{(3)}_{3,4}+w^{(2)}_{3,4}/(1+1/\gamma_{1,2}^{est(3)})$, giving the second desired weight.

The above procedure lends two practical improvements to the original MMA by (1) significantly reducing the operation count involving the $\{ I_{xn,c}' \}$ and (2) devising a \emph{numerically stable} scheme to efficiently update the weights. After the tail integral has converged, one computes $I_x'=I^{t'(N)}_{x1}+I_{x0}'$ to yield the final result.
\section{Results}
We now present a series of numerical results using the formulation presented above for
the analysis of (1) well-logging induction (resistivity) tools for geophysical prospection (compared against \cite{wei,anderson1,howard}) and (2) the field pattern generated by electric current sources supported on grounded dielectric substrates (compared against \cite{kong}). The layers are numbered starting with the layer at the highest elevation and $z_B$ contains the interface depth values.

 Induction tools are generally composed of a system of transmitter and receiver loop antennas that can be modeled as Hertzian magnetic dipoles. The parameter $L_m$ denotes the separation between the transmitter and $m$-th receiver (if all receivers are co-located, then $L=L_1$).

 The environmental parameter of interest is the resistivity of the surrounding Earth media, which can exhibit electrical anisotropy and planar-stratified inhomogeneity. Earth layers exhibiting reciprocal, electrical uniaxial anisotropy possess different resistivities on and transverse to their respective bedding planes, which are equal to $R_{hn}=1/\sigma_{hn}$ and $R_{vn}=1/\sigma_{vn}$ in layer $n$ (resp.). Furthermore, each such layer has a bedding plane with arbitrary misalignment w.r.t. to the $z$ axis, which for layer $n$ is characterized by a dip angle and a strike angle that are denoted as $\alpha_n$ and $\beta_n$ (resp.). $\alpha$ ($\beta$) refers to the tool's polar (azimuthal) rotation relative to the $z$ axis; see \cite{anderson1} for the formation dip/strike angle convention, which is the same as the tool dip/strike angle convention.

Note that for \emph{homogeneous} formations characterized by this type of anisotropy, we use the variable $\alpha$ to refer to the tool inclination angle relative to the $z$-directed optic axis or the tilting of the optic axis relative to the $z$ axis (with a $z$-directed tool) interchangeably; these definitions are equivalent in homogeneous formations exhibiting isotropy or reciprocal, electrical uniaxial anisotropy \cite{moran1}.

When displacement currents are non-negligible compared to induction currents, the anisotropy ratio of layer $n$, $\kappa_n$, is defined as
 \begin{equation} \kappa_n=\sqrt{(k_o\epsilon_{hn,r}+i\eta_o\sigma_{hn})/(k_o\epsilon_{vn,r}+i\eta_o\sigma_{vn})} \end{equation}
where $\epsilon_{hn,r}$  ($\epsilon_{vn,r}$) is the complex-valued dielectric constant parallel (orthogonal) to the layer's bedding plane \cite{howard}. This
reduces to  $\kappa_n=\sqrt{R_{vn}/R_{hn}}=\sqrt{\sigma_{hn}/\sigma_{vn}}$ \cite{anderson1}
when displacement currents are negligible compared to induction currents (i.e. at sufficiently low frequencies).
For later reference, we also state the approximate formula predicting the formation resistivity estimated by a standard coaxial induction tool in a homogeneous, uniaxial medium \cite{moran1}:
\begin{equation}\label{ReffHom} R_{ap} = \frac{\kappa R_h}{\sqrt{\sin^2\alpha+\kappa^2\cos^2\alpha}} \end{equation}
\subsection{\label{array}Arrayed Coaxial Sonde}
The first logging scenario simulated here is an arrayed, coaxial induction sonde with one transmitter and two receivers immersed in a homogeneous, uniaxial medium with $z$-directed optic axis \cite{howard}. We vary i) $\alpha$ and ii) $\kappa$ (i.e. fix $R_h,\epsilon_h$ and vary $R_v,\epsilon_v$).

To extract effective, homogeneous-medium resistivity information from the observed magnetic field data, we follow the approach explained in \cite{howard}, which we summarize here. First define the ratio of the two axial-directed magnetic field\footnote{That is, the magnetic field component directed along the sonde axis, normal to the area of the coaxial receiver loop antenna.} values, observed at the two receiver loop antennas spaced at distances $L_1$ and $L_2$ from the transmitter loop antenna (i.e. $H_{z1}$ and $H_{z2}$, resp.), as $g_{12}=H_{z1}/H_{z2}$. Also, for some complex-valued phasor quantity $F$, define its phase as $\angle F$ and its magnitude as $|F|$. Phase-apparent resistivity $R_{ap,Ph}$ is obtained by first generating a look-up table of $\angle g_{12}$, at a specified transmitter radiation frequency, as a function of conductivity present in a homogeneous, isotropic medium. Subsequently, when the sonde is immersed in a heterogeneous environment that may contain anisotropic media, one compares the actual observed $\angle g_{12}$ to the look-up table and extracts the effective conductivity. This is finally inverted to obtain phase-apparent resistivity. Similar applies for magnitude-apparent resistivity $R_{ap,Amp}$, except now working with $|g_{12}|$ rather than $\angle g_{12}$.

\begin{figure}[H]
\centering
\subfloat[\label{HFig2a}]{\includegraphics[width=2.25in]{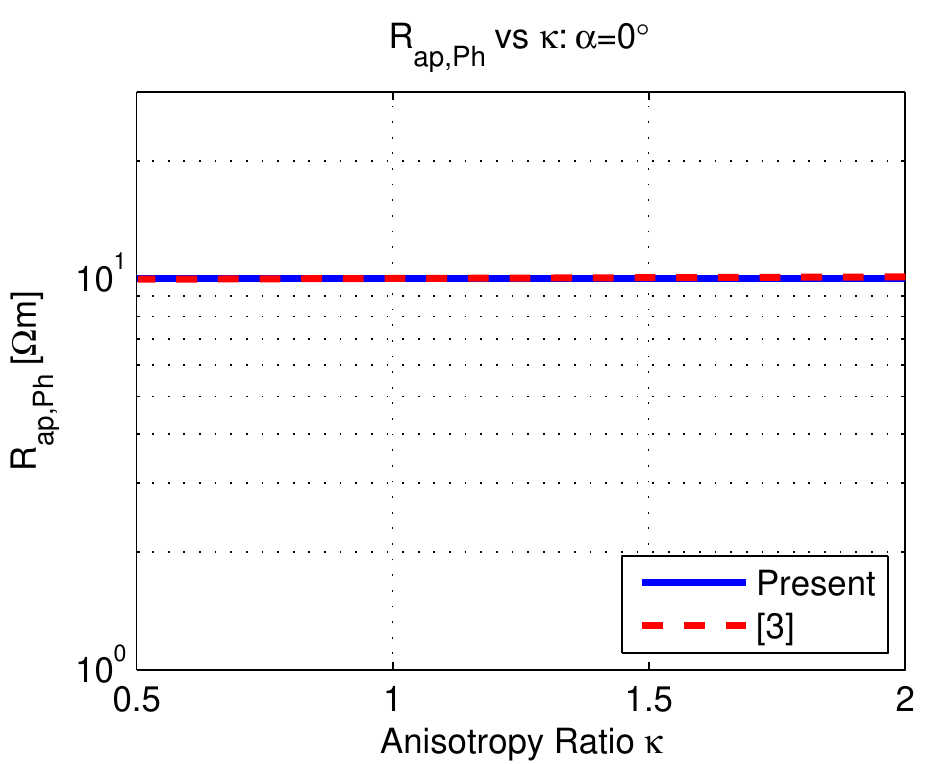}}
\subfloat[\label{HFig2c}]{\includegraphics[width=2.25in]{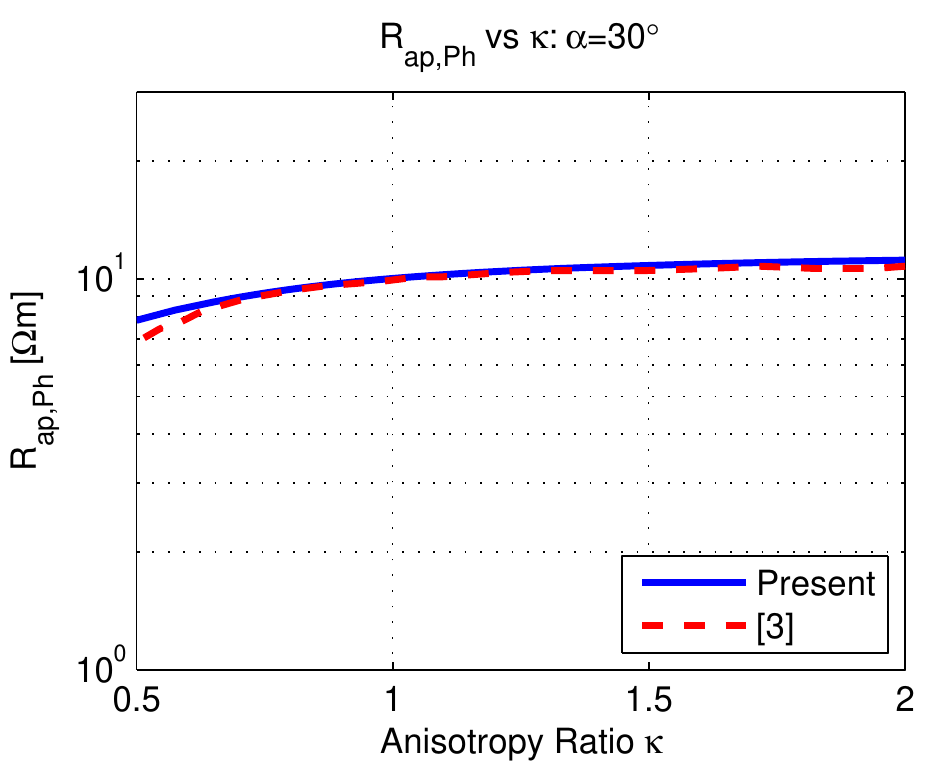}}
\subfloat[\label{HFig2d}]{\includegraphics[width=2.25in]{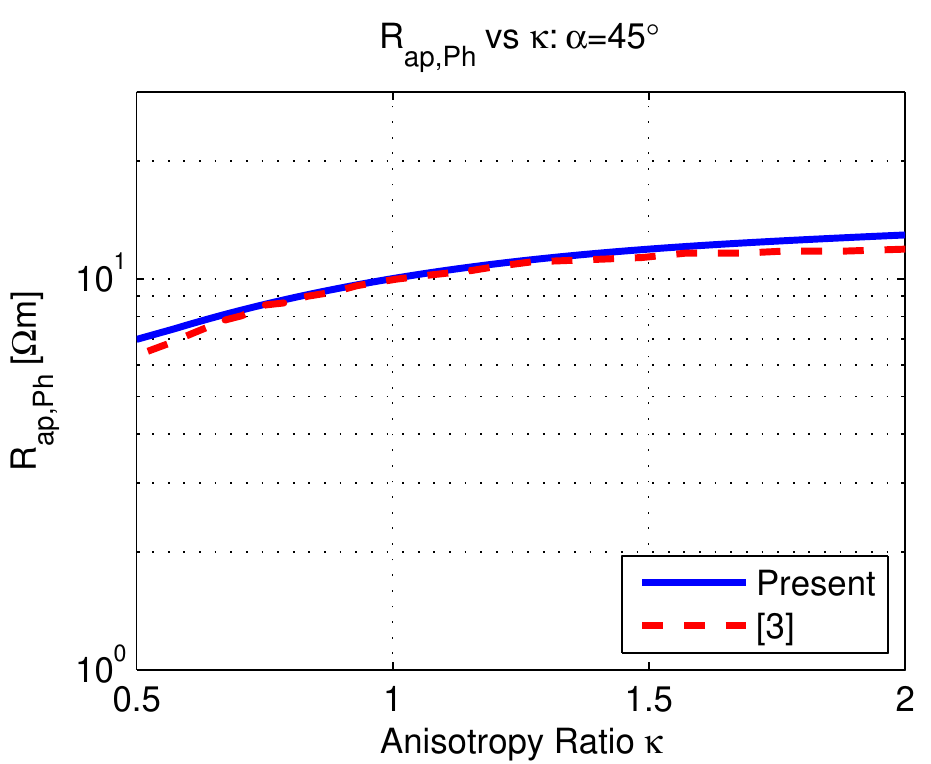}}

\subfloat[\label{HFig2e}]{\includegraphics[width=2.25in]{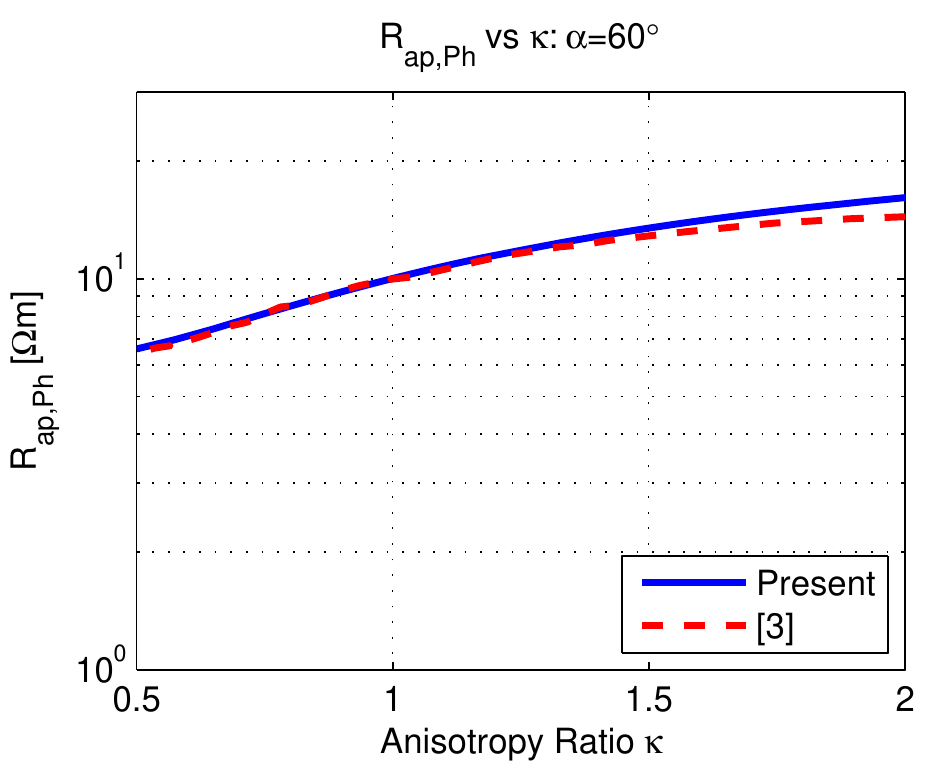}}
\subfloat[\label{HFig2f}]{\includegraphics[width=2.25in]{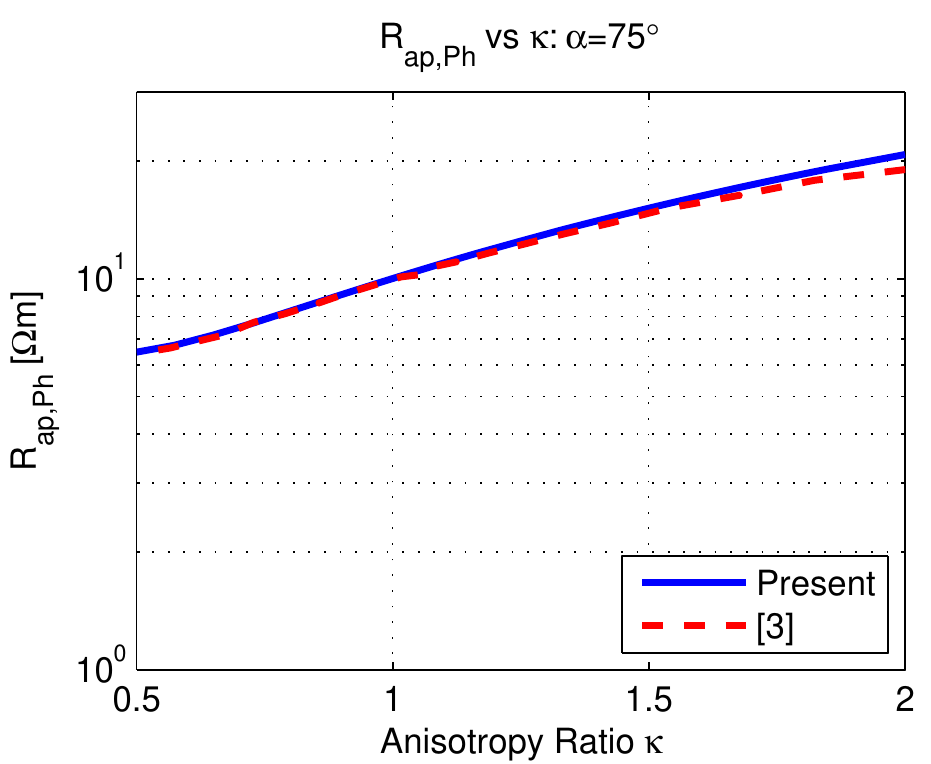}}
\subfloat[\label{HFig2g}]{\includegraphics[width=2.25in]{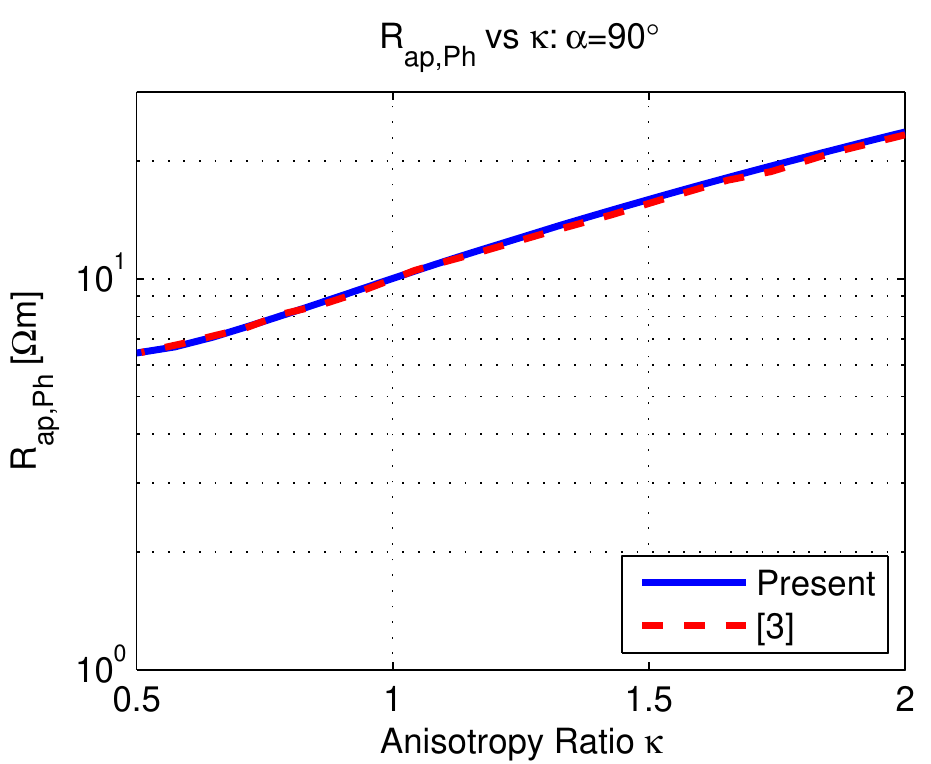}}
\caption{\label{HFig2}\small (Color online) Phase-apparent resistivity log comparison with Figure 2 of \cite{howard} (homogeneous medium): $R_{h}=10\Omega \mathrm{m},\beta=0^{\circ},f=2\mathrm{MHz},L_1=25\mathrm{in},L_2=31\mathrm{in}$.}
    \end{figure}

\begin{figure}[H]
\centering
\subfloat[\label{HFig3a}]{\includegraphics[width=2.25in]{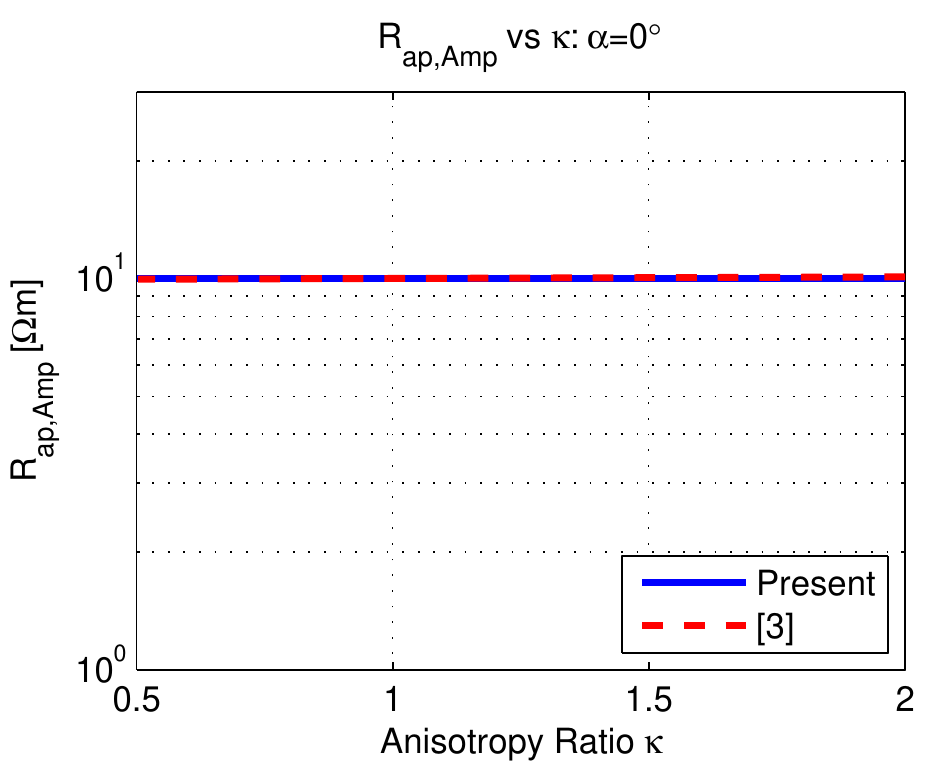}}
\subfloat[\label{HFig3c}]{\includegraphics[width=2.25in]{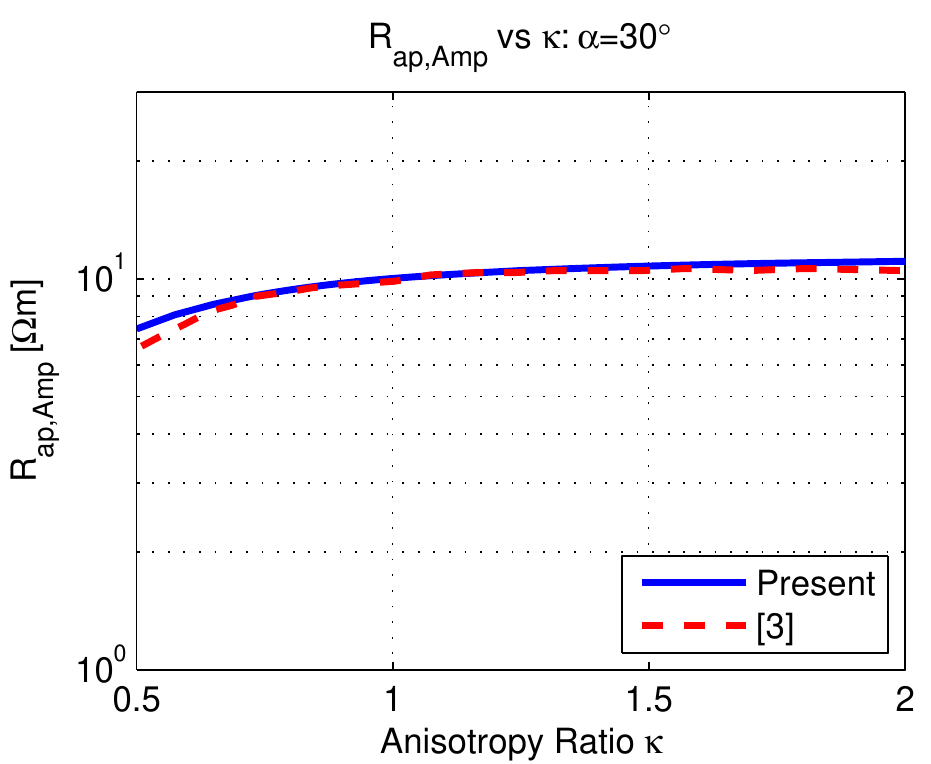}}
\subfloat[\label{HFig3d}]{\includegraphics[width=2.25in]{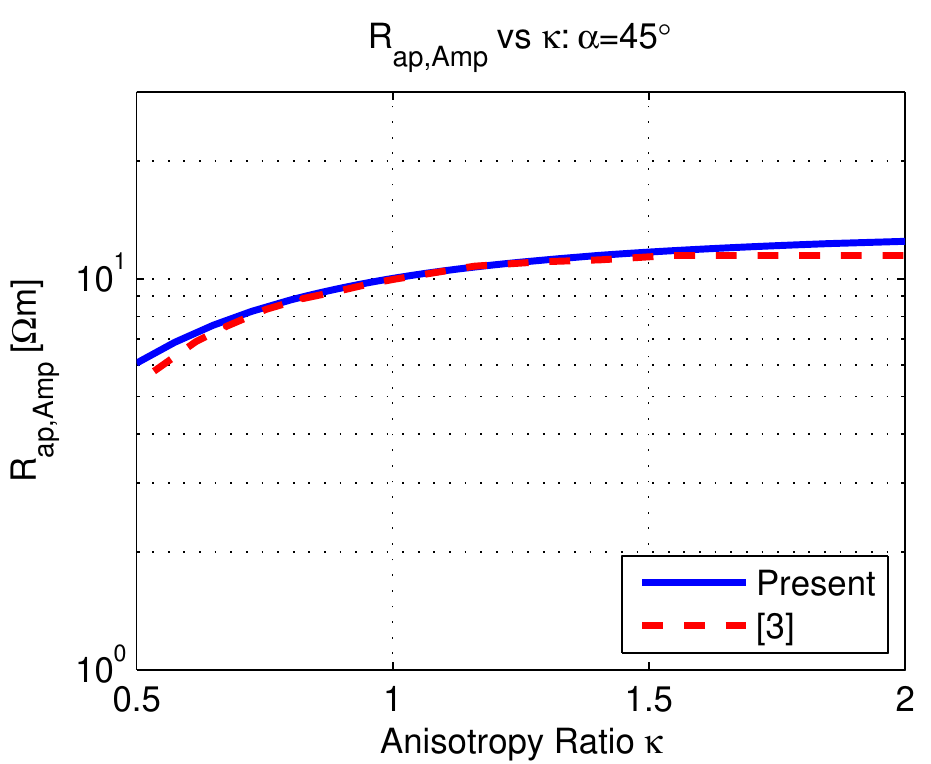}}

\subfloat[\label{HFig3e}]{\includegraphics[width=2.25in]{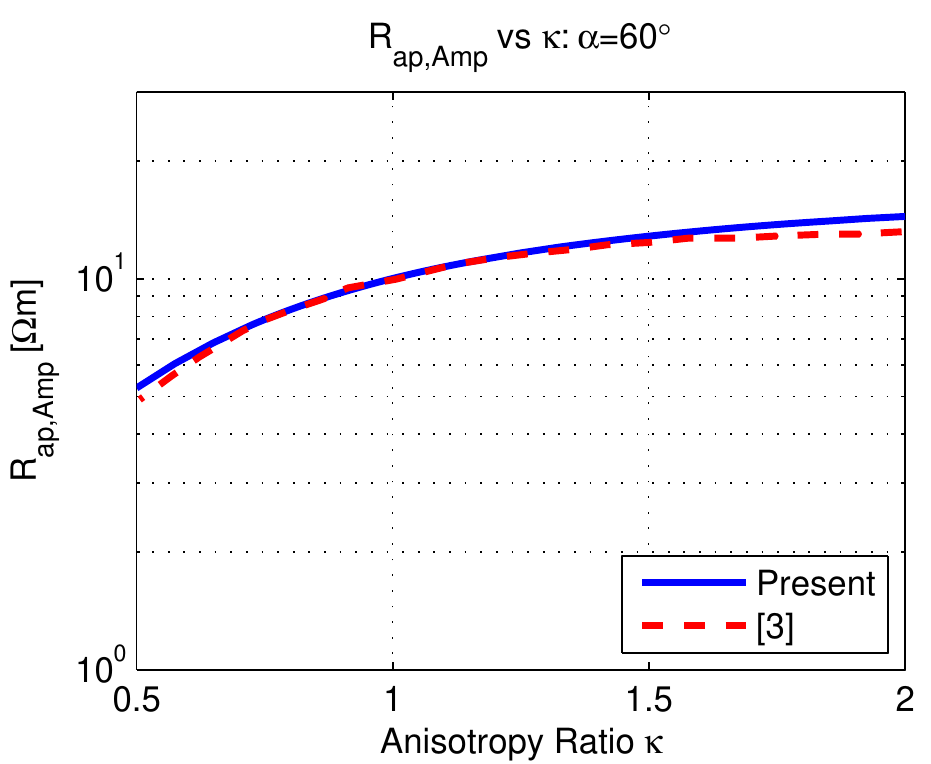}}
\subfloat[\label{HFig3f}]{\includegraphics[width=2.25in]{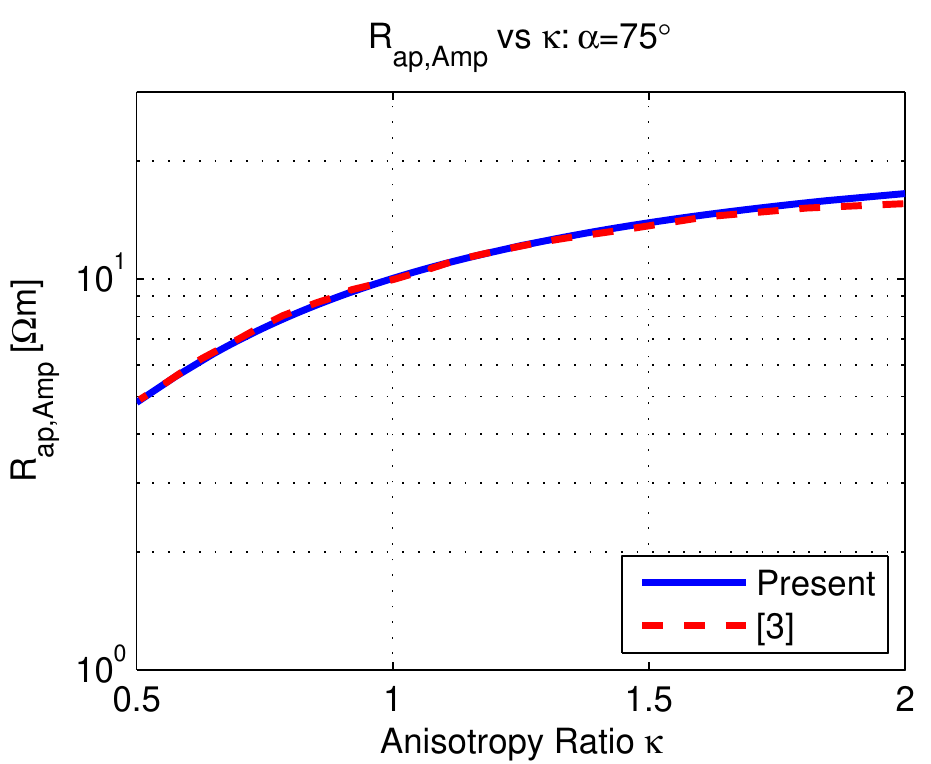}}
\subfloat[\label{HFig3g}]{\includegraphics[width=2.25in]{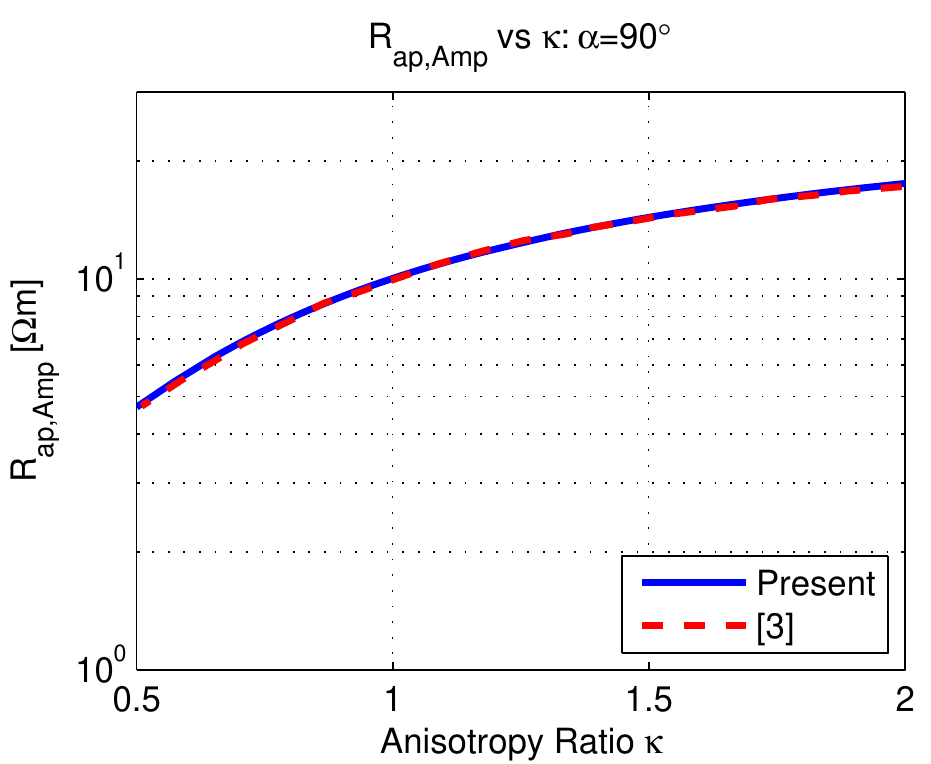}}
\caption{\label{HFig3}\small (Color online) Magnitude-apparent resistivity log comparison with Figure 3 of \cite{howard} (homogeneous medium): $R_{h}=10\Omega \mathrm{m},\beta=0^{\circ},f=2\mathrm{MHz},L_1=25\mathrm{in},L_2=31\mathrm{in}$.}
    \end{figure}

We see that throughout Figures \ref{HFig2}-\ref{HFig3}, agreement is consistently strong. Note that in Figures \ref{HFig2a} and \ref{HFig3a}, where $\alpha=0^\circ$, the sensed resistivity is insensitive to $\kappa$. This is because when $\alpha=0^\circ$ in a homogeneous, uniaxial medium, the coaxial sonde produces only $H$-mode plane wave spectra with electric field confined to the bedding plane \cite{felsen}. Furthermore, since the anisotropy ratio $\kappa$ is swept by keeping $R_h$ and $\epsilon_h$ constant while varying $R_v$ and $\epsilon_v$, it is expected that the received signal is independent of $\kappa$.
\subsection{Triaxial Induction Sonde}
The next logging scenarios involve a triaxial induction sonde with three mutually orthogonal, co-located transmitters and, spaced apart by a distance $L$, three mutually orthogonal, co-located receivers (see \cite{zhdanov} and Fig. 1 of \cite{wei}). To invert apparent conductivity from the received magnetic field, formula (18) of \cite{wang} is used.

\begin{figure}[H]
\centering
\subfloat[\label{WFig2a}]{\includegraphics[width=2.25in]{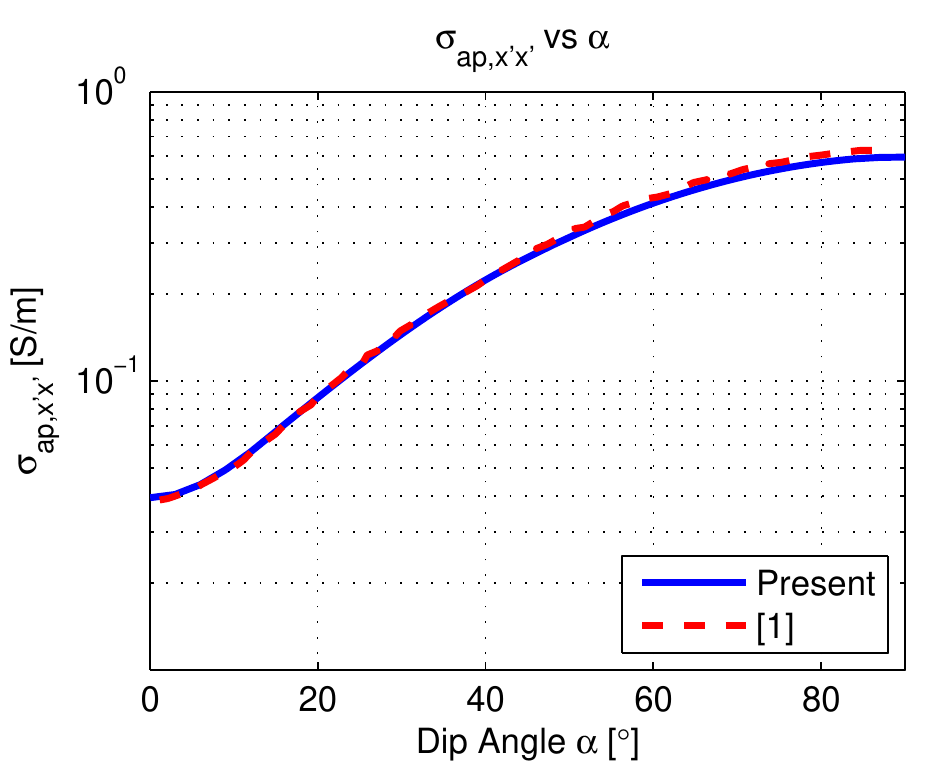}}
\subfloat[\label{WFig2b}]{\includegraphics[width=2.25in]{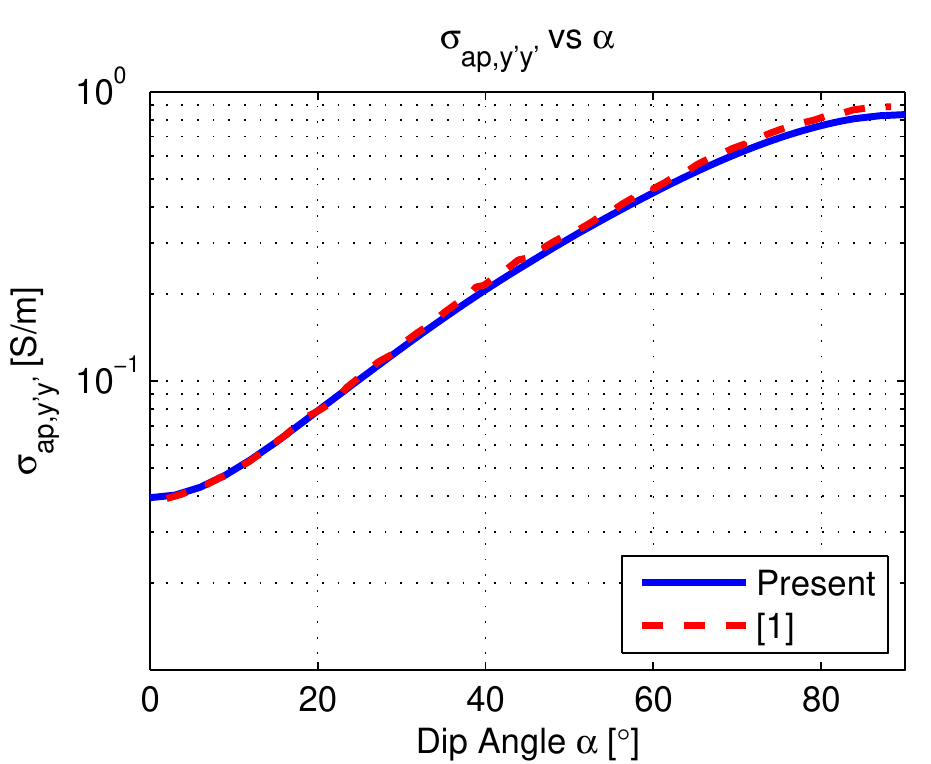}}
\subfloat[\label{WFig2c}]{\includegraphics[width=2.25in]{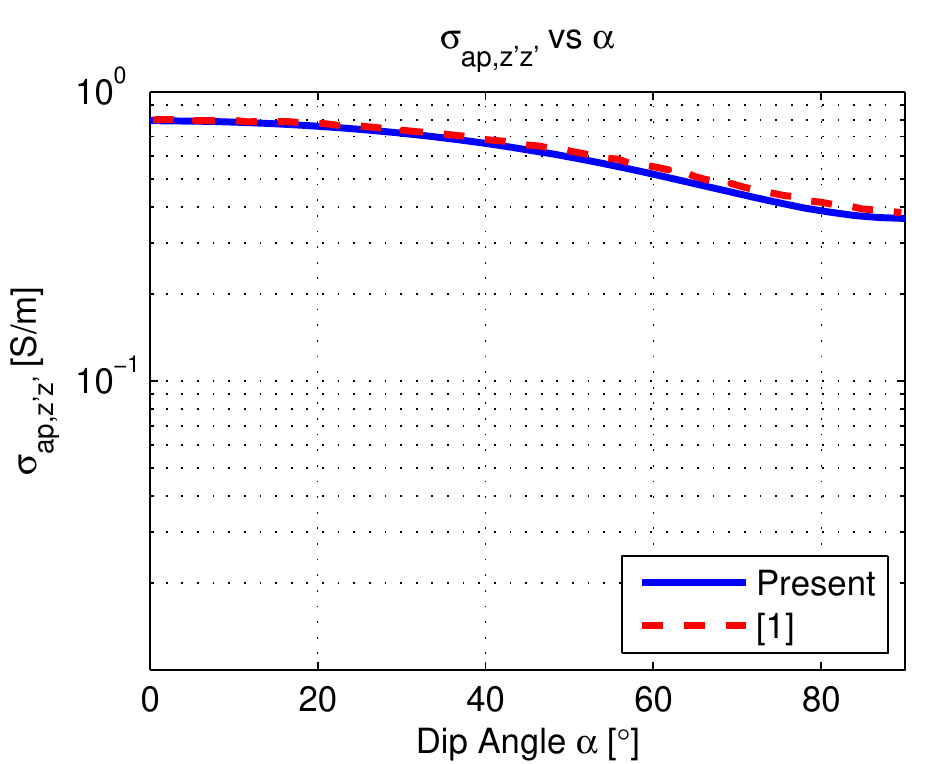}}
\caption{\label{WFig2}\small (Color online) Apparent conductivity log comparison with Figure 2 of \cite{wei} (homogeneous medium). $\kappa=\sqrt{5},R_{h}=1\Omega \mathrm{m},\beta=0^{\circ},f=25\mathrm{kHz},L=1\mathrm{m}$.}
\end{figure}

\begin{figure}[H]
\centering
\subfloat[\label{WFig3a}]{\includegraphics[width=2.25in]{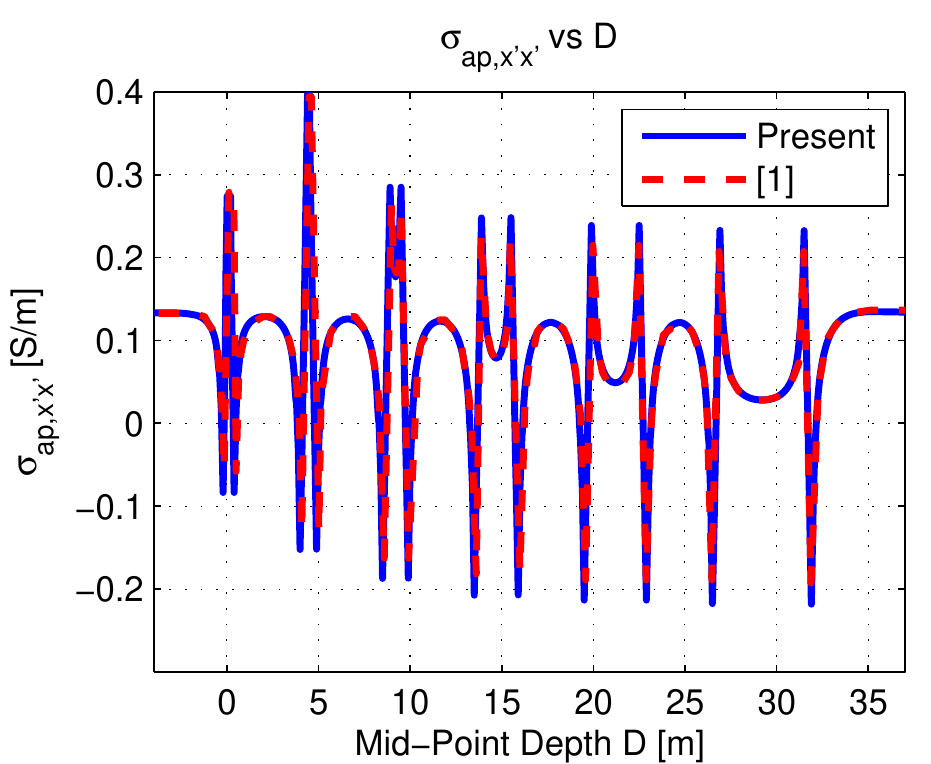}}
\subfloat[\label{WFig3b}]{\includegraphics[width=2.25in]{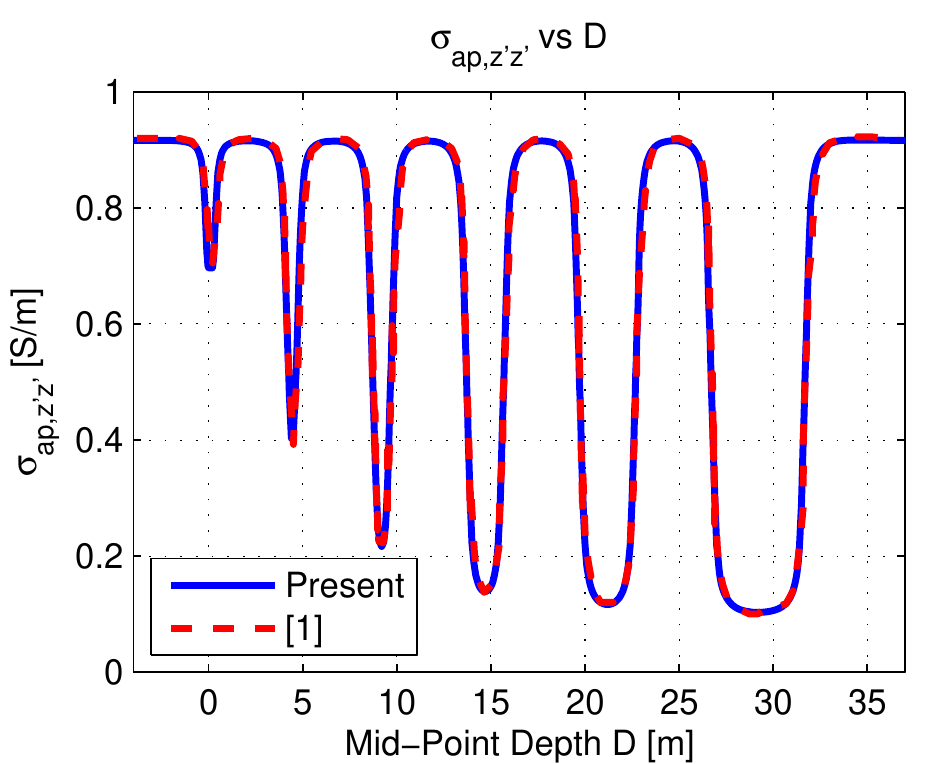}}
\caption{\label{WFig3}\small (Color online) Apparent conductivity log comparison with Figure 3 of \cite{wei}. $\{ \kappa_n \}$=$\sqrt{5}$ and $\{ \alpha_n \}$=$\{ \beta_n \}$=$0^{\circ}$ in all beds; $f$=25kHz, $L$=0.4m, $\sigma_h$=$\{$1.0, 0.1, 1.0, 0.1, 1.0, 0.1, 1.0, 0.1, 1.0, 0.1, 1.0, 0.1, 1.0$\}$S/m, $z_B$=$\{$0.0, 0.2, 4.2, 4.7, 8.7, 9.7, 13.7, 15.7, 19.7, 22.7, 26.7, 31.7$\}$m.}
    \end{figure}

Figure \ref{WFig2} corresponds to the sonde in a homogeneous, uniaxial medium with varying $\alpha$; agreement is excellent. Figure \ref{WFig3} corresponds to a thirteen-layer environment with $\alpha=\beta=0^\circ$. Note that our depth convention here corresponds to the half-way depth between the transmitters and receivers. Excellent agreement is observed between the results. For the coil separation used, $L$=0.4m, we notice that the coaxial ($\sigma_{a,z'z'}$) and co-planar ($\sigma_{a,x'x'}$) measurements provide marked resolution of even the thinnest bed present (0.2m thick); see the first spike and first valley from the left edge of Figures \ref{WFig3a} and \ref{WFig3b} (resp.).
    \newpage
 \subsection{Coaxial Sonde and Cross-bedding Anisotropy}
The next logging scenario simulated corresponds to a 2MHz coaxial sonde vertically traversing inhomogeneous environments. We compare our results against those presented in~\cite{anderson1}. It is important to note that there is an ambiguity in the resistivity inversion method and data post-processing used in ~\cite{anderson1} and hence only a \emph{qualitative} comparison is made here. Since the inversion method was not stated explicitly in \cite{anderson1}, we tried different inversion methods and found that the method corresponding to magnitude-apparent effective resistivity, specified in \cite{howard} and summarized above in section \ref{array}, produced the best-matching results with \cite{anderson1}. Also, \cite{anderson1} does not specify the depth convention in their plots (e.g. the transmitter depth). To render our data symmetric with respect to zero depth ($D=0$ft) in this case, we define the depth $D$ as mid-way between the receiver and transmitter.

\begin{figure}[h]
\centering
\subfloat[\label{AFig6a}]{\includegraphics[width=2.25in]{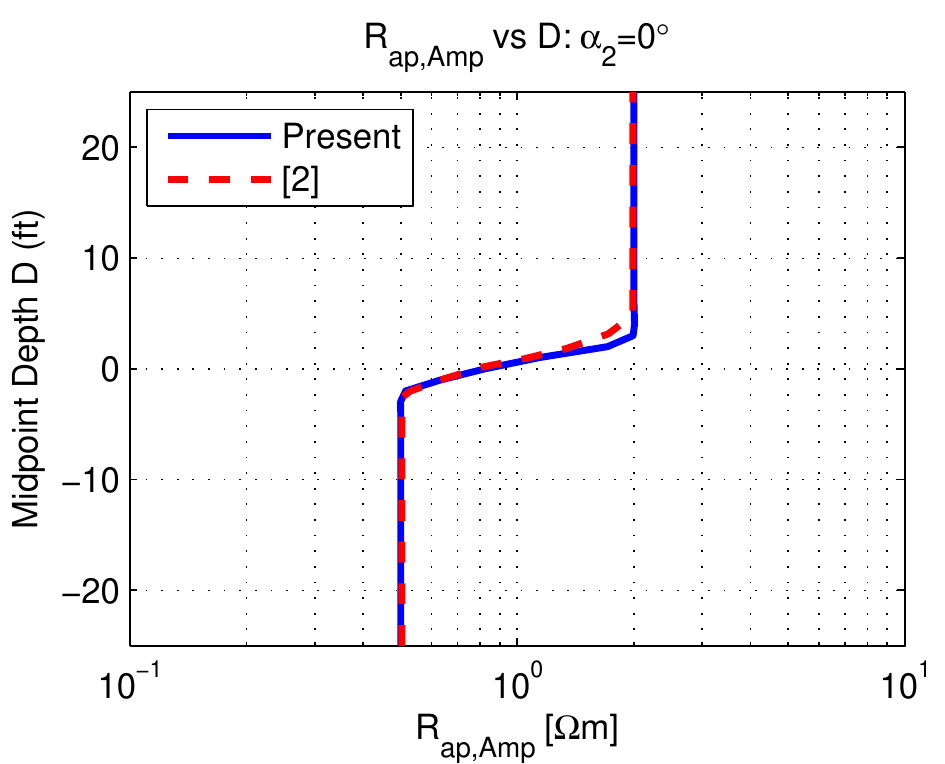}}
\subfloat[\label{AFig6b}]{\includegraphics[width=2.25in]{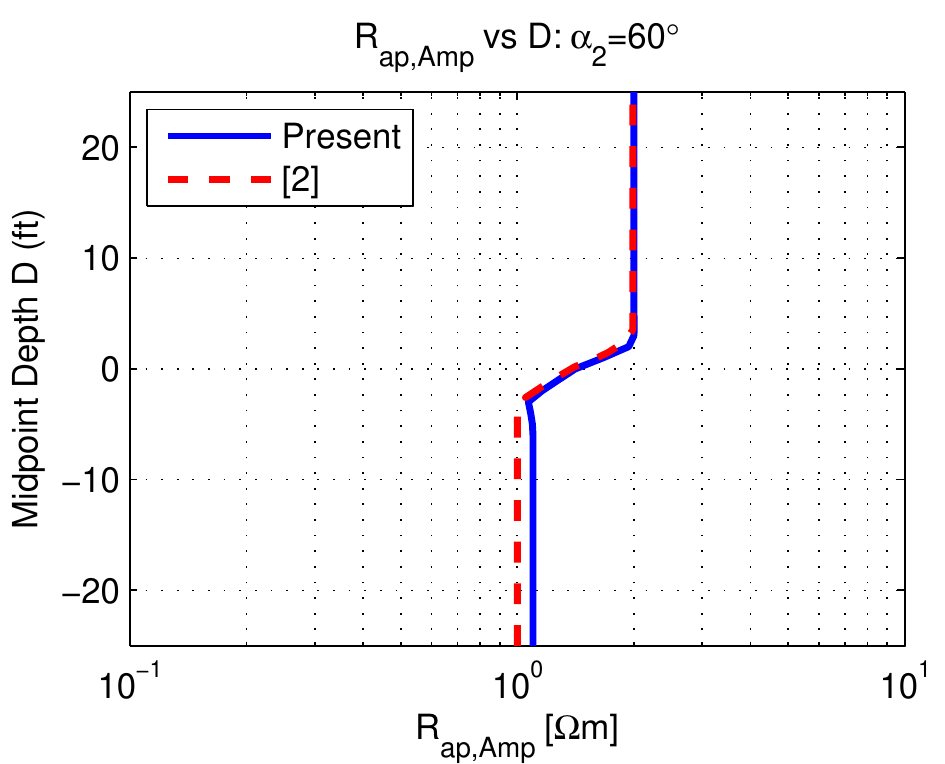}}
\caption{\label{AFig6}\small (Color online) Magnitude-apparent resistivity log comparison with Figure 6 of \cite{anderson1}: $\kappa_1=1,\kappa_2=\sqrt{20},R_{h1}=2\Omega \mathrm{m},R_{h2}=0.5\Omega \mathrm{m},\beta_2=0^{\circ},f=2\mathrm{MHz},L=40\mathrm{in},z_B=0\mathrm{ft}$.}
\end{figure}

In Figure \ref{AFig6}, where there is a low resistivity contrast of $R_{h1}=4R_{h2}$, observe that the agreement between the two data sets is good. Note that for Figures \ref{AFig6a}-\ref{AFig6b}, the effective resistivity in the top isotropic half-space levels off to that of the half space's actual resistivity, as is expected. Furthermore, note in Figure \ref{AFig6a} that deep within the bottom uniaxial half-space, the effective resistivity levels off to $R_{h2} \sim 0.5\Omega$m, which is consistent with Eq. (\ref{ReffHom}). This is because the transmitter antenna produces a primary (i.e. if $\boldsymbol{\bar{ \sigma}}=\bold{\bar{0}}$) $\boldsymbol{\hat{\phi}}$'-oriented electric field\footnote{The prime denotes the tool system \cite{zhdanov}.}. Being oriented perpendicular to the uniaxial medium's bedding plane, the loop only produces $H$-mode plane wave spectra \cite{felsen} and thus induces azimuthal currents parallel to the bedding plane possessing intensity affected solely by $R_{h2}$ and the top formation's resistivity. On the other hand, when $\alpha_2=60^\circ$, the transmitter loop's primary electric field now induces currents both parallel and perpendicular to the bedding plane. As a result, now the induced current and sensed resistivity $R_{ap,Amp}$ is also affected by $R_{v2}=\kappa_2^2R_{h2}$, leading to a higher value of $R_{ap,Amp}$ (as qualitatively corroborated by Eq. (\ref{ReffHom})).

\begin{figure}[H]
\centering
\subfloat[\label{AFig7a}]{\includegraphics[width=2.25in]{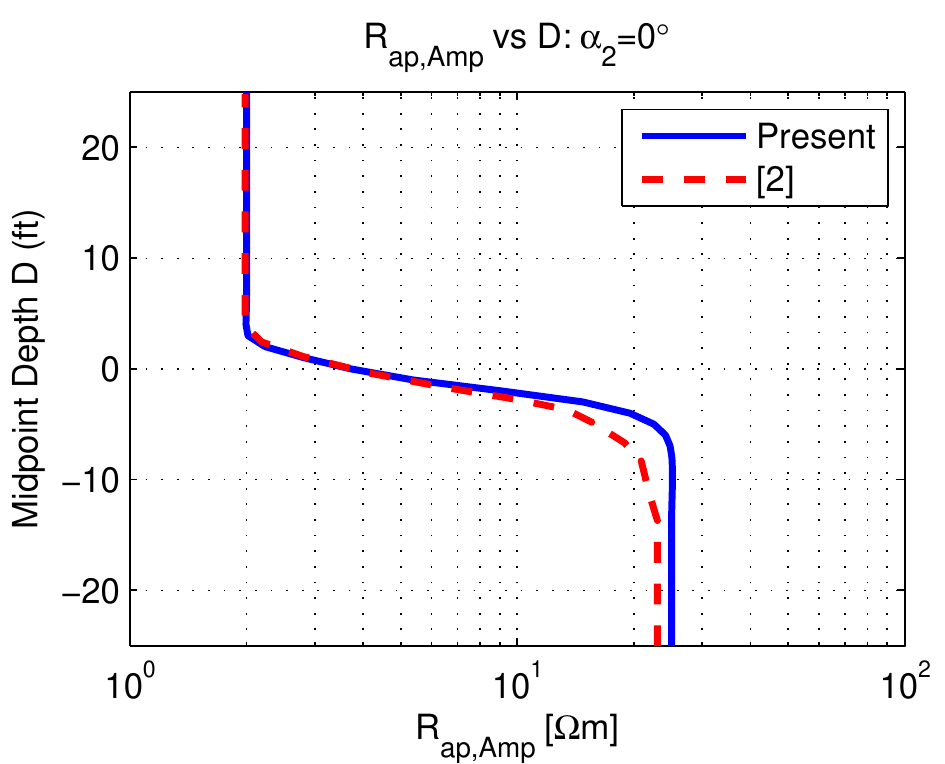}}
\subfloat[\label{AFig7b}]{\includegraphics[width=2.25in]{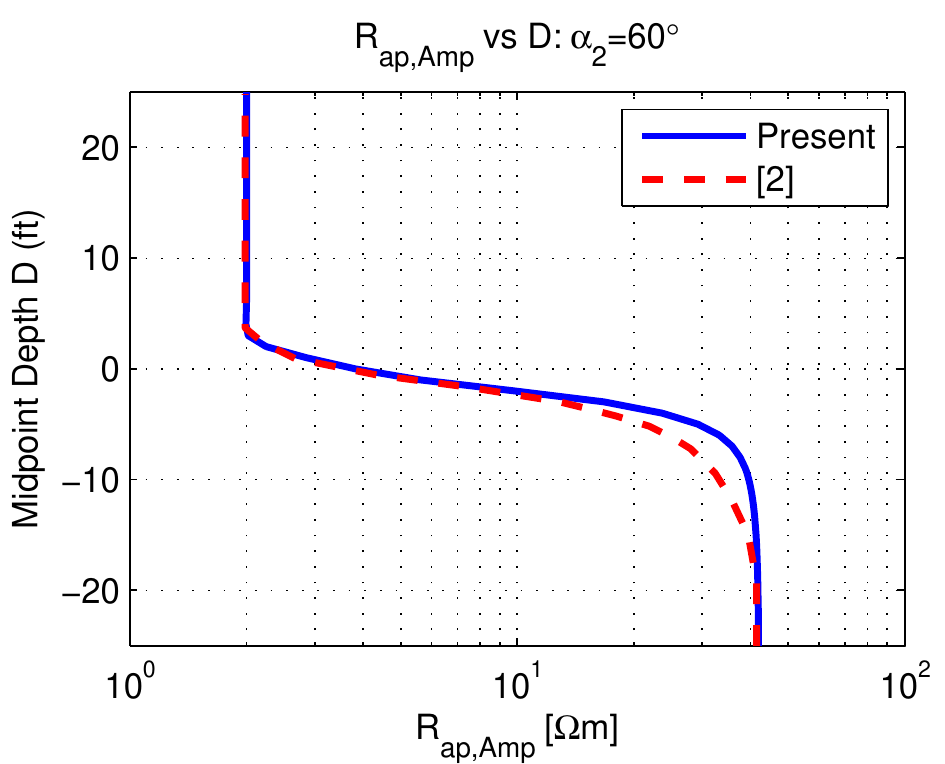}}
\caption{\label{AFig7}\small (Color online) Magnitude-apparent resistivity log comparison with Figure 7 of \cite{anderson1}: $\kappa_1=1,\kappa_2=\sqrt{20},R_{h1}=2\Omega \mathrm{m},R_{h2}=25\Omega \mathrm{m},\beta_2=0^{\circ},f=2\mathrm{MHz},L=40\mathrm{in},z_B=0\mathrm{ft}$.}
    \end{figure}

In Figure \ref{AFig7}, where there is a high resistivity contrast of $R_{h2}=12.5R_{h1}$, we notice a greater level of discrepancy. This is particularly so just beneath the interface at $z_B=0$ft, where the reflected fields are strongest. In the well-logging community, one refers to the phenomenon where conductive formations adversely reduce the apparent resistivity sensed in their resistive neighbors as the ``shoulder bed effect" \cite{chen}.

\begin{figure}[H]
\centering
\subfloat[\label{AFig8a}]{\includegraphics[width=2.25in]{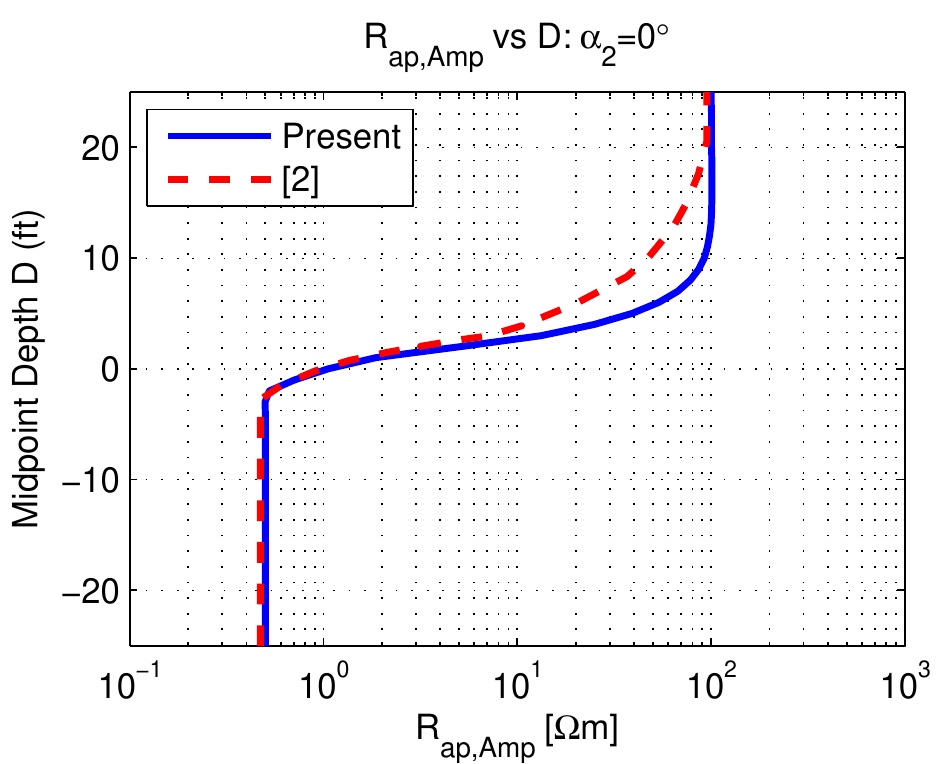}}
\subfloat[\label{AFig8b}]{\includegraphics[width=2.25in]{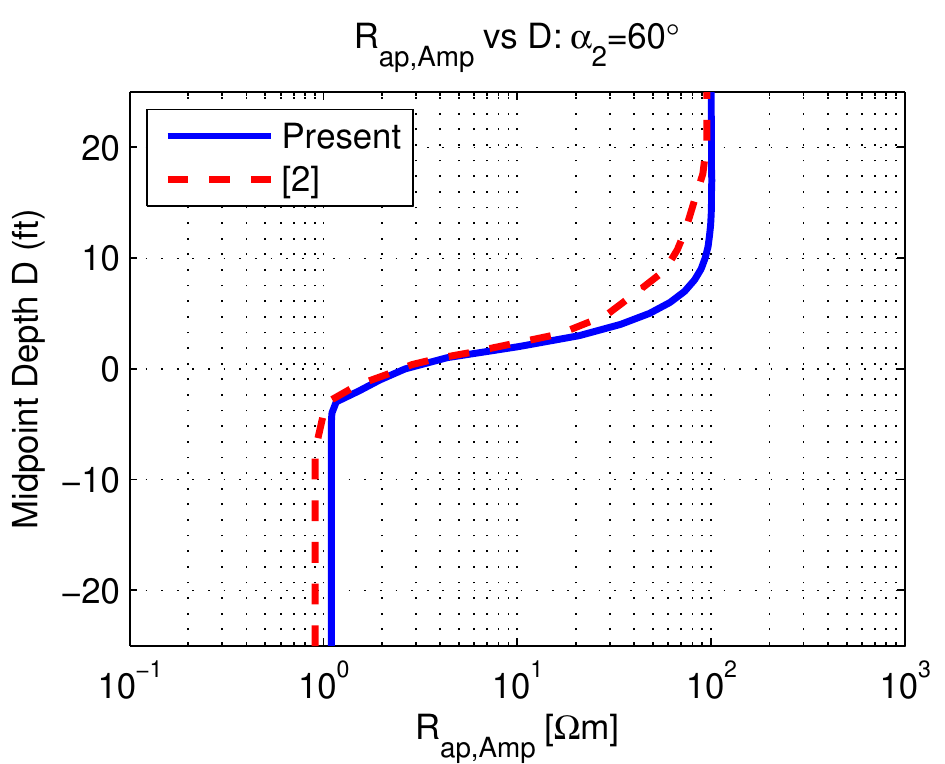}}
\caption{\label{AFig8}\small (Color online) Magnitude-apparent resistivity log comparison with Figure 8 of \cite{anderson1}: $\kappa_1=1,\kappa_2=\sqrt{20},R_{h1}=100\Omega \mathrm{m},R_{h2}=0.5\Omega \mathrm{m},\beta_2=0^{\circ},f=2\mathrm{MHz},L=40\mathrm{in},z_B=0\mathrm{ft}$.}
    \end{figure}

In Figure \ref{AFig8}, we again note a high resistivity contrast of $R_{h1}=200R_{h2}$. Comments dual to those made on Figure \ref{AFig7} apply here regarding (1) the resistivity log's notable deviation in the top isotropic region from the true resistivity of $100\Omega$m and (2) the greater disagreement versus \cite{anderson1}.

\begin{figure}[H]
\centering
\subfloat[\label{AFig9a}]{\includegraphics[width=2.25in]{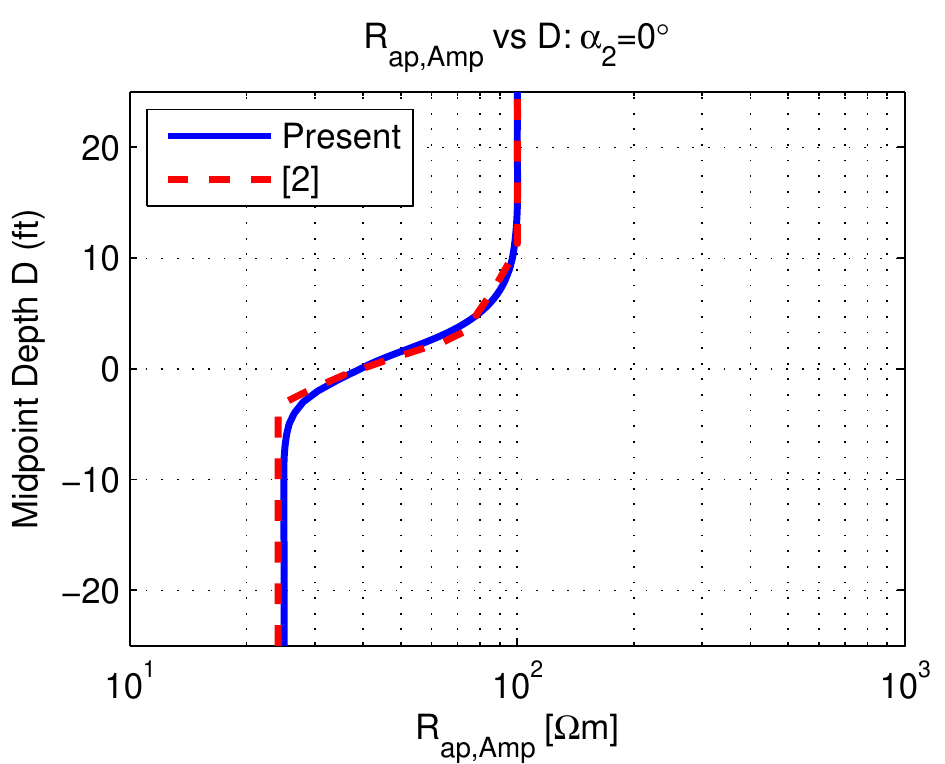}}
\subfloat[\label{AFig9b}]{\includegraphics[width=2.25in]{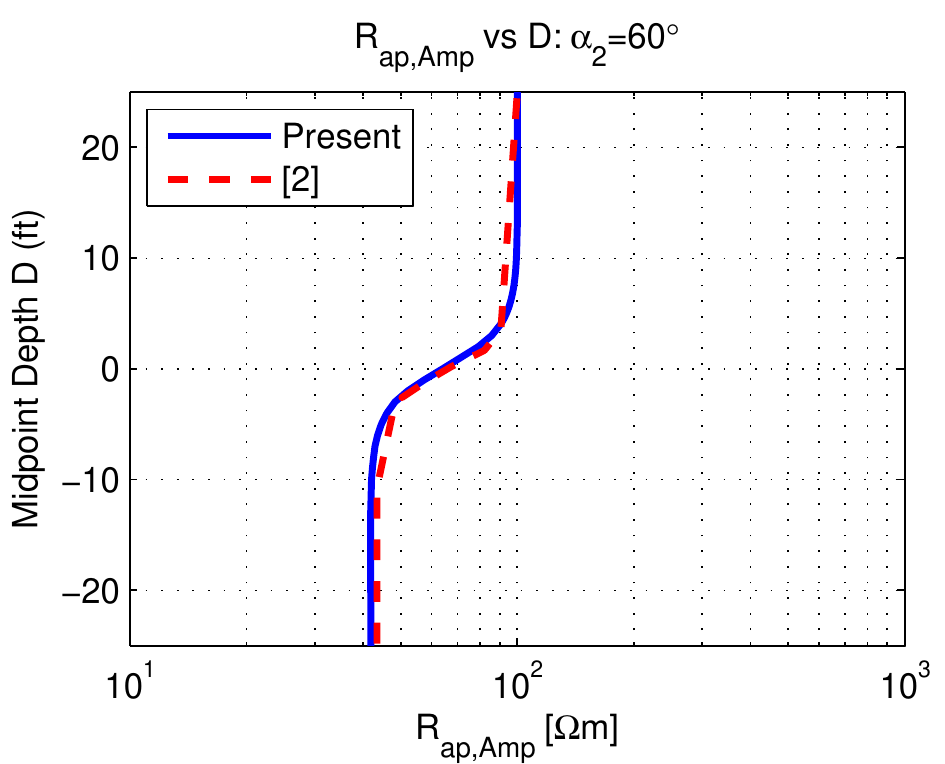}}
\caption{\label{AFig9}\small (Color online) Magnitude-apparent resistivity log comparison with Figure 9 of \cite{anderson1}: $\kappa_1=1,\kappa_2=\sqrt{20},R_{h1}=100\Omega \mathrm{m},R_{h2}=25\Omega \mathrm{m},\beta_2=0^{\circ},f=2\mathrm{MHz},L=40\mathrm{in},z_B=0\mathrm{ft}$.}
    \end{figure}

In Figure \ref{AFig9}, the resistivity contrast is low ($R_{h1}=4R_{h2}$). Like in Figure \ref{AFig6}, we note that there is excellent agreement.

\begin{figure}[H]
\centering
\subfloat[\label{AFig11a}]{\includegraphics[width=2.25in]{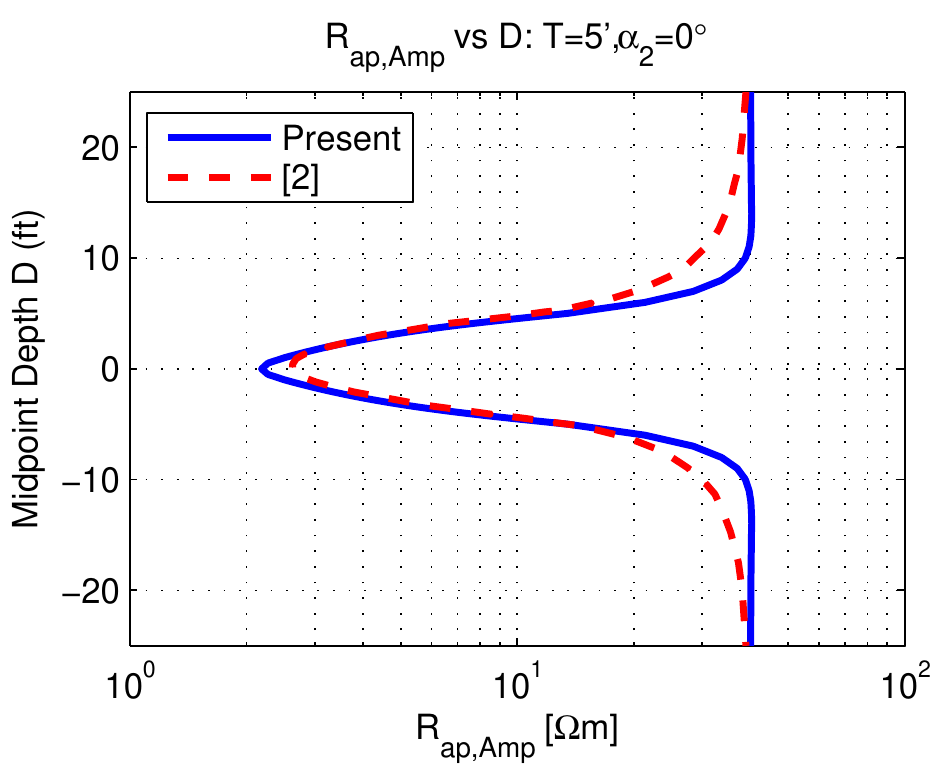}}
\subfloat[\label{AFig11b}]{\includegraphics[width=2.25in]{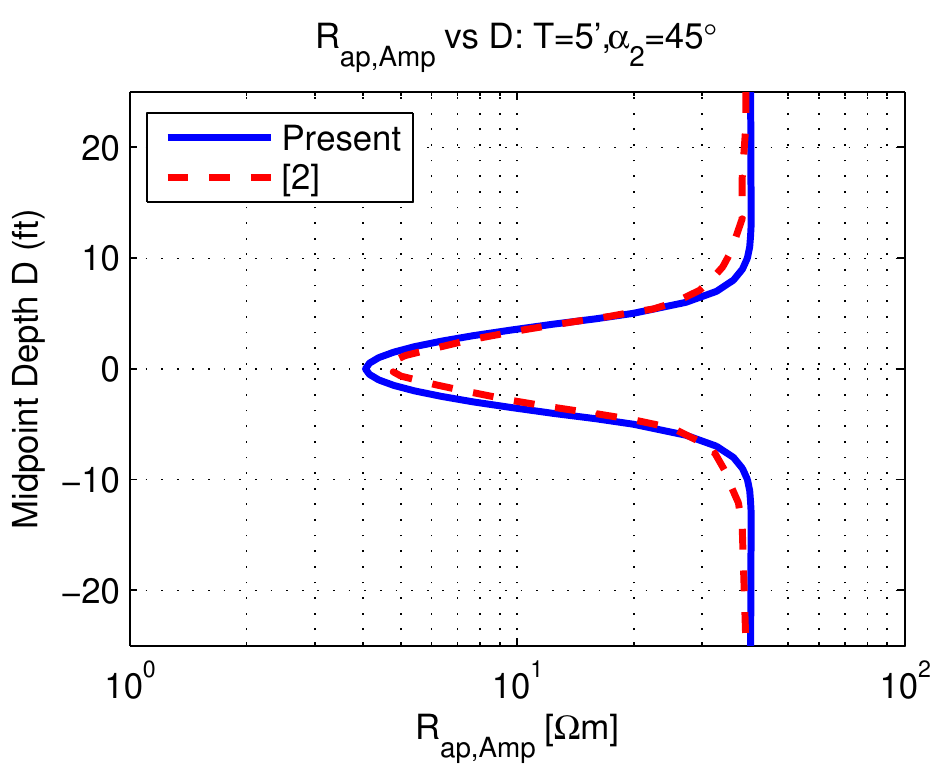}}

\subfloat[\label{AFig11c}]{\includegraphics[width=2.25in]{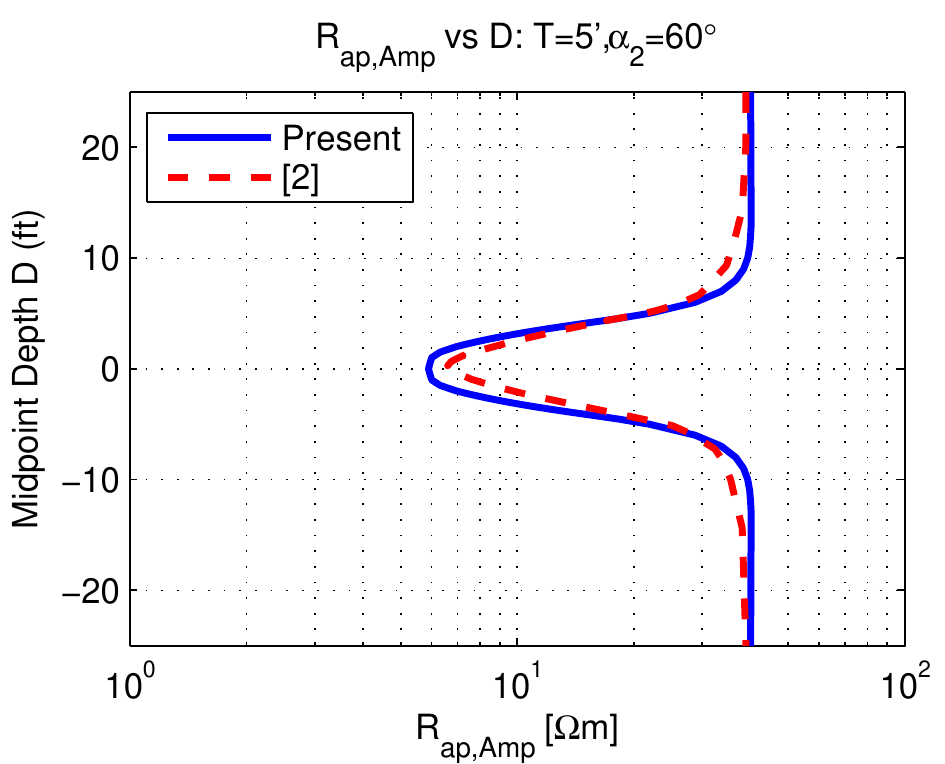}}
\subfloat[\label{AFig11d}]{\includegraphics[width=2.25in]{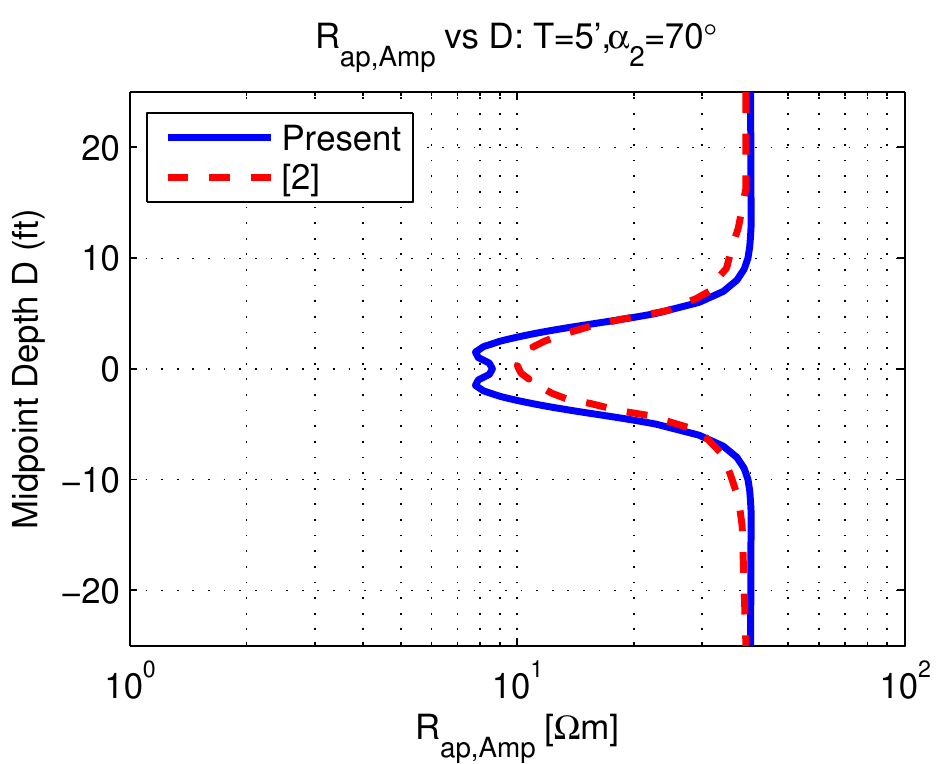}}

\subfloat[\label{AFig11e}]{\includegraphics[width=2.25in]{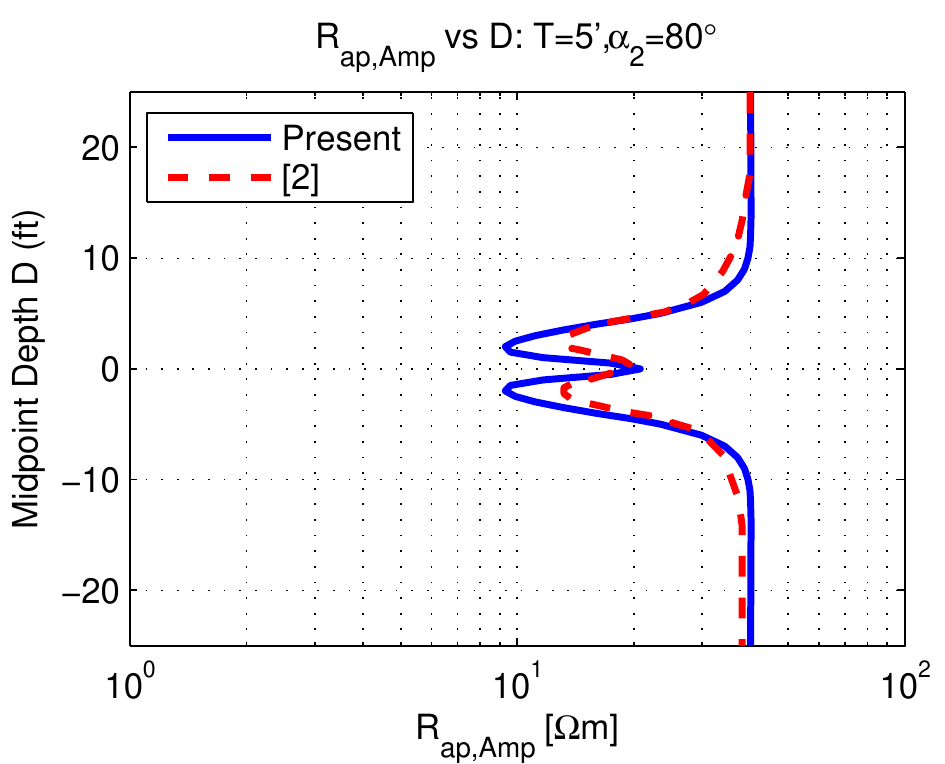}}
\subfloat[\label{AFig11f}]{\includegraphics[width=2.25in]{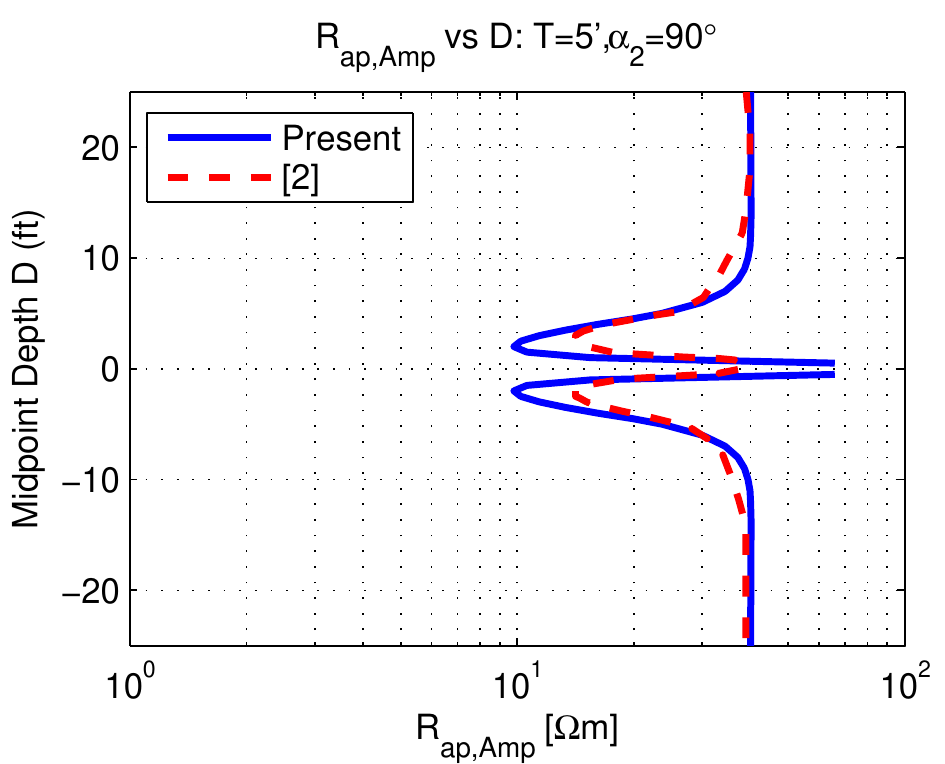}}
\caption{\label{AFig11}\small (Color online) Magnitude-apparent resistivity log comparison with Figure 11 of \cite{anderson1}: $\kappa_1=\kappa_3=1,\kappa_2=5,R_{h1}=R_{h3}=40\Omega \mathrm{m},R_{h2}=2\Omega \mathrm{m},\beta_2=0^{\circ},f=2\mathrm{MHz},L=40\mathrm{in},z_B=\left \{2.5,-2.5\right \}\mathrm{ft}$.}
    \end{figure}

\begin{figure}[H]
\centering
\subfloat[\label{AFig13a}]{\includegraphics[width=2.25in]{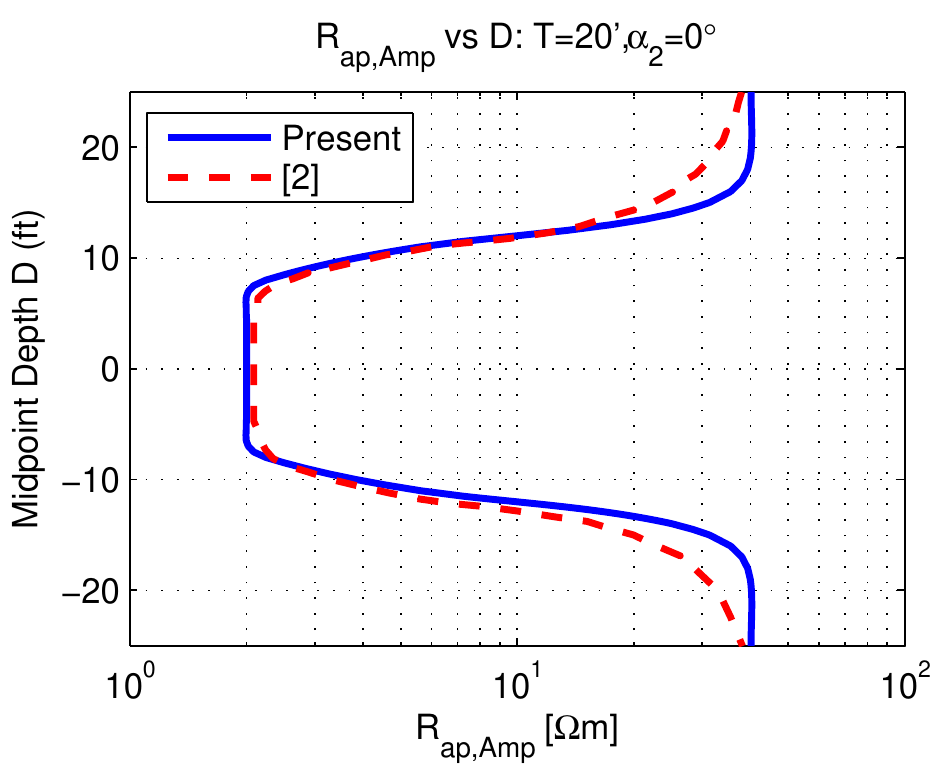}}
\subfloat[\label{AFig13b}]{\includegraphics[width=2.25in]{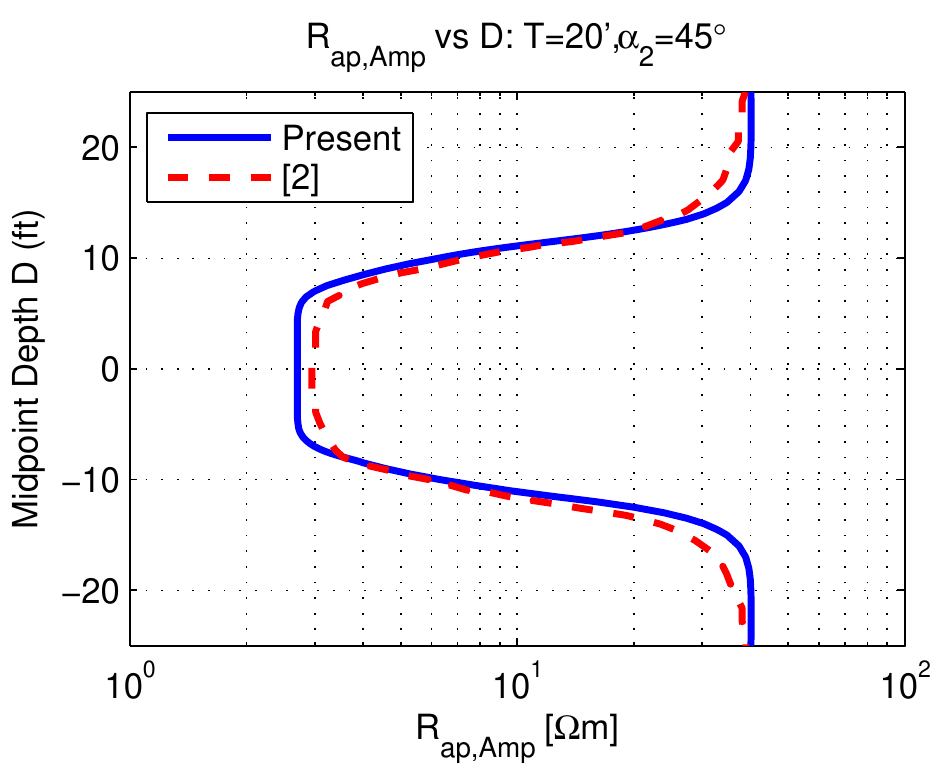}}

\subfloat[\label{AFig13c}]{\includegraphics[width=2.25in]{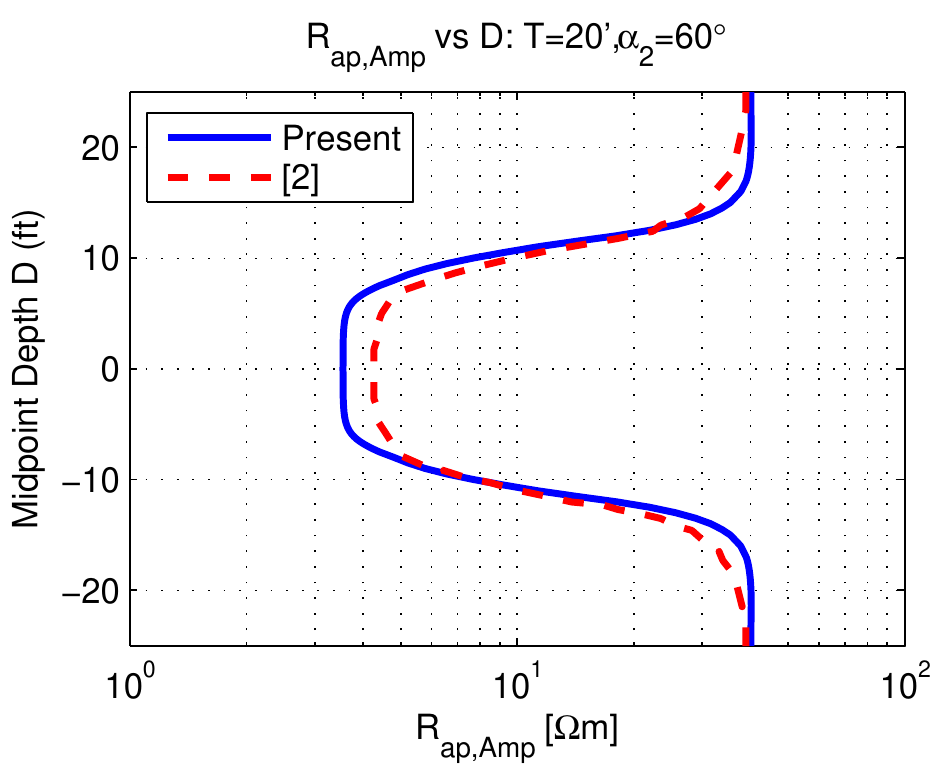}}
\subfloat[\label{AFig13d}]{\includegraphics[width=2.25in]{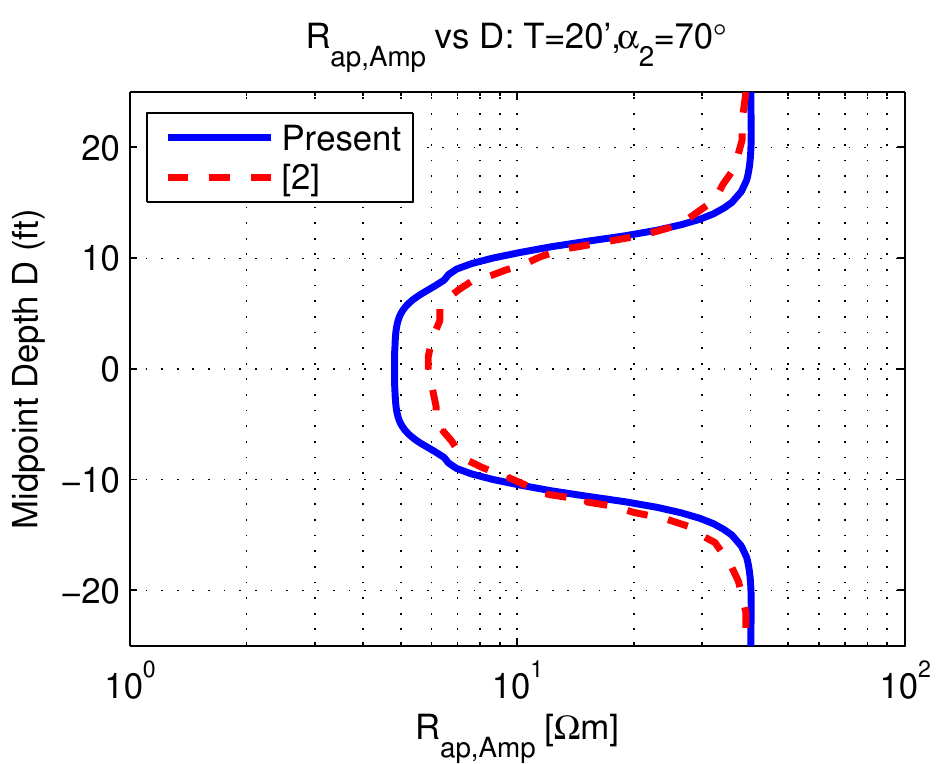}}

\subfloat[\label{AFig13e}]{\includegraphics[width=2.25in]{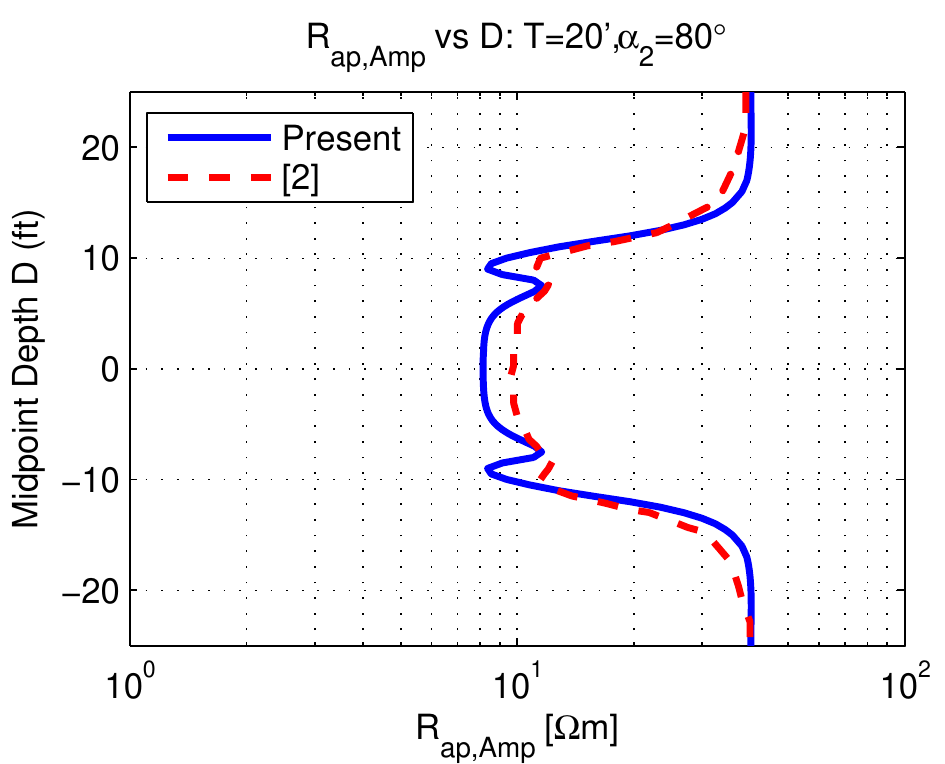}}
\subfloat[\label{AFig13f}]{\includegraphics[width=2.25in]{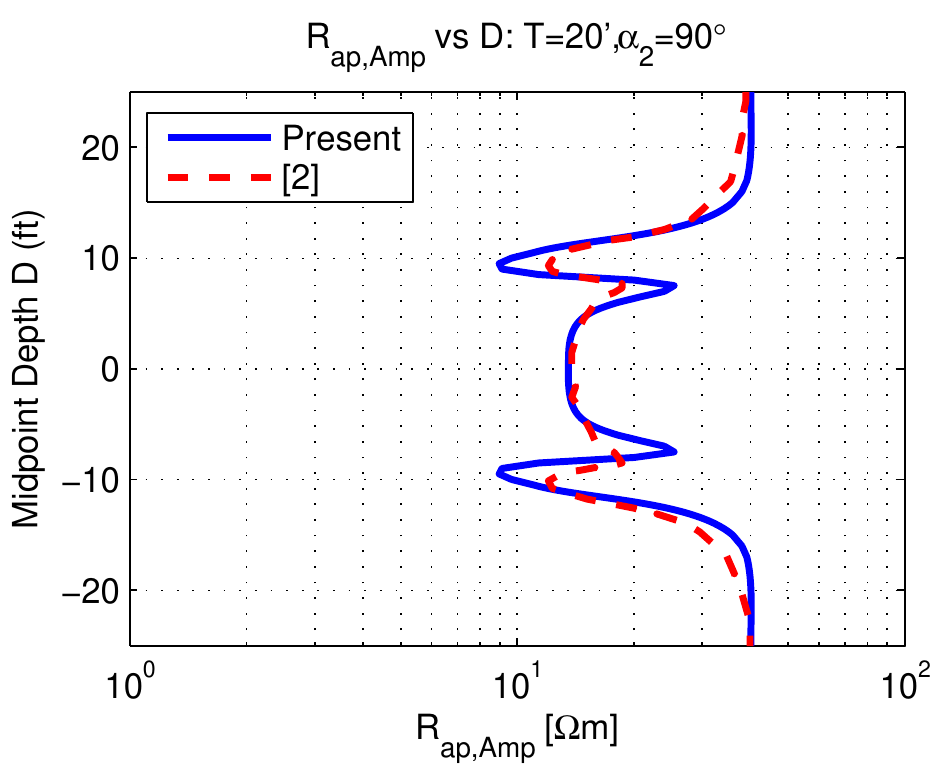}}
\caption{\label{AFig13}\small (Color online) Magnitude-apparent resistivity log comparison with Figure 13 of \cite{anderson1}: $\kappa_1=\kappa_3=1,\kappa_2=5,R_{h1}=R_{h3}=40\Omega \mathrm{m},R_{h2}=2\Omega \mathrm{m},\beta_2=0^{\circ},f=2\mathrm{MHz},L=40\mathrm{in},z_B=\left \{10,-10\right \}\mathrm{ft}$.}
    \end{figure}

Now we comment upon Figures \ref{AFig11}-\ref{AFig13}. The data from \cite{anderson1} suggest a very strong shoulder bed effect present in the top and bottom isotropic half-spaces when $\alpha_2=0^\circ$, leading to notable disagreement for Figures \ref{AFig11a} and \ref{AFig13a}. There is also notable discrepancy in modeling the formation interface ``horns" and resistivity valleys (see, in particular, the infinite-resistivity spike in Figure \ref{AFig11f}). However, the data sets in \cite{anderson1} are not free of infinite-resistivity spikes either (see Figures 14 and 23 in \cite{anderson1}), suggesting that the resistivity inversion and data post-processing methods used (and their differences between here and \cite{anderson1}) are causing the observed discrepancies. These quantitative discrepancies aside, however, we notice excellent \emph{qualitative} agreement in modeling the shoulder bed effect, as well as the interface horns and valleys due to the high-dipping-angle uniaxial bed.
\vspace{-5pt}
\newpage
\FloatBarrier
    \vspace{-20pt}
\FloatBarrier
    \vspace{-20pt}
\newpage
\subsection{Dipole Fields Near a PEC-Backed Microwave Substrate}
The last validation result concerns  a $y$-directed Hertzian electric dipole on top of a dielectric substrate supported by a metallic ground plane~\cite{kong}). The ground is modeled as a semi-infinite layer with conductivity $\sigma=10^9$ S/m. We compute the radiated $H_x$, $H_z$, and $E_y$ components. This environment is meant to highlight the algorithm's ability to simulate magnetic fields produced by an electric, rather than magnetic source and thus (from the duality theorem) its ability to compute magnetic \emph{and} electric fields from both electric \emph{and} magnetic sources. By simulating a case with $4\lambda_o \leq |x-x'| \leq 14\lambda_o$ ($\lambda_o = 37.5$m), we also provide here an example of the general-purpose nature of the algorithm in regards to the $\bold{r}-\bold{r}'$ geometry. We emphasize that this flexibility is primarily attributed to the \emph{adaptive} extension of the original MMA as discussed in Section \ref{sec3}.

Figure \ref{KongFig1_Hz} below shows excellent agreement, in the range $4.25\lambda_o \leq |x-x'| \leq 13.6\lambda_o$, with the available data from~\cite{kong}. Figures \ref{KongFig1_Hx} and \ref{KongFig1_Ey} shows similar results for the other two components. The oscillatory behavior results from interference effects caused by the ground plane. To facilitate easier comparison with \cite{kong} and exhibit the three field magnitude variations on identical scales, all three data sets were scaled such that their maximum magnitudes correspond to the maximum magnitude of $H_z$ in \cite{kong}.
\begin{figure}[H]
\centering
\subfloat[\label{KongFig1_Hx}]{\includegraphics[width=2.25in]{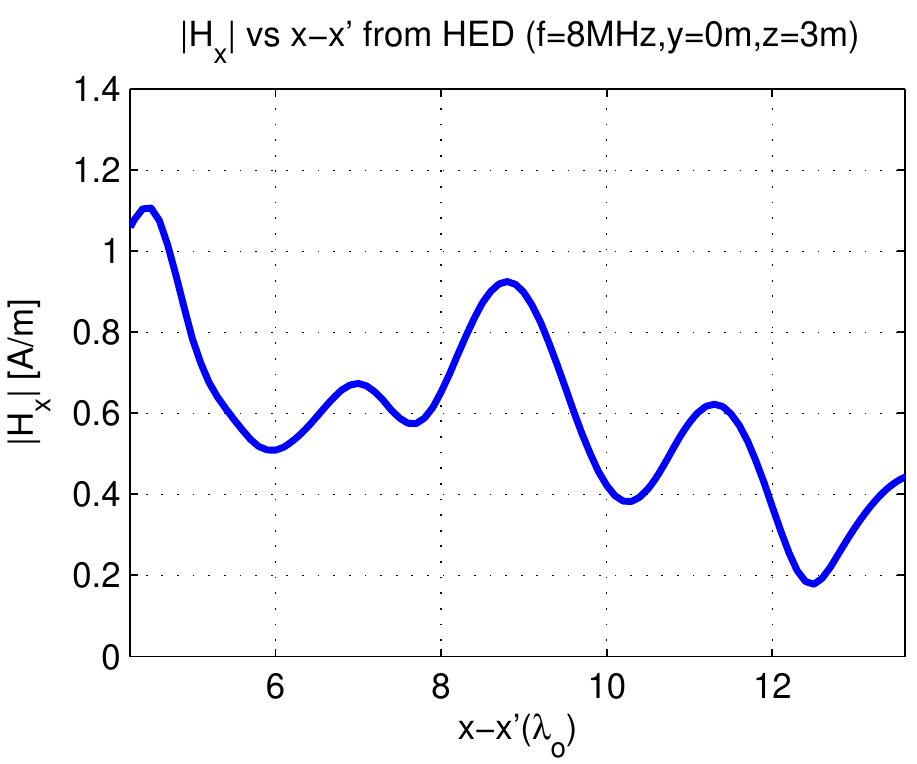}}
\subfloat[\label{KongFig1_Hz}]{\includegraphics[width=2.25in]{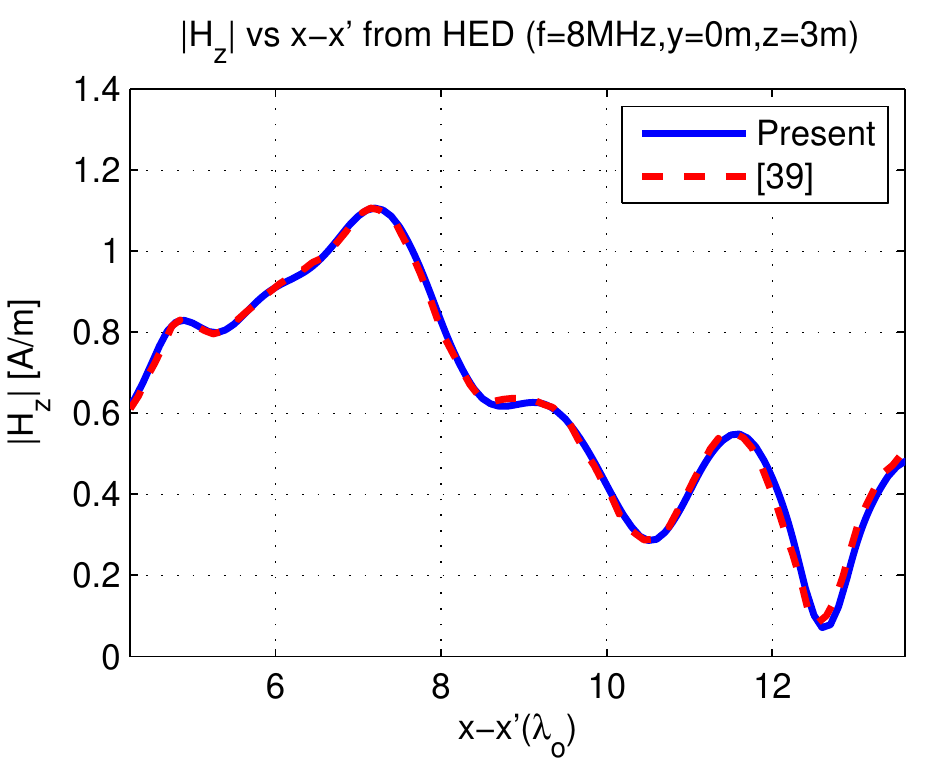}}
\subfloat[\label{KongFig1_Ey}]{\includegraphics[width=2.25in]{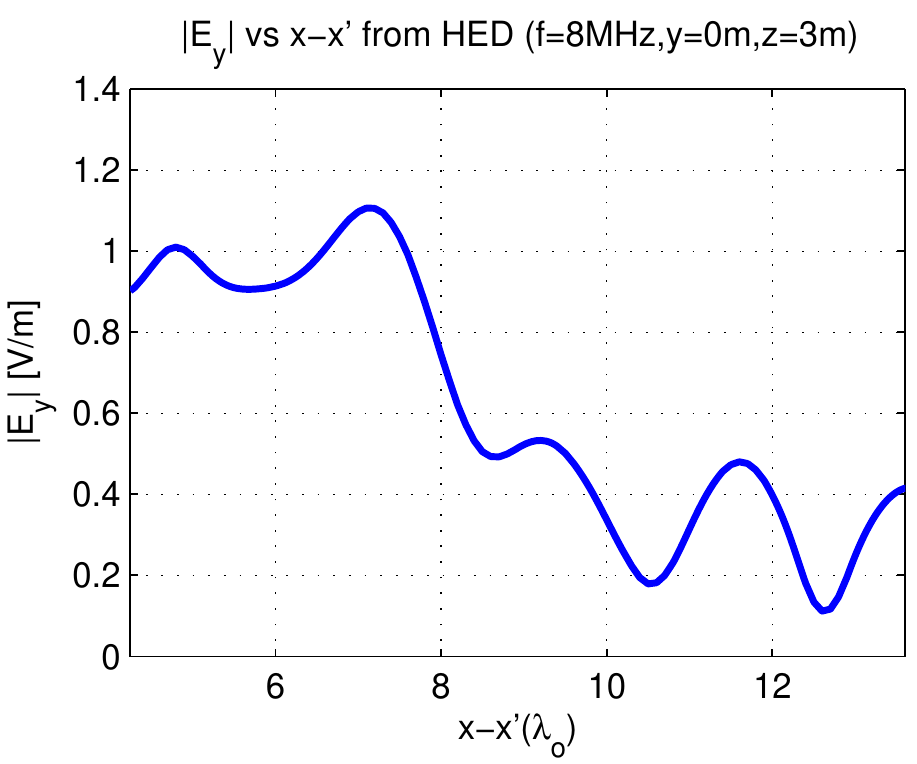}}
\caption{\label{KFig1}\small (Color online) Field component intensities from a $y$-directed Horizontal Electric Dipole (HED), which is radiating at $f=8$MHz ($\lambda_o=37.5$m), centered at the origin, and supported on a grounded dielectric substrate 4$\lambda_o$ thick with free space above. The substrate's dielectric constant is $\epsilon_r=3.3(1+0.01i)$. Only $|H_z|$ reference data was published in \cite{kong}.}
    \end{figure}
\section{Convergence Characteristics}
To characterize our numerical formulation's ability to converge towards the field solution, we present two case studies concerning the $z$-directed magnetic field component $H_z$ produced by a $z$-directed magnetic dipole radiating at $f$=2MHz in free space. The first case comprises a benign scenario in which $\bold{r}-\bold{r}'=(1,1,1)$m, while the second case represents a much more challenging scenario where $\bold{r}-\bold{r}'=(500,500,1)$m in which the integrand oscillates on the order of 500 times more rapidly than the first case. For both cases, we choose $x-x'=y-y'$ to ensure the code faces the same convergence challenges when evaluating both the $k_x$ and $k_y$ integrals. Furthermore, we set the pre-extrapolation region magnification factor $l_o$ (see section \ref{sec4a}) equal to ten and artificially set $\xi_1=2P_k$ to facilitate characterization of the interval sub-division factor $h$, with which one quantifies the sub-interval lengths after full interval sub-division, as $h \xi_1$.

For each case, we present results related to both the pre-extrapolation and extrapolation domain characteristics. To avoid mixing the numerical formulation's handling of the pre-extrapolation and extrapolation region sections of the $k_x$ and $k_y$ integration paths, the ``pre-extrapolation domain" (termed ``Region 1" below) refers to the region $(-\xi_1<k_x'<\xi_1) \cup (-\xi_1<k_y'<\xi_1)$. Similarly, the ``extrapolation domain" (termed ``Region 2" below) refers to the region $(k_x' > \xi_1) \cup (k_x' < -\xi_1) \cup (k_y' > \xi_1) \cup (k_y' < -\xi_1)$. Since one cannot obtain closed-form solutions to the pre-extrapolation and extrapolation domain contributions, reference field values from which one measures accuracy must be chosen; their computation details are provided in Figures \ref{ConvFig}-\ref{ConvFig2} below.

For the pre-extrapolation domain study, we exhibit the accuracy obtained versus ($h$) and the Patterson-Gauss quadrature order ($p$) used to integrate each sub-interval. We notice that for both cases, there is the expected increase in accuracy both as one reduces $h$ and increases $p$.

For the extrapolation domain contribution, we make the typical assumption \cite{mosig1,mich2,mich1} that the integrand is well-behaved in this portion of the spectral domain and thus do not perform interval sub-division. Instead, we set the $k_x$ and $k_y$ plane extrapolation region interval lengths to be half the spectral period of the Fourier kernels exp[$ik_x(x-x')$] and exp[$ik_y(y-y')$] (resp.), as suggested in \cite{mich2}, and examine the variation of accuracy versus the number of extrapolation region intervals employed ($B$) and the Legendre-Gauss quadrature order used ($LGQ$) to integrate each interval\footnote{$B$ intervals are used in both the $k_x'>0$ and $k_x'<0$ integration path half-tails; the same applies for the $k_y$ path half-tails.}. For the extrapolation domain field contribution, we notice that as one increases $LGQ$ and $B$ there is the expected decay in error. In particular, for small $B$ ($B \sim 3$ for both cases) we notice that tail integral truncation effects dominate the region two error. On the other hand, after a certain value of $B$ ($B \sim 10$ for case 1 and $B \sim 6$ for case 2), we find that aliasing/sampling effects dominate the error.

Note that for all figures below, errors below -150dB were coerced to equal -150dB. This is because error levels below approximately -150dB do not represent error levels attained due to the convergence characteristic of the formulation itself, but instead represent instances wherein the given and reference answers are equal in all the digits \emph{available} using finite, double-precision arithmetic.
\begin{figure}[H]
\centering
\subfloat[\label{Case1PreExt}]{\includegraphics[width=3.25in]{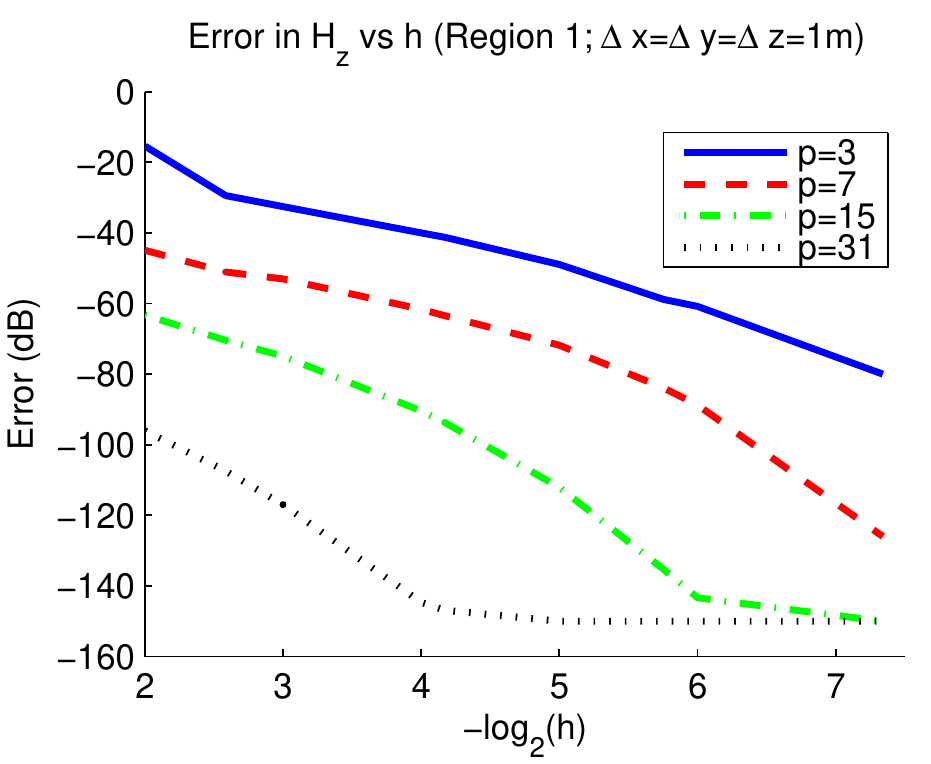}}
\subfloat[\label{Case2PreExt}]{\includegraphics[width=3.25in]{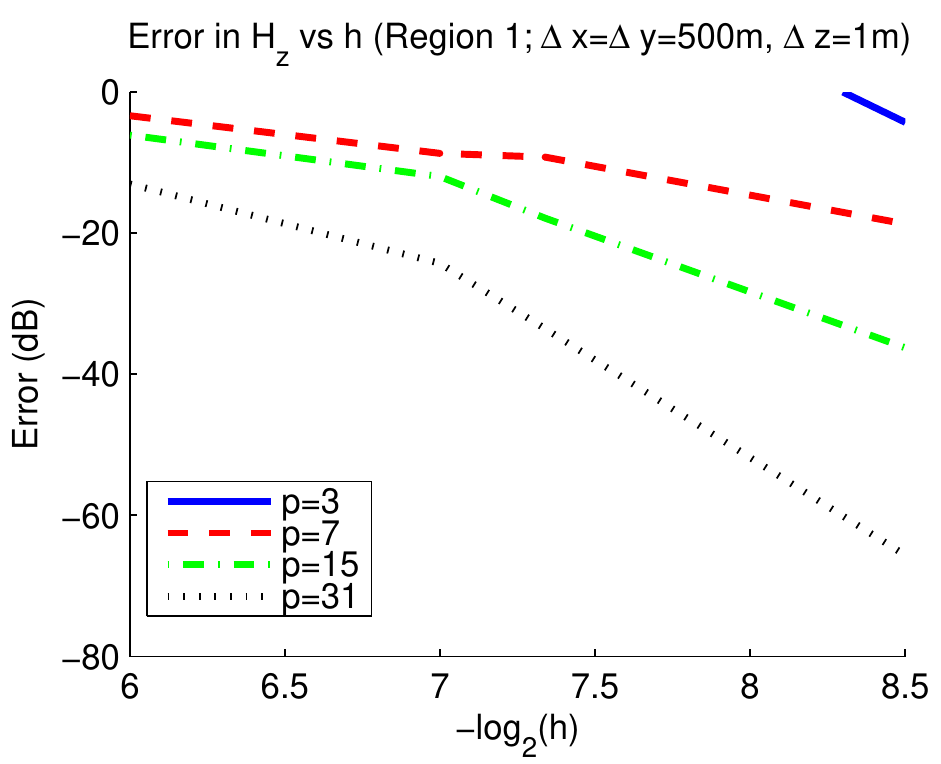}}
\caption{\label{ConvFig}\small (Color online) Convergence towards the solution comprising the field contribution from ``Region 1". The reference field values are computed using $p$=31 for both figures, as well as \\ -log$_2$(h)=9 for Figure \ref{Case1PreExt} and -log$_2$(h)=11 for Figure \ref{Case2PreExt}. The reference field values computed for Figures \ref{Case1PreExt} and \ref{Case2PreExt} use different $h$ because in the latter scenario, $H_z$ converges more slowly and thus necessitates smaller $h$ values in the non-reference field results to show a meaningful decay in error. As a result, one also requires an even smaller $h$ for the reference field result from which the relative error is computed.}
    \end{figure}
\begin{figure}[H]
\centering
\subfloat[\label{Case1Ext}]{\includegraphics[width=3.25in]{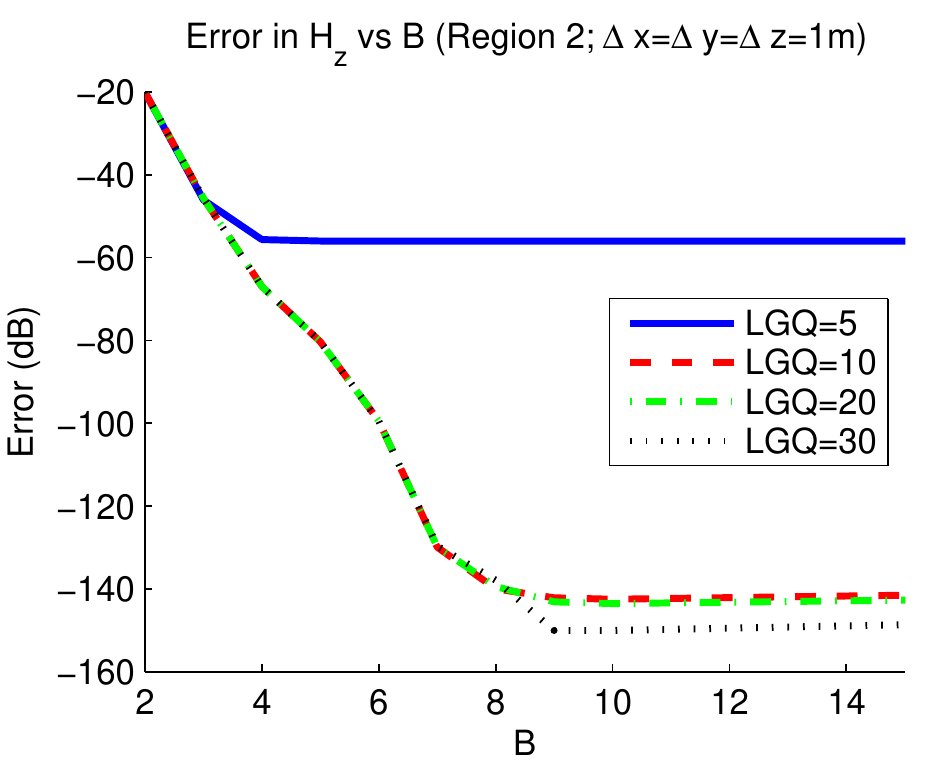}}
\subfloat[\label{Case2Ext}]{\includegraphics[width=3.25in]{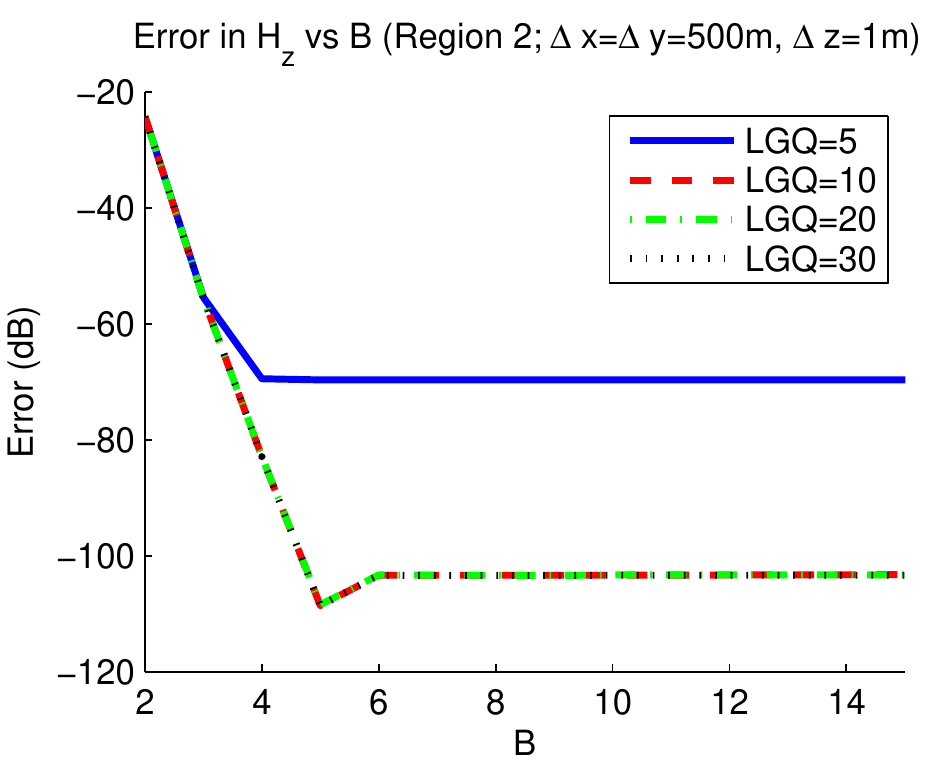}}
\caption{\label{ConvFig2}\small (Color online) Convergence towards the solution comprising the field contribution from ``Region 2". The reference field values are computed using LGQ=30 for both figures, as well as $B=150$ for Figure \ref{Case1Ext} and $B=1000$ for Figure \ref{Case2Ext}. The reference field values computed for Figures \ref{Case1Ext} and \ref{Case2Ext} use different $B$. This is because in the latter scenario, as can be observed, $H_z$ converges more slowly; indeed, while $H_z$ in case two \emph{levels off} more rapidly than in case one, it fails to reach accuracy near to machine precision within the same range of $B$ exhibited for both cases. Thus similar reasoning applies as that behind using smaller $h$ for the reference and non-reference field results in Figure \ref{Case2PreExt} (versus Figure \ref{Case1PreExt}).}
    \end{figure}
    \newpage
\section{Conclusion}
We have presented a general-purpose and efficient pseudo-analytical formulation to compute electromagnetic fields from dipole sources in planar-stratified environments with \emph{arbitrary} anisotropy, loss, and $\bold{r}-\bold{r}'$ geometries. The formulation is based on embedding spectral Green's Function kernels within Fourier-type integrals to compute the space-domain fields. Some of the salient features that are combined here to yield a robust algorithm are: (a) judicious selection of a numerically robust integration path, (b) re-casting of critical formulae to facilitate accurate field computations and obviate numerical overflow, (c) adaptive integration along the pre-extrapolation region of the integrals, and (d) \emph{adaptive} extension of the original MMA, applied to environments containing media with anisotropy and loss, both to accelerate the tail integral's convergence and to endow error control to its evaluation. The formulation's accuracy has been validated through four sets of numerical data and its convergence properties characterized.

\begin{acknowledgements}

We thank Halliburton Energy Services for the permission to publish this work. In particular, the authors would like to thank Dr. Paul Rodney for his review and comment on this paper.

\end{acknowledgements}
\bibliographystyle{apsrev4-1}
\bibliography{reflist_APS}
\appendix*

\newpage
\appendix
\section{Conventions and Notation}
The following definitions are used throughout this paper:
\begin{enumerate}
\item $i$ is the unit-magnitude imaginary number.
\item $c_o$ is the speed of light in free space.
\item \(\mu_o \ \mathrm{[H/m]} \) is the free space magnetic permeability.
\item \(\epsilon_o = \frac{1}{\mu_o c_o^2 } \ \mathrm{[F/m]} \) is the free space electric permittivity.
\item \(\omega=2 \pi f \ \mathrm{[rad/sec]}\) is the angular frequency at which the source radiates.
\item $ \boldsymbol{\bar{\epsilon}}_c= \mathrm{Re}\left(\boldsymbol{\bar{\epsilon}}_c\right)+i\mathrm{Im}\left(\boldsymbol{\bar{\epsilon}}_c\right)$ is the 3 $\times$ 3 complex permittivity tensor.
\item  \(\boldsymbol{\bar{\epsilon}}_r= \boldsymbol{\bar{\epsilon}}_c/\epsilon_o \) is the relative permittivity tensor.
\item $\boldsymbol{\bar{\mu}}_c= \mathrm{Re}\left(\boldsymbol{\bar{\mu}}_c\right)+i\mathrm{Im}\left(\boldsymbol{\bar{\mu}}_c\right)$ is the 3 $\times$ 3 complex permeability tensor.
\item \(\boldsymbol{\bar{\mu}}_r= \boldsymbol{\bar{\mu}}_c/\mu_o\) is the relative magnetic permeability tensor.
\item \(k_o = \omega \sqrt{\epsilon_o \mu_o} \ \mathrm{[m^{-1}]} \) is the wave number of free space.
\item $\eta_o=\sqrt{\mu_o/\epsilon_o} \ \mathrm{[\Omega]}$ is the characteristic wave impedance of free space.
\item $\lambda_o=\frac{c_o}{f}$ is the free-space wavelength corresponding to $f$.
\item The time convention exp($-i \omega t$) is assumed and suppressed.
\item $\bold{r}=(x,y,z)$ is the observation location.
\item $\bold{r}'=(x',y',z')$ is the source location.
\item $\delta\left(\bold{r}-\bold{r}'\right)=\delta\left(x-x'\right)\delta\left(y-y'\right)\delta\left(z-z'\right)$ is the Dirac delta function.
\item $u(\cdot)$ represents the Heaviside unit step function.
\end{enumerate}
Note that individual material tensor components mentioned refer to \emph{relative} values. We also make some notational comments (exceptions are defined when they arise):
\begin{enumerate}
\item Vector, matrix, and tensor quantities have boldface script.
\item Unit-magnitude vectors have an over-hat symbol.
 \item Matrix/tensor quantities have an over-bar.
\item Field quantities (besides $\bold{k}$) exhibiting purely spectral dependence have an over-tilde and are denoted spectral quantities.
\item Field quantities with ($k_x,k_y,z$) are denoted mixed-domain quantities.
\item Field quantities exhibiting purely spatial dependence have calligraphic script and are denoted spatial quantities.
\item $z$ can refer to the observation depth or the coordinate, depending on context.
\item Modal (non-modal) field quantities appear in lower (upper) case.
\item The up-going mode eigenvectors and eigenvalues are block-represented as $\bold{\tilde{\bar{S}}}_m^+=\begin{bmatrix} \bold{\tilde{\hat{s}}}_{m,1} & \bold{\tilde{\hat{s}}}_{m,2}\end{bmatrix}$ and $\bold{\bar{\Lambda}}_m^+(z)=\exp \left(\mathrm{diag}\left[i\tilde{k}_{m,1z}z,  \, i\tilde{k}_{m,2z}z \right] \right)$ (resp.).
\item The down-going mode eigenvectors and eigenvalues are block-represented as $\bold{\tilde{\bar{S}}}_m^-=\begin{bmatrix} \bold{\tilde{\hat{s}}}_{m,3} & \bold{\tilde{\hat{s}}}_{m,4}\end{bmatrix}$ and $\bold{\bar{\Lambda}}_m^-(z)=  \exp \left(\mathrm{diag}\left[i\tilde{k}_{m,3z}z, \, i\tilde{k}_{m,4z}z \right] \right)$ (resp.).
\item Fields ``phase-referenced" to $z^*$ possess a exp($i\tilde{k}_z(z-z^*)$) $z$-dependence.
\item The $n \times n$ identity matrix is denoted $\bold{\bar{I}}_n$.
\end{enumerate}
    \end{document}